%% file: main.tex
\documentclass[letterpaper,twocolumn,10pt]{article}
\usepackage{usenix-2020-09}

\usepackage{tikz}
\usepackage{amsmath,amssymb}

\usepackage{filecontents}

\usepackage{nicefrac}
\usepackage{siunitx}
\usepackage{array,framed}
\usepackage{booktabs}
\usepackage{
  color,
  float,
  epsfig,
  wrapfig,
  graphics,
  graphicx,
  subcaption
}
\usepackage{setspace}
\usepackage{latexsym,fancyhdr,url}
\usepackage{enumerate}
\usepackage[algo2e]{algorithm2e}
\usepackage{algorithmic}
\usepackage{algorithm}
\usepackage{graphics}
\usepackage{xparse} 
\usepackage{xspace}
\usepackage{multirow}
\usepackage{csvsimple}
\usepackage{balance}
\usepackage{mathrsfs}
\usepackage{amsfonts}
\usepackage{epsfig}
\usepackage{pifont}

\usepackage{enumitem}
\usepackage{mathtools}
\usepackage{tikz}
\usepackage{pgfplots} 
\usepackage{pgfplotstable}
\pgfplotsset{compat = newest}
\usetikzlibrary{arrows}
\usetikzlibrary{patterns}
\usepackage{stackrel}
\usepackage{pstricks}
\usepackage{subcaption}
\usepackage{wrapfig}
\usepackage[utf8]{inputenc}
\usepackage{booktabs}
\usepackage{expl3}
\usepackage{collcell}
\usepackage{colortbl} 
\usepackage{xurl}
\usepackage{hyperref}
\usepackage{rotating}
\usepackage{soul}
\usetikzlibrary{
  shapes.geometric,
  arrows,
  external,
  pgfplots.groupplots,
  matrix
}

\DeclareMathAlphabet{\mathcal}{OMS}{cmsy}{m}{n}

\DeclareGraphicsExtensions{%
    .png,.PNG,%
    .pdf,.PDF,%
    .jpg,.mps,.jpeg,.jbig2,.jb2,.JPG,.JPEG,.JBIG2,.JB2}

\input{macros}
\begin{document}

\date{}

\title{\Large \bf The Impact of Exposed Passwords on Honeyword Efficacy}

\author{
{\rm Zonghao Huang}\\
Duke University
\and
{\rm Lujo Bauer}\\
Carnegie Mellon University
\and
{\rm Michael K. Reiter}\\
Duke University
} 

\maketitle

\input{abstract.tex}

\input{introduction.tex}


\input{related_work.tex}


\input{background.tex}


\input{user_chosen.tex}

\input{alg_generated.tex}


\input{discussion.tex}

\input{conclusion.tex}


\section*{Acknowledgements}
The work described in this paper was supported in part by NSF grants 2113345 and 2131859.  
We are also grateful to Bhuwan Dhingra for helpful discussions.

\bibliographystyle{plain}
\bibliography{full, main}

\input{appendix.tex}


\end{document}

%% file: macros.tex
\newcommand{\myparagraph}[1]{\smallskip\noindent\textbf{#1}:\xspace}

\newcounter{note}[section]

\newcommand{\chkmrk}{\ding{52}\xspace}

\newcommand{\secref}[1]{\mbox{Sec.~\ref{#1}}\xspace}

\newcommand{\secrefstatic}[1]{\mbox{Sec.~{#1}}}
\newcommand{\figref}[1]{\mbox{Fig.~\ref{#1}}\xspace}

\newcommand{\figrefstatic}[1]{\mbox{Fig.~{#1}}}

\newcommand{\tblref}[1]{\mbox{Table~\ref{#1}}\xspace}

\newcommand{\appref}[1]{\mbox{App.~\ref{#1}}\xspace}
\newcommand{\eqnref}[1]{\mbox{(\ref{#1})}\xspace}

\DeclareMathOperator*{\avg}{avg}

\newcommand{\defeq}{\stackrel{\mathclap{\tiny\mbox{def}}}{=}}

\newlength{\figureheight}

\NewDocumentCommand{\genericNat}{ g }{\ensuremath{z\IfNoValueF{#1}{_{#1}}{}}\xspace}
\newcommand{\setSize}[1]{\ensuremath{\left|{#1}\right|}\xspace}

\NewDocumentCommand{\prob}{o g}{\ensuremath{\mathbb{P}\mathopen{}\IfNoValueTF{#1}{\left}{#1}({#2}\IfNoValueTF{#1}{\right}{#1})\mathclose{}}\xspace}

\newcommand{\cset}[3]{\ensuremath{#1\{}{#2}\ensuremath{\;#1|} \ifmmode{\;}\fi {#3}\ensuremath{#1\}}\xspace}
\newcommand{\cprob}[3]{\ensuremath{\mathbb{P}\mathopen{}#1({#2}\;#1|\;{#3}#1)\mathclose{}}\xspace}
\newcommand{\cexpv}[3]{\ensuremath{\mathbb{E}#1(}{#2}\ensuremath{\;#1|} \ifmmode{\;}\fi {#3}\ensuremath{#1)}\xspace}
\newcommand{\setNotation}[1]{\ensuremath{\MakeUppercase{#1}}\xspace}

\newcommand{\algNotation}[1]{\ensuremath{\mathcal{#1}}\xspace}
\newcommand{\arrComponent}[2]{\ensuremath{\big[{#1}\big]_{#2}}\xspace}
\newcommand{\arr}{\ensuremath{v}\xspace}
\newcommand{\arrIdx}{\ensuremath{j}\xspace}

\newcommand{\passwordDomain}{\ensuremath{\{0,1\}^{\ast}}\xspace}
\newcommand{\auxiliaryInfoDomain}{\ensuremath{\{0,1\}^{\ast}}\xspace}
\newcommand{\passwordSet}[1]{\ensuremath{\setNotation{S}_{#1}}\xspace}
\NewDocumentCommand{\password}{ g }{\ensuremath{p\IfNoValueF{#1}{_{#1}}}\xspace}
\newcommand{\passwordAlt}{\ensuremath{p'}\xspace}
\newcommand{\passwordAltAlt}{\ensuremath{p''}\xspace}
\newcommand{\passwordAltAltAlt}{\ensuremath{p'''}\xspace}
\newcommand{\auxiliaryInfo}{\ensuremath{\setNotation{x}}\xspace}

\newcommand{\honeywordSet}{\ensuremath{\setNotation{H}}\xspace}

\newcommand{\passwordGuessSet}{\ensuremath{\setNotation{G}}\xspace}
\newcommand{\pwdDistance}{\ensuremath{d}\xspace}

\newcommand{\breachAttacker}{\ensuremath{\algNotation{B}}\xspace}
\newcommand{\alarmAttacker}{\ensuremath{\algNotation{A}}\xspace}

\newcommand{\experiment}[2]{\ensuremath{\mathsf{Expt}^{#1}_{#2}}\xspace}
\newcommand{\falsePositiveExpt}{\textsc{fpp}\xspace}
\newcommand{\falseNegativeExpt}{\textsc{fnp}\xspace}
\newcommand{\falseNegativeProb}[1]{\ensuremath{\mathsf{FNP}_{#1}}\xspace}
\newcommand{\falsePositiveProb}[1]{\ensuremath{\mathsf{FPP}_{#1}}\xspace}
\newcommand{\fpExptWParams}{\ensuremath{\experiment{\falsePositiveExpt}{\choosePassword, \chooseHoneywords{\nmbrHoneywords}, \alarmThreshold, \attackBudget}}\xspace}
\newcommand{\fnExptWParams}{\ensuremath{\experiment{\falseNegativeExpt}{\choosePassword, \chooseHoneywords{\nmbrHoneywords}, \alarmThreshold}}\xspace}
\newcommand{\fpProbWParams}{\ensuremath{\falsePositiveProb{\choosePassword, \chooseHoneywords{\nmbrHoneywords}, \alarmThreshold, \attackBudget}}\xspace}
\newcommand{\fnProbWParams}{\ensuremath{\falseNegativeProb{\choosePassword, \chooseHoneywords{\nmbrHoneywords}, \alarmThreshold}}\xspace}
\newcommand{\fpProbWOParams}{\ensuremath{\falsePositiveProb{}}\xspace}
\newcommand{\fnProbWOParams}{\ensuremath{\falseNegativeProb{}}\xspace}
\NewDocumentCommand{\choosePassword}{ g }{\ensuremath{\algNotation{U}\IfNoValueF{#1}{_{#1}}}\xspace}
\newcommand{\chooseHoneywords}[1]{\ensuremath{\algNotation{H}_{#1}}\xspace}
\newcommand{\nmbrHoneywords}{\ensuremath{n}\xspace}
\newcommand{\alarmThreshold}{\ensuremath{\alpha}\xspace}
\newcommand{\attackBudget}{\ensuremath{\beta}\xspace}

\newcommand{\pwdAlphabet}{\ensuremath{\Sigma}\xspace}

\newcommand{\modelBatch}{\ensuremath{\setNotation{B}}\xspace}
\newcommand{\cosineSim}{\ensuremath{\mathsf{sim}}\xspace}
\newcommand{\emailAddr}{\ensuremath{\mathsf{addr}}\xspace}

\newcommand{\multiplicity}[1]{\ensuremath{m\mathopen{}\left({#1}\right)\mathclose{}}\xspace}
\newcommand{\letters}{\ensuremath{\setNotation{L}}\xspace}
\newcommand{\nonletters}{\ensuremath{\overline{\letters}}\xspace}
\newcommand{\digits}{\ensuremath{\setNotation{D}}\xspace}
\newcommand{\nondigits}{\ensuremath{\overline{\digits}}\xspace}
\newcommand{\symbols}{\ensuremath{\setNotation{S}}\xspace}

\newcommand{\regexp}{\ensuremath{\mathit{regexp}}\xspace}
\newcommand{\specialtag}{\texttt{pwd\_str}\xspace}

\newcommand{\reals}{\ensuremath{\mathbb{R}}\xspace}
\newcommand{\Latentdim}{\ensuremath{d}\xspace}

\NewDocumentCommand{\Dataset}{ g g }{\ensuremath{\setNotation{D}^{#1}\IfNoValueF{#2}{_{#2}}}\xspace}
\newcommand{\UserChosenTag}{\ensuremath{\mathsf{u}}\xspace}
\newcommand{\AlgGenTag}{\ensuremath{\mathsf{a}}\xspace}
\newcommand{\TrainTag}{\ensuremath{\mathsf{tr}}\xspace}
\newcommand{\EvalTag}{\ensuremath{\mathsf{va}}\xspace}
\newcommand{\TestTag}{\ensuremath{\mathsf{te}}\xspace}

\newcommand{\DigitSeg}[1]{\ensuremath{\mathsf{D}_{#1}}\xspace}
\newcommand{\LetterSeg}[1]{\ensuremath{\mathsf{L}_{#1}}\xspace}

\NewDocumentCommand{\Symbol}{g}{\ensuremath{\mathsf{SYMBOL}\IfNoValueF{#1}{^{#1}}}\xspace}

\newcommand{\character}[1]{\ensuremath{c_{#1}}\xspace}
\newcommand{\MarkovOrder}{\ensuremath{s}\xspace}
\newcommand{\PositionIndex}{\ensuremath{i}\xspace}

\newcommand{\PasswordLength}{\ensuremath{\ell}\xspace}
\newcommand{\TweakParameter}{\ensuremath{t}\xspace}

\newcommand{\List}{\ensuremath{\mathsf{List}}\xspace}
\newcommand{\PCFG}{\ensuremath{\mathsf{PCFG}}\xspace}
\newcommand{\Markov}{\ensuremath{\mathsf{Markov}}\xspace}
\newcommand{\RNN}{\ensuremath{\mathsf{RNN}}\xspace}
\newcommand{\Combo}{\ensuremath{\mathsf{Combo}}\xspace}
\newcommand{\Listpwd}{\ensuremath{\mathsf{List*}}\xspace}
\newcommand{\PCFGpwd}{\ensuremath{\mathsf{PCFG*}}\xspace}
\newcommand{\Markovpwd}{\ensuremath{\mathsf{Markov*}}\xspace}
\newcommand{\RNNpwd}{\ensuremath{\mathsf{RNN*}}\xspace}
\newcommand{\Combopwd}{\ensuremath{\mathsf{Combo*}}\xspace}
\newcommand{\TarList}{\ensuremath{\mathsf{List\#}}\xspace}
\newcommand{\TarPCFG}{\ensuremath{\mathsf{PCFG\#}}\xspace}
\newcommand{\TarMarkov}{\ensuremath{\mathsf{Markov\#}}\xspace}
\newcommand{\TarRNN}{\ensuremath{\mathsf{RNN\#}}\xspace}
\newcommand{\TarCombo}{\ensuremath{\mathsf{Combo\#}}\xspace}
\newcommand{\CGPT}{\ensuremath{\mathsf{CGPT3}}\xspace}
\newcommand{\CBTr}{\ensuremath{\mathsf{CBT*}}\xspace}
\newcommand{\CHM}{\ensuremath{\mathsf{CHM}}\xspace}
\newcommand{\Tweak}{\ensuremath{\mathsf{Tweak}}\xspace}
\newcommand{\PasstoPath}{\ensuremath{\mathsf{P2P}}\xspace}
\NewDocumentCommand{\CBT}{ g }{\ensuremath{\mathsf{CBT}\IfNoValueF{#1}{{#1}}}\xspace}

\newcommand{\FixedGenerator}{\ensuremath{\mathsf{FXED}}\xspace}
\newcommand{\RandomGenerator}{\ensuremath{\mathsf{RAND}}\xspace}
\newcommand{\ClassifiedGenerator}{\ensuremath{\mathsf{CLSD}}\xspace}

\newcommand{\Model}{\ensuremath{f}\xspace}
\newcommand{\SymbolOfEnd}{\ensuremath{\textsf{eos}}\xspace}
\newcommand{\SymbolOfStart}{\ensuremath{\textsf{sos}}\xspace}

\newcommand{\Insertion}{\ensuremath{\mathsf{INS}}\xspace}
\newcommand{\Substitute}{\ensuremath{\mathsf{SUB}}\xspace}
\newcommand{\Deletion}{\ensuremath{\mathsf{DEL}}\xspace}
\NewDocumentCommand{\TransformationUnit}{ g }{\ensuremath{z}\IfNoValueF{#1}{_{#1}}\xspace}
\newcommand{\TransformationOperation}{\ensuremath{o}\xspace}
\newcommand{\TransformationCharacter}{\ensuremath{c}\xspace}
\newcommand{\TransformationIndex}{\ensuremath{k}\xspace}
\newcommand{\TransformationLength}{\ensuremath{q}\xspace}
\newcommand{\TransformationSet}{\ensuremath{\setNotation{T}}\xspace}
\newcommand{\EmptySymbol}{\ensuremath{\bot}\xspace}
\newcommand{\TransformationSeq}[2]{\ensuremath{\setNotation{T}_{{#1}\rightarrow{#2}}\xspace}}

\NewDocumentCommand{\TokenSet}{ g g }{\ensuremath{\Sigma_{#1}\IfNoValueF{#2}{^{(#2)}}}\xspace}

\NewDocumentCommand{\token}{ g g }{\ensuremath{\sigma_{#1}\IfNoValueF{#2}{^{(#2)}}}\xspace}

\newcommand{\Classifier}{\ensuremath{f}\xspace}
\newcommand{\NumGenerators}{\ensuremath{Y}\xspace}
\newcommand{\GeneratorIndex}{\ensuremath{y}\xspace}
\newcommand{\FixedIndex}{\ensuremath{y_{\mathsf{fx}}}\xspace}
\newcommand{\AttackerPrediction}[1]{\ensuremath{y}_{#1}\xspace}

\newcommand{\resets}{\ensuremath{\kappa}\xspace}

\newcommand{\PercentRandom}{\ensuremath{\pi}\xspace}
\newcommand{\mixedTag}{\ensuremath{\mathsf{m}}\xspace}

\ExplSyntaxOn
\cs_new:Npn \intensity #1
   {\fp_eval:n {1.0 - (\MinIntensity + (1.0-\MinIntensity) * ((({#1}-\value{MinNumber})/(\value{MaxNumber}-\value{MinNumber}))))}
   }
\ExplSyntaxOff
\newcommand{\MinIntensity}{0.0}   
\newcounter{MinNumber}
\newcounter{MaxNumber}

\newcommand{\ApplyGradientX}[1]{\cellcolor[gray]{\intensity{#1}}}
\newcolumntype{X}{>{\collectcell\ApplyGradientX}p{1.25em}<{\endcollectcell}}


%% file: abstract.tex
\begin{abstract}
Honeywords are decoy passwords that can be added to a credential
database; if a login attempt uses a honeyword, this indicates that the
site's credential database has been leaked.  In this paper we explore
the basic requirements for honeywords to be effective, in a threat
model where the attacker knows passwords for the same users at other
sites.  First, we show that for user-chosen (vs. algorithmically
generated, i.e., by a password manager) passwords, existing
honeyword-generation algorithms do not simultaneously achieve
false-positive and false-negative rates near their ideals of
$\approx 0$ and $\approx \frac{1}{1+\nmbrHoneywords}$, respectively,
in this threat model, where \nmbrHoneywords is the number of
honeywords per account. Second, we show that for users leveraging
algorithmically generated passwords, state-of-the-art methods for
honeyword generation will produce honeywords that are not sufficiently
deceptive, yielding many false negatives.  Instead, we find that only
a honeyword-generation algorithm that uses the \textit{same} password
generator as the user can provide deceptive honeywords in this case.
However, when the defender's ability to infer the generator from the
(one) account password is less accurate than the attacker's ability to
infer the generator from potentially many, this deception can again
wane.  Taken together, our results provide a cautionary note for the
state of honeyword research and pose new challenges to the field.
\end{abstract}

%% file: introduction.tex
\section{Introduction}
\label{sec:intro}
	
Credential database breaches have long been a widespread security
problem and are only becoming more so.  The 2022 Verizon Data Breach
Investigations Report places credentials as one of the two most often
breached types of confidential data, since they are so useful for
attackers to masquerade as legitimate users on the
system~\cite[p.~18]{verizon2022:dbir}.  Credential database breaches
are the largest source of compromised passwords used in credential
stuffing campaigns~\cite{thomas2017:credential}.  In turn, credential
stuffing campaigns are the cause of the vast majority of account
takeovers~\cite{shape2018:spill}.  Unfortunately, there is usually a
significant delay between the breach of a credential database and its
discovery.  Estimates of the average delay to detect a data breach
range from six to eight months in a 2022
report~\cite[\figrefstatic{45}]{ibm2022:breach}. 
The resulting window of vulnerability gives attackers the
opportunity to crack the passwords offline, and then sell them or
leverage them directly~\cite{thomas2017:credential, shape2018:spill}.
	
A strategy to accelerate the detection of credential database
breaches, suggested by Juels and Rivest nearly a decade
ago~\cite{juels2013:honeywords}, is for a site to store decoy
passwords, or \textit{honeywords}, alongside real passwords in its
credential database, so that if the attacker breaches the database,
the correct passwords are hidden among the honeywords.  The entry of a
honeyword in a login attempt then alerts the site to its breach, since
the legitimate user does not know the honeyword.  In the time since
this proposal, researchers have proposed various algorithms for
generating honeywords (see \secref{sec:background}) to meet two
central criteria: (i) that it be difficult for an attacker who has
breached a site's credential database to distinguish the legitimate
password for an account from that account's honeywords, and (ii) that
it be difficult for an attacker who has not breached a site's
credential database to guess honeywords for an account, since such
guesses will induce false breach alarms.
	
The tendency of users to reuse passwords across sites
(e.g.,~\cite{wash2016:reuse, pearman2017:habitat}) presents a
challenge for honeywords, since an attacker can stuff an account's
breached passwords at \textit{other} sites where the same user has
accounts, thereby discovering the legitimate password as the one that
works at another site.  As such, previous advances in honeyword system
designs~\cite{wang2021:amnesia} provide a mechanism by which one site
can monitor for the entry of another site's honeywords in local login
attempts.  Still, however, for an account for which the attacker can
obtain the user's passwords on other sites (e.g., by breaching these
other sites, or by phishing their passwords), the attacker will likely
need not resort to credential stuffing to differentiate the legitimate
password from its honeywords.  While this might seem like an
unnecessarily challenging threat model, it is unfortunately realistic:
A July 2020 report found more than 15 billion credentials in
circulation in cybercriminal
marketplaces~\cite{digitalshadows2020:15billion}, or an average of
more than two for every person on the planet.
	
In this paper we conduct the first critical analysis of
honeyword-generation algorithms in this setting, i.e., wherein the
attacker knows legitimate passwords at other sites for the users
represented in a database it is targeting.  There is reason to suspect
that this threat model would pose significant challenges to honeyword
efficacy for user-chosen (versus algorithmically generated) passwords.
On the one hand, if the honeyword-generation algorithm used to
populate the targeted database generates honeywords that are all
dissimilar from the user-chosen password, then the known password(s)
for the same user might enable the attacker to distinguish the
user-chosen password from its honeywords with high probability.  If
so, the false-negative probability (the probability that the site
fails to detect the breach) would be high. On the other hand, if the
honeyword-generation algorithm generates some honeywords that are
similar to the user-chosen password, then this might make it easier
for an attacker who has \textit{not} breached the database to guess
and enter honeywords in login attempts, thereby inducing a false
breach alarm (false positive).

Through a systematic analysis of current honeyword-generation
algorithms, we quantify this tension and, by doing so, show that there
appears to be no known algorithm providing a good tradeoff for
accounts with user-chosen passwords.  We additionally applied two
password tweaking techniques from password guessing to improve
honeyword generation.  While these two algorithms relieve this tension
by providing slightly lower false-negative probability, they still
induce a high false-positive probability.  Therefore, it remains far
from clear that there is \textit{any} honeyword-generation algorithm
that ensures low false-negative probability and provides adequate
resistance to false breach alarms (i.e., a false-positive rate
  near $0$).

We then turn our attention to accounts with algorithmically generated
passwords, as might be generated by a password manager.  The critical
finding that we uncover in this case is that honeyword-generation
algorithms that do not take into account the method by which the
legitimate password was generated will yield high false-negative
probability.  For example, if the user employs a password manager that
generates passwords to fit a user-configured specification, and if the
passwords exposed for that user permit the attacker to infer this
specification, then the attacker can discard any honeywords not fitting
that pattern.  We will quantify the ability of the adversary to do so
against existing honeyword-generation algorithms, most of which do not
guarantee honeywords of the same pattern as the legitimate password.
We then consider the possibility that the honeyword-generation
algorithm itself leverages a password manager to generate honeywords
whenever the user does.  However, due to the numerous generator
configurations that users might adopt, doing so is not foolproof.  In
particular, if the attacker knows potentially more passwords for the
same user's accounts elsewhere, it can classify the user's typical
configuration better than the defender can.  This advantage thus
implies an increase in false negatives, which we will demonstrate in
certain cases.

To summarize, our contributions are as follows:
\begin{itemize}[nosep,leftmargin=1em,labelwidth=*,align=left]
\item We formalize the false-positive and false-negative rates of
  honeywords in a model in which the attacker possesses passwords for
  the same user at other sites (obtained by, e.g., breaching those
  sites or phishing the user).
\item Using these definitions and empirical datasets of compromised
  passwords, we show that existing honeyword-generation algorithms
  (and two honeyword-generation methods adapted from password-guessing
  attacks) exhibit poor tradeoffs between false negatives and false
  positives in this threat model.  All the analyzed methods have a
  false-negative rate much higher than random guessing (i.e., it is
  often easy for false-negative attackers to distinguish the account
  password from honeywords) or a false-positive rate much higher than
  zero (i.e., it is often easy for false-positive attackers to induce
  false breach alarms).
\item We conduct the first study of using honeywords to protect
  algorithmically generated passwords.  Though relevant only for sites
  that reversibly encrypt their password databases (since password
  hashing, which is best practice, should render algorithmically
  generated passwords irrecoverable to an attacker who breaches the
  database), our study provides interesting findings in this setting.
  Using passwords gathered from popular password managers, we show
  that introducing honeywords without attention to the account's
  password being algorithmically generated offers little protection
  for existing honeyword-generation algorithms.  We further explore
  the use of automatic password generators to generate honeywords when
  the account password is identified as being algorithmically
  generated itself, but find that the myriad configurations of these
  generators can be a pitfall for honeyword generation.
\end{itemize}
Source code for conducting our experiments is available from
\url{https://github.com/zonghaohuang007/honeywords-analysis}.

%% file: related_work.tex
\section{Related Work}
\label{sec:related-work}

\subsection{Honeywords}
Since honeywords were first proposed~\cite{juels2013:honeywords},
there have been several research efforts on designing
honeyword-generation techniques~\cite{erguler2016:flatness,
akshima2019:honeywords, dionysiou2021:honeygen, wang2022:honeywords,
chakraborty2022:honeyword} or evaluating their security, mostly
against attacks trying to access a breached site's accounts without
alerting the site to its breach. In their original proposal, Juels and
Rivest defined an abstract model of a honeyword system and proposed
several legacy-UI methods including \textit{chaffing-by-tweaking} and
\textit{chaffing-with-a-password-model (modeling syntax)}, and one
modified-UI method. The modified-UI method requires the authentication
system to guide the user in the selection of her account password and
thus has inherent usability challenges, and so we do not consider it
in this paper. The legacy-UI methods use random replacement of
characters in the account password. We use one of them in this paper
to represent this class of techniques, as discussed in
\secref{sec:background}.  We also consider a method, called the
``\List'' model in \secref{sec:background}, that utilizes existing
passwords as the honeywords for the site's
accounts~\cite{wang2018:honeywords} (similar to
Erguler~\cite{erguler2016:flatness}). A proposal by Dionysiou, et
al.~\cite{dionysiou2021:honeygen} leverages a machine learning model
to search for similar passwords in the system and then generates
\nmbrHoneywords honeywords by tweaking the searched passwords randomly
(e.g., by the chaffing-by-tweaking method), also described in
\secref{sec:background}.  More recently, Yu and
Martin~\cite{yu2023:honeygpt} proposed to leverage the Generative
Pre-trained Transformer 3 Model (GPT-3)~\cite{brown2020:gpt3} to
generate honeywords. Their method includes two steps: first, a
password-specific segmentation technique called
PwdSegment~\cite{xu2021:chunk} is used to extract chunks from the
input password, and second, a prompt including the chunk information
is provided as the input to GPT-3, which returns a list of passwords
similar to the input password, used as honeywords. Wang and
Reiter~\cite{wang2024:bernoulli} proposed a honeyword-selection
mechanism based on a Bernoulli selection that achieves tunable false
positives.
     
Recent works have investigated the security of honeywords under
\textit{targeted-guessing attacks} where the attacker has personal
identity information (PII) about the users. Wang, et
al.~\cite{wang2018:honeywords} performed the first security analysis
on honeywords under such attacks, but they focused only on the
legacy-UI methods proposed by Juels and Rivest, empirically showing
that these methods fail to achieve low false-negative rates.  More
recently, Wang, et al.~\cite{wang2022:honeywords} considered both PII
and registration order (the time when the user accounts were created)
as the auxiliary information available to the attacker.  They proposed
leveraging this auxiliary information in a password model like the
List model, probabilistic context-free grammars
(PCFG)~\cite{chomsky1956:pcfg}, a Markov
model~\cite{rabiner1986:markov}, or a combination thereof, to generate
honeywords.  Their proposed methods achieved low false-negative rates
under the threat model considered in their
work~\cite{wang2022:honeywords}. However, our empirical results
demonstrate that existing honeyword-generation techniques, including
those considered by Wang, et al., have a high false-negative
probability in our threat model.  Setting a larger number of
honeywords per account, as suggested by Wang, et al., generally lowers
false-negative rates but increases false-positive rates. We are the
first to systematically analyze the trade-off, showing that existing
honeyword-generation methods suffer from high false-positive or
false-negative rates under a threat model where the passwords of the
same user from the other sites are exposed to attackers.  

\subsection{Password Guessing}
A related topic to honeyword generation is password guessing, which is
used to crack passwords~\cite{weir2009:pcfg-cracking, durmuth2015:omen, xu2023:improving}
in an online or offline manner or used to evaluate their
strength~\cite{kelley2012:guess, dell2015:monte, golla2018:psm}. Since
a honeyword is simply a decoy password, it is reasonable that
honeyword research will benefit from the development of password
guessing techniques.  Weir, et al.~\cite{weir2009:pcfg-cracking}
proposed the first method to utilize a probabilistic model to generate
passwords. They designed the model using PCFGs trained on a training
set of passwords and empirically demonstrated the effectiveness
compared with word-mangling rule-based methods.  Ma, et
al.~\cite{ma2014:markov-cracking} leveraged a Markov model to learn
the distribution of passwords. They showed that their proposed method
achieved slightly better performance than PCFGs in password cracking
when normalization and smoothing were used. Melicher, et
al.~\cite{melicher2016:fla-pwdmeter} designed a password model using a
recurrent neural network~\cite{rumelhart1986:rnn}, which achieves
improved accuracy in password strength measurement. Pasquini, et
al.~\cite{pasquini2021:password-gan} utilized Generative Adversarial
Networks (GAN)~\cite{goodfellow2020:gan} to train a password
generative model. They showed that the trained model can be used to
produce passwords more effectively if a password template is known,
due to the strong locality in the latent space of the generative
model. Most recently, Xu, et al.~\cite{xu2023:improving} improved password guessing by learning
a bi-direction transformer model.

Recent works showed that password guessing can be improved by
utilizing account holder PII and passwords used by the same user at
other sites. Wang, et al.~\cite{wang2016:tarpcfg-cracking} proposed a
PCFG model named \textit{TarGuess} where PII is considered in the
model training. Pal, et al.~\cite{pal2019:credential-stuffing} studied
the case that attacker utilized the passwords used by the same users
leaked from \textit{another} site to crack password, known as
\textit{credential tweaking}. They trained a \textit{Pass2Path} model
by a recurrent neural network to simulate credential tweaking, which
compromised at least $16\%$ of user accounts in their tests.  He, et
al.~\cite{he2022:passtrans} considered a similar threat model but
improved the compromising rate using a deep neural
transformer~\cite{vaswani2017:attention}. Recently, Wang, et
al.~\cite{wang2023:pass2edit} modeled password reuse behavior by a
multi-step generative model, which improved the password guessing. In
this paper, we adapted some of these techniques from
\textit{credential tweaking} (\Tweak and \PasstoPath as described in
\secref{sec:background}) for honeyword generation.

\subsection{Honeyword-Based Systems}

Our study is agnostic to system designs leveraging honeywords, whether
they be symmetric or asymmetric.  Asymmetric designs are ones that
detect honeyword entry using a secret that the attacker is presumed to
be unable to capture in the breach.  For example, the original
honeyword-system design~\cite{juels2013:honeywords} leverages a
trusted server called a \textit{honeychecker} that holds the index of
the legitimate password for each account, which the login server
consults to determine whether a login attempt uses a honeyword or the
legitimate password.  This honeychecker is assumed to keep its indices
secret despite the login server's breach.  Other asymmetric designs
include ErsatzPasswords~\cite{almeshekah2015:ersatz} and
Lethe~\cite{dionysiou2022:lethe}.

By contrast, a symmetric design is one where the attacker is allowed
to capture all state used for honeyword-entry detection when he
breaches the site.  An example of a symmetric design is
Amnesia~\cite{wang2021:amnesia}.  In this design, the attacker
captures all the information needed to undetectably access an
account---possibly using a honeyword---at a site it breaches.
However, the act of doing so configures the site to learn of its
breach once the legitimate user accesses the site subsequently, using
a different password.  That is, in Amnesia, the use of two different
passwords to enter an account is what alerts the site to its breach,
since one must be a honeyword.

%% file: background.tex
\section{Background}
\label{sec:background}

\begin{figure}[t]
  \begin{subfigure}[t]{0.475\columnwidth}
  \begin{tabbing}
  ***\=***\=\kill
  Experiment $\fpExptWParams(\alarmAttacker)$ \\
  \> $(\password, \auxiliaryInfo) \gets \choosePassword()$ \\
  \> $\honeywordSet \gets \chooseHoneywords{\nmbrHoneywords}(\password)$ \\
  \> $\passwordGuessSet \gets \alarmAttacker(\password)$ \\
  \> if $\setSize{\passwordGuessSet} \le \attackBudget \wedge \setSize{\passwordGuessSet \cap \honeywordSet} \ge \alarmThreshold$ \\
  \> \> then return $1$ \\
  \> \> else return $0$ \\[15pt]
  $\fpProbWParams(\alarmAttacker) \defeq$ \\
  \> $\prob{\experiment{\falsePositiveExpt}{\choosePassword, \chooseHoneywords{\nmbrHoneywords}, \alarmThreshold, \attackBudget}(\alarmAttacker)=1}$ \\
  $\fpProbWParams \defeq$ \\
  \> $\displaystyle \max_{\alarmAttacker} \fpProbWParams(\alarmAttacker)$
  \end{tabbing}
  \vspace{-0.1in}
  \caption{False-positive probability}
  \label{fig:advantage:fpp}
  \end{subfigure}
  \hfill
  \begin{subfigure}[t]{0.475\columnwidth}
  \begin{tabbing}
  ***\=***\=\kill
  Experiment $\fnExptWParams(\breachAttacker)$ \\
  \> $(\password, \auxiliaryInfo) \gets \choosePassword()$ \\
  \> $\honeywordSet \gets \chooseHoneywords{\nmbrHoneywords}(\password)$ \\
  \> $\passwordGuessSet \gets \breachAttacker(\honeywordSet \cup \{\password\}, \auxiliaryInfo)$ \\
  \> if $\setSize{\passwordGuessSet \cap \honeywordSet} < \alarmThreshold \wedge \password \in \passwordGuessSet$ \\
  \> \> then return $1$ \\
  \> \> else return $0$ \\[15pt]
  $\fnProbWParams(\breachAttacker) \defeq$\\
  \> $\prob[\Big]{\experiment{\falseNegativeExpt}{\choosePassword, \chooseHoneywords{\nmbrHoneywords}, \alarmThreshold}(\breachAttacker)=1}$ \\
  $\fnProbWParams \defeq$\\
  \> $\displaystyle\max_{\breachAttacker} \fnProbWParams(\breachAttacker)$
  \end{tabbing}
  \vspace{-0.1in}
  \caption{False-negative probability}
  \label{fig:advantage:fnp}
  \end{subfigure}
  \vspace{0.1in}
  \caption{Measures for breach detection by honeywords. A
      false-positive attacker \alarmAttacker tries to trick an
      unbreached site into detecting that it has been breached 
      (\figref{fig:advantage:fpp}).  A false-negative attacker
      \breachAttacker attempts to access an account after breaching
      the site, without alerting the site to its breach
      (\figref{fig:advantage:fnp}).}
  \vspace{0.1in}
  \label{fig:advantage}
  \end{figure}

\subsection{Definitions}
\label{sec:background:definitions}

Honeywords are decoy passwords added to each account entry in a
credential database.  The principle behind honeywords is that since
the legitimate user does not know the honeywords generated for her
account, the only party who is able to enter those honeywords is an
attacker who discovered them by breaching the credential database.  As
such, login attempts using honeywords should be taken as compelling
evidence of a database breach.

To make this principle precise, we define the false-positive and
false-negative probabilities of a honeyword scheme in a way that
abstracts away the details of the system leveraging them.  We do so
using the experiments shown in \figref{fig:advantage} and
described in text below.  In these experiments, a random user is
modeled by a randomized algorithm \choosePassword, by which the user
selects her password $\password \in \{0,1\}^{\ast}$ for a site.  The
invocation $\choosePassword()$ outputs not only the password
\password, but also auxiliary information $\auxiliaryInfo \subset
\{0,1\}^\ast$ that is correlated with \password and that the attacker
might learn.  In this work, \auxiliaryInfo will be passwords set by
the same user at other sites, though other works have considered other
types of auxiliary information (e.g.,~\cite{wang2018:honeywords}).
Given \password, the site selects honeywords for this account using
the randomized algorithm \chooseHoneywords{\nmbrHoneywords}, which
outputs a set \honeywordSet where $\setSize{\honeywordSet} =
\nmbrHoneywords$ and $\password \not\in \honeywordSet$.
	
A \textit{false-positive attacker} \alarmAttacker attempts to trigger
a breach alarm at this site even though it has not breached the site,
by leveraging its knowledge of \password and
\chooseHoneywords{\nmbrHoneywords} to guess honeywords in
\honeywordSet. In this work, we consider the worst case where
\alarmAttacker is permitted to know \password since \alarmAttacker
might represent a legitimate user of this site or because it might
represent an outsider who, say, phished \password.  \alarmAttacker
might know \auxiliaryInfo but \auxiliaryInfo does not help in guessing
$\honeywordSet$ if $\password$ is already known.  \alarmAttacker
is provided knowledge of the honeyword-generation algorithm
\chooseHoneywords{\nmbrHoneywords} to provide a conservative
analysis.\footnote{Allowing \alarmAttacker knowledge of
\chooseHoneywords{\nmbrHoneywords} conforms with general security
design principles; e.g., ``Do not rely on secret designs, attacker
ignorance, or \textit{security by
obscurity}.''~\cite[p.~21]{vanoorschot2021:tools}.  In our context
specifically, if the attacker knows only that
\chooseHoneywords{\nmbrHoneywords} is one of several alternatives,
it can try each alternative via a different account.  In this sense,
our measure is analogous to the \textit{min auto} approach to
measuring password strength~\cite{ur2015:measuring,
mukherjee2023:mascara}, in which the strength of a password is
measured by the number of tries to guess it, under the guessing
strategy (from among several) that minimizes that number.}
\alarmAttacker's probability of triggering an alarm is defined in
\figref{fig:advantage:fpp}, where $\alarmThreshold \ge 1$ is the
number of honeywords whose entry will trigger a breach alarm and where
$\attackBudget \ge 1$ denotes the number of login attempts
\alarmAttacker is permitted to attempt for this account.  In words,
given \password (along with \choosePassword,
\chooseHoneywords{\nmbrHoneywords}, \alarmThreshold, and
\attackBudget, which are public parameters of the experiment),
\alarmAttacker wins by outputting a set \passwordGuessSet that it can
enter in its budget of login attempts ($\setSize{\passwordGuessSet}
\le \attackBudget$) and that will trigger an alarm
($\setSize{\passwordGuessSet \cap \honeywordSet} \ge
\alarmThreshold$).  Traditionally the threshold for raising a breach
alarm has typically been set to $\alarmThreshold = 1$, though this
definition permits other values; A larger \alarmThreshold implies
  a more stringent condition for raising an alarm.  \alarmAttacker's
\textit{false-positive probability} $\fpProbWParams(\alarmAttacker)$
is the probability that \alarmAttacker wins, and the overall
false-positive probability \fpProbWParams is
$\fpProbWParams(\alarmAttacker)$ for the attacker algorithm
\alarmAttacker that maximizes that probability.  When the
parameters \choosePassword, \chooseHoneywords{\nmbrHoneywords},
\alarmThreshold, and \attackBudget are clear from context, we will
abbreviate $\fpProbWParams(\alarmAttacker)$ and \fpProbWParams as
$\fpProbWOParams(\alarmAttacker)$ and \fpProbWOParams, respectively,
to simplify notation.
	
In contrast, a \textit{false-negative attacker} \breachAttacker is an
attacker who attempts to access this user's account after breaching
the site but without alerting the site that it has been breached.
This adversary's advantage in doing so is defined in
\figref{fig:advantage:fnp}.  In words, \breachAttacker obtains the set
$\honeywordSet \cup \{\password\}$, sometimes called the
\textit{sweetwords} for this account, as well as auxiliary information
\auxiliaryInfo.  The set $\honeywordSet \cup \{\password\}$ is
sweetwords are recovered by the attacker from the salted hash
file.  \breachAttacker then wins if it outputs a set \passwordGuessSet
that will not trigger an alarm ($\setSize{\passwordGuessSet \cap
  \honeywordSet} < \alarmThreshold$) and that permits it to access the
account ($\password \in \passwordGuessSet$).  Here we presume that
$\passwordGuessSet \subseteq \honeywordSet \cup \{\password\}$, since
passwords other than the sweetwords outside $\honeywordSet
  \cup \{\password\}$ offer no help for \breachAttacker to achieve
his goals.  Consequently, we drop \attackBudget as a parameter of the
experiment; since $\attackBudget \ge \alarmThreshold \ge
\setSize{\passwordGuessSet}$, it does not constrain \breachAttacker's
choice of \passwordGuessSet.  Again, traditionally the threshold for
raising a breach alarm has been set to $\alarmThreshold = 1$, in which
case the probability with which \breachAttacker guesses \password from
the sweetwords $\honeywordSet \cup \{\password\}$ on the
first try (i.e., $\setSize{\passwordGuessSet} = 1$) is called the
\textit{flatness} of the honeyword scheme.  \breachAttacker's
\textit{false-negative probability} $\fnProbWParams(\breachAttacker)$
is the probability that \breachAttacker wins, and the overall
false-negative probability \fnProbWParams is
$\fnProbWParams(\breachAttacker)$ for the attacker \breachAttacker
that maximizes that probability. When the parameters
\choosePassword, \chooseHoneywords{\nmbrHoneywords}, and
\alarmThreshold are clear from context, we will abbreviate
$\fnProbWParams(\breachAttacker)$ and \fnProbWParams as
$\fnProbWOParams(\breachAttacker)$ and \fnProbWOParams,
respectively, to simplify notation.

A honeyword-generation algorithm \chooseHoneywords{\nmbrHoneywords}
can at best achieve $\fpProbWOParams \approx 0$ and $\fnProbWOParams =
\frac{\alarmThreshold}{\nmbrHoneywords + 1}$.  Our research evaluates
the extent to which known honeyword-generation algorithms, described
in \secref{sec:background:algorithms}, approach this ideal.  When
considering false-negative attackers, we will evaluate an attacker who
prioritizes accounts by its perceived likelihood of success in
guessing the account password \password, by refining \choosePassword
to represent ``easy'' users for whom $(\honeywordSet \cup
\{\password\}) \cap \auxiliaryInfo \neq \emptyset$, likely due to
exact password reuse across accounts; ``medium'' users who are not
``easy'' but for whom there are elements of $\honeywordSet \cup
\{\password\}$ and \auxiliaryInfo that are close to one another (in a
sense we will define later), likely because the user set passwords at
her other accounts that are similar to \password (partial password
reuse); or ``hard'' users for whom neither condition holds.

\subsection{Honeyword-Generation Algorithms}
\label{sec:background:algorithms}

In this section, we introduce honeyword-generation algorithms, some of
which have been introduced in previous
works~\cite{juels2013:honeywords, dionysiou2021:honeygen,
wang2022:honeywords}.  Generally, honeyword-generation algorithms
can be classified into two groups: \textit{password-independent}
algorithms and \textit{password-dependent} algorithms.

\subsubsection{Password-Independent Honeyword Generation}
Password-independent algorithms generate honeywords independently of
the account passwords. They do so by sampling password candidates from
\textit{password models} pretrained on a multiset of passwords. In
this work, we consider four widely used password models: list
model~\cite{wang2018:honeywords}, probabilistic context-free grammar
model~\cite{weir2009:pcfg-cracking}, Markov
model~\cite{ma2014:markov-cracking}, and recurrent neural
network~\cite{melicher2016:fla-pwdmeter}, and their
combination~\cite{wang2022:honeywords}. The detailed descriptions of
these password models are included in
\appref{app:honeyword-generation:independent}.  We denote these
generation methods as \List, \PCFG, \Markov, \RNN, and \Combo,
respectively.

\subsubsection{Password-Dependent Honeyword Generation}

Password-dependent algorithms generate honeywords that are dependent
on the account passwords. These algorithms include
password-strength-dependent methods and password-context-dependent
methods.

Password-strength-dependent methods generate honeywords whose strength
is equal or similar to the input password \password. These methods
still leverage password models such as \List, \PCFG, \Markov, \RNN, or
their combination but select a sampled candidate as a honeyword if and
only if its strength is equal to that of the input password.  However,
if the input password is weak, it might be difficult to generate
\nmbrHoneywords honeywords with equal password strength, under the
hypothesis that user-chosen passwords follow a Zipf distribution
(e.g.,~\cite{wang2017:zipf}).  So, in this work, we relax this
requirement so that a sampled candidate will be used as a honeyword if
its length equals the length of the input password.  We denote this
algorithm for generating honeywords from \List, \PCFG, \Markov, \RNN,
or a combined method by \Listpwd, \PCFGpwd, \Markovpwd, \RNNpwd,
and \Combopwd.

Password-context-dependent methods generate honeywords by modifying
the input password.  Here we consider four types of techniques:
targeted password model-based generation, LLM-based generation, random
replacement-based tweaking, and DNN-based tweaking.

\myparagraph{Targeted password model-based generation}
These methods generate honeywords from password models that learn a distribution of
\textit{password templates}~\cite{wang2022:honeywords}.  Here a
password template is a pattern describing passwords set by the same
user at different sites, wherein common substrings are indicated in
the template using a special tag \specialtag.  For example, the
template ``\specialtag z'' might be generated from ``bike123z'' and
``bike123'' if these passwords were set by the same user at two
different sites.  Password models like PCFG are pretrained on a
multiset of password templates, as targeted password models.  Then,
honeywords are generated by sampling templates from the targeted
password models and replacing \specialtag in the templates with the
input password. We denote these generation methods from \List, \PCFG,
\Markov, \RNN, or a combined method by \TarList, \TarPCFG, \TarMarkov,
\TarRNN, and \TarCombo.

\myparagraph{LLM-based generation}
These techniques generate honeywords by querying a large language
model like GPT-3~\cite{brown2020:gpt3} with prompts based on the input
password. We consider a recently proposed method, chunk-level GPT-3
(\CGPT)~\cite{yu2023:honeygpt}. The detailed description of \CGPT is
included in \appref{app:honeyword-generation:dependent}.

\myparagraph{Random replacement-based tweaking}
These techniques generate honeywords by randomly changing some
characters of the input password or similar passwords. We consider
chaffing-by-tweaking or
\CBT{\TweakParameter}~\cite{juels2013:honeywords}, which generates
honeywords by randomly replacing the last \TweakParameter characters
of the input password with characters of the same type;
\CBTr~\cite{dionysiou2021:honeygen}, which generates honeywords by
similarly replacing all the characters; and chaffing-by-a-hybrid-model
(\CHM~\cite{dionysiou2021:honeygen}).  Detailed descriptions of
\CBT{\TweakParameter}, \CBTr, and \CHM are included in
\appref{app:honeyword-generation:dependent}.

\myparagraph{DNN-based tweaking}
DNN-based tweaking techniques leverage DNNs to tweak the chosen
password to generate its honeywords.  We consider a deep tweak model
(\Tweak)~\cite{he2022:passtrans} and tweaking path model
(\PasstoPath)~\cite{pal2019:credential-stuffing}, which are adapted
from similar constructions originally developed to crack
passwords~\cite{he2022:passtrans, pal2019:credential-stuffing}.  The
deep tweak model is a DNN that, on input a password, outputs a tweaked
password.  The tweaking path model inputs a password and outputs an
edit path that is used to change the input password. More
descriptions of these two techniques are included in
\appref{app:honeyword-generation:dependent}.

%% file: user_chosen.tex
\section{User-Chosen Passwords}
\label{sec:user-selected}
	
The first case we consider is when \choosePassword is an algorithm
implemented by an average human user, and \auxiliaryInfo is a multiset
of passwords chosen by the same user at other sites.  In this case, we
show that the field has yet to identify \textit{any}
honeyword-generation algorithm that achieves small \fnProbWOParams and
\fpProbWOParams simultaneously.  Intuitively, this is true because when
a user selects passwords without automated help (i.e., \choosePassword
is an average user), then an attacker who guesses passwords
\passwordGuessSet that are similar to passwords in \auxiliaryInfo will
be highly effective in either inducing false detections (a high
$\fpProbWOParams(\alarmAttacker)$) or avoiding true detection (a high
$\fnProbWOParams(\breachAttacker)$).  On the one hand, if
$\chooseHoneywords{\nmbrHoneywords}(\password)$ outputs honeywords
dissimilar to \password, then since users often choose \password
similar to elements of \auxiliaryInfo, it will be relatively easy for
an attacker \breachAttacker to select \password from $\honeywordSet
\cup \{\password\}$ as the one most similar to passwords in
\auxiliaryInfo.  So, for $\fnProbWOParams(\breachAttacker)$ to be
small, $\chooseHoneywords{\nmbrHoneywords}(\password)$ must output at
least some honeywords that are similar to \password.  On the other
hand, the more it does so, the easier it is for an attacker
\alarmAttacker to induce false detections by guessing passwords
\passwordGuessSet similar to passwords in \auxiliaryInfo.

\subsection{Attack Strategies}
\label{sec:user-selected:attack-strategy}

In this section, we introduce the false-positive attacker
\alarmAttacker and the false-negative attacker \breachAttacker that we
use in the evaluation of $\fpProbWOParams(\alarmAttacker)$ and
$\fnProbWOParams(\breachAttacker)$, respectively.

\myparagraph{False-positive attacker \alarmAttacker}
In the evaluation of $\fpProbWOParams(\alarmAttacker)$, recall that
\alarmAttacker is given access to \password.  The attacker
\alarmAttacker leverages the honeyword-generation algorithm
\chooseHoneywords{} on input $\password$ to generate a set of
honeyword candidates. Then, if applicable, it sorts the candidates by
the probabilities assigned by the honeyword-generation algorithm and
uses the top \attackBudget candidates as the guessed honeywords
\passwordGuessSet; otherwise, it picks \attackBudget candidates
uniformly at random without replacement as \passwordGuessSet.

\myparagraph{False-negative attacker \breachAttacker}
We evaluate $\fnProbWOParams(\breachAttacker)$ for user-chosen
passwords as follows. Given passwords \auxiliaryInfo, \breachAttacker
leverages a metric function $\pwdDistance(\cdot): \passwordDomain
\times \passwordDomain \rightarrow \reals$ to measure the similarity
between the elements of \auxiliaryInfo and the sweetwords
$\honeywordSet{} \cup \{\password\}$, and ranks each sweetword based
on its similarity to the most similar element of \auxiliaryInfo.  The
top \alarmThreshold ranked sweetwords are used to guess \password.

\subsection{Model to Measure Password Similarity}
\label{sec:user-selected:sim-model}

In the evaluation of $\fnProbWOParams(\breachAttacker)$, we need to
define a metric function that inputs a pair of passwords and returns a
score reflecting the similarity between the inputs. To formulate such
a metric function, we designed a similarity model $\Model(\cdot):
\passwordDomain \rightarrow \reals^{\Latentdim}$ by a deep neural
network, which takes as input a password \password and outputs its
\textit{latent representation} such that the cosine similarity between
any two latent representations $\Model(\password)$ and
$\Model(\passwordAlt)$ grows with the probability that
$\choosePassword()$ would have output both, i.e., with
\cprob{\big}{\passwordAlt \in \auxiliaryInfo}{(\password,
\auxiliaryInfo) \gets \choosePassword()}.

The similarity model is used to learn the embedding of
passwords. Learning such an embedding of passwords into a latent space
is essentially a \textit{metric learning}
problem~\cite{wang2017:deep-metric, oh2016:deep-metric}. Therefore, we
applied contrastive learning, which is one of the most widely used
frameworks to train a model to perform this embedding so as to
maximizing cosine similarity between positive (similar) pairs while
minimizing cosine similarity of negative (dissimilar)
pairs~\cite{chen2020:simclr}. Training a contrastive model is
performed in \textit{batches}, each a multiset $\modelBatch \subseteq
\passwordDomain \times \auxiliaryInfoDomain$.  Each $(\password,
\passwordAlt) \in \modelBatch$ consists of similar passwords
(intuitively, for which \cprob{\big}{\passwordAlt \in
  \auxiliaryInfo}{(\password, \auxiliaryInfo) \gets \choosePassword()}
is high), whereas for any $(\passwordAltAlt, \passwordAltAltAlt) \in
\modelBatch \setminus \{(\password, \passwordAlt)\}$, \password and
\passwordAltAlt are presumed to be dissimilar, as are \passwordAlt and
\passwordAltAltAlt.  Training for a contrastive learning model of
password similarity, therefore, updates \Model to minimize a
\textit{loss function}, which typically would take the form
\begin{align}
\avg_{(\password, \passwordAlt) \in \modelBatch}
-\log \frac{\exp(\cosineSim(\Model(\password), \Model(\passwordAlt)))}%
{\displaystyle\sum_{\substack{\mathclap{(\passwordAltAlt, \passwordAltAltAlt) \in \modelBatch:} \\
\mathclap{(\passwordAltAlt, \passwordAltAltAlt) \neq (\password, \passwordAlt)}}}
\hspace{1.5em}\left(
\begin{array}{@{}l@{}}
\exp(\cosineSim(\Model(\password), \Model(\passwordAltAlt)))~+ \\
\exp(\cosineSim(\Model(\passwordAlt), \Model(\passwordAltAltAlt)))
\end{array}
\right)
}
\label{eqn:loss}
\end{align}
where \cosineSim denotes cosine similarity (see Chen, et
al.~\cite{chen2020:simclr}). Such updates with all the data samples
from the training dataset passed through the trained model constitute
one \textit{epoch}.  The design and training of the similarity model
are described in \appref{app:sim_model}.

\subsection{Evaluation}
\label{sec:user-selected:evaluation}

In this subsection, we detail our evaluation of the user-chosen
password case, which includes the used dataset, and
the experimental results for $\fpProbWOParams(\alarmAttacker)$ and
$\fnProbWOParams(\breachAttacker)$.

\subsubsection{The Dataset}
\label{sec:user-selected:evaluation:dataset}

\begin{wrapfigure}{r}{0.5\columnwidth}
  \vspace{-3ex}
  \resizebox{0.5\columnwidth}{!}{\input{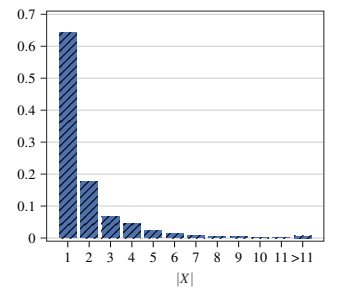}}
  \caption{Distribution of \setSize{\auxiliaryInfo}}
  \label{fig:distribution_auxiliaryInfo}
  \vspace{-0ex}
\end{wrapfigure}

The dataset we used in the case of user-chosen passwords is the
\textit{4iQ} dataset~\cite{casal2017:4iq}, consisting of $1.4$ billion
(email, password) pairs, of which $1.1$ billion emails and $463$
million passwords are unique. Others attribute the \textit{4iQ}
dataset to various leaks from LinkedIn, Myspace, Badoo, Yahoo,
Twitter, Zoosk, and Neopet, and have used it to analyze users' choices
of passwords across sites~\cite{pal2019:credential-stuffing} (despite
the possibility of some being automatically generated).  Our use of 
leaked passwords was approved by our
IRB, which specified protections in our handling of this data (who
could access the data, what results could be reported, etc.).  In
order to use \textit{4iQ}, we preprocessed the dataset by referring to
previous works (e.g.,~\cite{pal2019:credential-stuffing}).

\begin{itemize}[nosep,leftmargin=1em,labelwidth=*,align=left]
\item \textbf{Cleaning:} We removed any (email, password) pairs that
  satisfied any of the following conditions: the password contained
  non-ASCII characters, the space character, or a substring of $20$
  (or more) hex characters; the password had a length of less than $4$
  or more than $30$; or the email contained non-ASCII characters or
  the space character.

\item \textbf{Joining by email and username:} For each email address
  \emailAddr appearing in the dataset, we collected the passwords
  appearing with that email address into a multiset
  \passwordSet{\emailAddr}.  Then we merged some password multisets
  \passwordSet{\emailAddr} as follows: two multisets were merged if
  they contained at least one password in common and if the username
  parts of their email addresses were the same.  We then eliminated
  each \passwordSet{\emailAddr} containing only one password or $>
   1{,}000$ passwords.  In the resulting dataset, around $48\%$ of
  users reused passwords, which is within the range between $43\%$ and
  $51\%$ estimated by previous work
  (e.g.~\cite{das2014:tangled}). More statistics about the resulting
  dataset are shown in \tblref{table:dataset}.

\item \textbf{Splitting into training and testing sets:} Of the
  $195{,}894{,}983$ password multisets that remained, $80\%$
  ($156{,}722{,}455$ multisets with $451{,}020{,}019$ passwords) were
  set aside as training sets \Dataset{\TrainTag}{\UserChosenTag} used
  to train models. The
  other $20\%$ ($39{,}172{,}528$ multisets with $112{,}723{,}111$
  passwords) of the password multisets were set aside as testing sets
  \Dataset{\TestTag}{\UserChosenTag}. When evaluating
  $\fnProbWOParams(\breachAttacker)$ and
  $\fpProbWOParams(\alarmAttacker)$, the algorithm \choosePassword was
  implemented by choosing \password and the members of \auxiliaryInfo
  without replacement from a single multiset \passwordSet{\emailAddr}
  chosen uniformly at random from the testing sets, and returning
  $(\password, \auxiliaryInfo)$ as the result with $\auxiliaryInfo =
  \passwordSet{\emailAddr} \setminus \{\password\}$.
  \setSize{\auxiliaryInfo} represents the amount of attacker's
  knowledge about this user's passwords at other sites. Its
  distribution in $\Dataset{\TestTag}{\UserChosenTag}$ is shown in
  \figref{fig:distribution_auxiliaryInfo}.
\end{itemize}

\begin{table}
\small
\centering
\begin{tabular}{lr}
\toprule
Statistic & Value \\
\midrule
Total number of users & $195{,}894{,}983$ \\
Total number of passwords & $563{,}743{,}130$ \\
Average passwords per user & $2.877$ \\
Average distinct passwords per user & $1.961$ \\
Percentage of users reusing passwords & $48.507\%$ \\
\bottomrule
\end{tabular}
\vspace*{1ex}
\caption{Statistics of the preprocessed dataset}
\label{table:dataset}
\end{table}

\begin{figure*}[t!]
  \begin{subfigure}[b]{\textwidth}
    \centering
    \resizebox{!}{2em}{\input{user_chosen_trade-off_legend.tex}}
  \end{subfigure}
  
  \begin{subfigure}[b]{0.179\textwidth}
    \setlength\figureheight{1.9in}
    \centering
    \caption{\List}
    \vspace{1.25ex}
    \resizebox{\textwidth}{!}{\input{figure/user_chosen/List/trade-off_20}}
    \label{fig:uc:to20:list}
  \end{subfigure}
  \hfill
  \begin{subfigure}[b]{0.141\textwidth}
    \setlength\figureheight{1.9in}
    \centering
    \caption{\PCFG}
    \vspace{2ex}
    \resizebox{\textwidth}{!}{\input{figure/user_chosen/PCFG/trade-off_20}}
    \label{fig:uc:to20:pcfg}
  \end{subfigure}
  \hfill
  \begin{subfigure}[b]{0.141\textwidth}
    \setlength\figureheight{1.9in}
    \centering
    \caption{\Markov}
    \vspace{2ex}
    \resizebox{\textwidth}{!}{\input{figure/user_chosen/Markov/trade-off_20}}
    \label{fig:uc:to20:markov}
  \end{subfigure}
  \hfill
  \begin{subfigure}[b]{0.141\textwidth}
    \setlength\figureheight{1.9in}
    \centering
    \caption{\Combo}
    \vspace{2ex}
    \resizebox{\textwidth}{!}{\input{figure/user_chosen/combination/trade-off_20}}
    \label{fig:uc:to20:combo}
  \end{subfigure}
  \hfill
  \begin{subfigure}[b]{0.141\textwidth}
    \setlength\figureheight{1.9in}
    \centering
    \caption{\Listpwd}
    \vspace{2ex}
    \resizebox{\textwidth}{!}{\input{figure/user_chosen/List_/trade-off_20}}
    \label{fig:uc:to20:list-pwd}
  \end{subfigure}
  \hfill
  \begin{subfigure}[b]{0.141\textwidth}
    \setlength\figureheight{1.9in}
    \centering
    \caption{\PCFGpwd}
    \vspace{2ex}
    \resizebox{\textwidth}{!}{\input{figure/user_chosen/PCFG_/trade-off_20}}
    \label{fig:uc:to20:pcfg-pwd}
  \end{subfigure}
  \\[-2.5ex]
  \begin{subfigure}[b]{\textwidth}
    \setlength\figureheight{2in}
      \centering
      \resizebox{!}{1.5em}{\input{xlabel.tex}}
  \end{subfigure}
  \vspace{1ex}
  \\
  \begin{subfigure}[b]{0.179\textwidth}
    \setlength\figureheight{1.9in}
    \centering
    \caption{\Markovpwd}
    \vspace{1.25ex}
    \resizebox{\textwidth}{!}{\input{figure/user_chosen/Markov_/trade-off_20}}
    \label{fig:uc:to20:markov-pwd}
  \end{subfigure}
  \hfill
  \begin{subfigure}[b]{0.141\textwidth}
    \setlength\figureheight{1.9in}
    \centering
    \caption{\Combopwd}
    \vspace{2ex}
    \resizebox{\textwidth}{!}{\input{figure/user_chosen/combination_/trade-off_20}}
    \label{fig:uc:to20:combo-pwd}
  \end{subfigure}
  \hfill
  \begin{subfigure}[b]{0.141\textwidth}
    \setlength\figureheight{1.9in}
    \centering
    \caption{\TarList}
    \vspace{2ex}
    \resizebox{\textwidth}{!}{\input{figure/user_chosen/List_target/trade-off_20}}
    \label{fig:uc:to20:list-tgt}
  \end{subfigure}
  \hfill
  \begin{subfigure}[b]{0.141\textwidth}
    \setlength\figureheight{1.9in}
    \centering
    \caption{\TarPCFG}
    \vspace{2ex}
    \resizebox{\textwidth}{!}{\input{figure/user_chosen/PCFG_target/trade-off_20}}
    \label{fig:uc:to20:pcfg-tgt}
  \end{subfigure}
  \hfill
  \begin{subfigure}[b]{0.141\textwidth}
    \setlength\figureheight{1.9in}
    \centering
    \caption{\TarMarkov}
    \vspace{2ex}
    \resizebox{\textwidth}{!}{\input{figure/user_chosen/Markov_target/trade-off_20}}
    \label{fig:uc:to20:markov-tgt}
  \end{subfigure}
  \hfill
  \begin{subfigure}[b]{0.141\textwidth}
    \setlength\figureheight{1.9in}
    \centering
    \caption{\TarCombo}
    \vspace{2ex}
    \resizebox{\textwidth}{!}{\input{figure/user_chosen/combination_target/trade-off_20}}
    \label{fig:uc:to20:combo-tgt}
  \end{subfigure}
  \\[-2.5ex]
  \begin{subfigure}[b]{\textwidth}
    \setlength\figureheight{2in}
      \centering
      \resizebox{!}{1.5em}{\input{xlabel.tex}}
  \end{subfigure}
  \vspace{1ex}
  \\
  \begin{subfigure}[b]{0.179\textwidth}
    \setlength\figureheight{1.9in}
    \centering
    \caption{\CGPT}
    \vspace{1.25ex}
    \resizebox{\textwidth}{!}{\input{figure/user_chosen/gpt/trade-off_20}}
    \label{fig:uc:to20:gpt}
  \end{subfigure}
  \hfill
  \begin{subfigure}[b]{0.141\textwidth}
    \setlength\figureheight{1.9in}
    \centering
    \caption{\CBT{4}}
    \vspace{2ex}
    \resizebox{\textwidth}{!}{\input{figure/user_chosen/chaffing4/trade-off_20}}
    \label{fig:uc:to20:cbt4}
  \end{subfigure}
  \hfill
  \begin{subfigure}[b]{0.141\textwidth}
    \setlength\figureheight{1.9in}
    \centering
    \caption{\CBTr}
    \vspace{2ex}
    \resizebox{\textwidth}{!}{\input{figure/user_chosen/CBTr/trade-off_20}}
    \label{fig:uc:to20:cbtr}
  \end{subfigure}
  \hfill
  \begin{subfigure}[b]{0.141\textwidth}
    \setlength\figureheight{1.9in}
    \centering
    \caption{\CHM}
    \vspace{2ex}
    \resizebox{\textwidth}{!}{\input{figure/user_chosen/CHM/trade-off_20}}
    \label{fig:uc:to20:chm-pwd}
  \end{subfigure}
  \hfill
  \begin{subfigure}[b]{0.141\textwidth}
    \setlength\figureheight{1.9in}
    \centering
    \caption{\Tweak}
    \vspace{2ex}
    \resizebox{\textwidth}{!}{\input{figure/user_chosen/Tweak/trade-off_20}}
    \label{fig:uc:to20:tweak}
  \end{subfigure}
  \hfill
  \begin{subfigure}[b]{0.141\textwidth}
    \setlength\figureheight{1.9in}
    \centering
    \caption{\PasstoPath}
    \vspace{2ex}
    \resizebox{\textwidth}{!}{\input{figure/user_chosen/pass2path/trade-off_20}}
    \label{fig:uc:to20:pass2path}
  \end{subfigure}
  \\[-2.5ex]
    \begin{subfigure}[b]{\textwidth}
      \setlength\figureheight{2in}
        \centering
        \resizebox{!}{1.5em}{\input{xlabel.tex}}
    \end{subfigure}
    
  \caption{$\fpProbWOParams(\alarmAttacker)$
    vs.\ $\fnProbWOParams(\breachAttacker)$ as \alarmThreshold is varied,
    for the case of user-chosen passwords ($\nmbrHoneywords=19$,
    $\attackBudget=1000$). The best $\fnProbWOParams(\breachAttacker)$
    are $0.54$ (\PasstoPath, \figref{fig:uc:to20:pass2path}), $0.56$
    (\Tweak, \figref{fig:uc:to20:tweak}), $0.57$
    (\CHM, \figref{fig:uc:to20:chm-pwd}), and $0.58$
    (\CGPT, \figref{fig:uc:to20:gpt}); all others have
    $\fnProbWOParams(\breachAttacker) > 0.59$. All suffer
    $\fpProbWOParams(\alarmAttacker) > 0.27$ at $\alarmThreshold = 1$.
    Those that reach $\fpProbWOParams(\alarmAttacker) \approx 0$ do so
    with $\fnProbWOParams(\breachAttacker) > 0.81$.}
  \label{fig:uc:to20} 
  \vspace*{2ex}
  \end{figure*}

  \begin{figure*}[t!]
    \begin{subfigure}[b]{\textwidth}
      \centering
      \resizebox{!}{2em}{\input{user_chosen_trade-off_legend.tex}}
    \end{subfigure}
    
    \begin{subfigure}[b]{0.179\textwidth}
      \setlength\figureheight{1.9in}
      \centering
      \caption{\List}
      \vspace{1.25ex}
      \resizebox{\textwidth}{!}{\input{figure/user_chosen/List/trade-off_100}}
      \label{fig:uc:to100:list}
    \end{subfigure}
    \hfill
    \begin{subfigure}[b]{0.141\textwidth}
      \setlength\figureheight{1.9in}
      \centering
      \caption{\PCFG}
      \vspace{2ex}
      \resizebox{\textwidth}{!}{\input{figure/user_chosen/PCFG/trade-off_100}}
      \label{fig:uc:to100:pcfg}
    \end{subfigure}
    \hfill
    \begin{subfigure}[b]{0.141\textwidth}
      \setlength\figureheight{1.9in}
      \centering
      \caption{\Markov}
      \vspace{2ex}
      \resizebox{\textwidth}{!}{\input{figure/user_chosen/Markov/trade-off_100}}
      \label{fig:uc:to100:markov}
    \end{subfigure}
    \hfill
    \begin{subfigure}[b]{0.141\textwidth}
      \setlength\figureheight{1.9in}
      \centering
      \caption{\Combo}
      \vspace{2ex}
      \resizebox{\textwidth}{!}{\input{figure/user_chosen/combination/trade-off_100}}
      \label{fig:uc:to100:combo}
    \end{subfigure}
    \hfill
    \begin{subfigure}[b]{0.141\textwidth}
      \setlength\figureheight{1.9in}
      \centering
      \caption{\Listpwd}
      \vspace{2ex}
      \resizebox{\textwidth}{!}{\input{figure/user_chosen/List_/trade-off_100}}
      \label{fig:uc:to100:list-pwd}
    \end{subfigure}
    \hfill
    \begin{subfigure}[b]{0.141\textwidth}
      \setlength\figureheight{1.9in}
      \centering
      \caption{\PCFGpwd}
      \vspace{2ex}
      \resizebox{\textwidth}{!}{\input{figure/user_chosen/PCFG_/trade-off_100}}
      \label{fig:uc:to100:pcfg-pwd}
    \end{subfigure}
    \\[-2.5ex]
    \begin{subfigure}[b]{\textwidth}
      \setlength\figureheight{2in}
        \centering
        \resizebox{!}{1.5em}{\input{xlabel.tex}}
    \end{subfigure}
    \vspace{1ex}
    \\
    \begin{subfigure}[b]{0.179\textwidth}
      \setlength\figureheight{1.9in}
      \centering
      \caption{\Markovpwd}
      \vspace{1.25ex}
      \resizebox{\textwidth}{!}{\input{figure/user_chosen/Markov_/trade-off_100}}
      \label{fig:uc:to100:markov-pwd}
    \end{subfigure}
    \hfill
    \begin{subfigure}[b]{0.141\textwidth}
      \setlength\figureheight{1.9in}
      \centering
      \caption{\Combopwd}
      \vspace{2ex}
      \resizebox{\textwidth}{!}{\input{figure/user_chosen/combination_/trade-off_100}}
      \label{fig:uc:to100:combo-pwd}
    \end{subfigure}
    \hfill
    \begin{subfigure}[b]{0.141\textwidth}
      \setlength\figureheight{1.9in}
      \centering
      \caption{\TarList}
      \vspace{2ex}
      \resizebox{\textwidth}{!}{\input{figure/user_chosen/List_target/trade-off_100}}
      \label{fig:uc:to100:list-tgt}
    \end{subfigure}
    \hfill
    \begin{subfigure}[b]{0.141\textwidth}
      \setlength\figureheight{1.9in}
      \centering
      \caption{\TarPCFG}
      \vspace{2ex}
      \resizebox{\textwidth}{!}{\input{figure/user_chosen/PCFG_target/trade-off_100}}
      \label{fig:uc:to100:pcfg-tgt}
    \end{subfigure}
    \hfill
    \begin{subfigure}[b]{0.141\textwidth}
      \setlength\figureheight{1.9in}
      \centering
      \caption{\TarMarkov}
      \vspace{2ex}
      \resizebox{\textwidth}{!}{\input{figure/user_chosen/Markov_target/trade-off_100}}
      \label{fig:uc:to100:markov-tgt}
    \end{subfigure}
    \hfill
    \begin{subfigure}[b]{0.141\textwidth}
      \setlength\figureheight{1.9in}
      \centering
      \caption{\TarCombo}
      \vspace{2ex}
      \resizebox{\textwidth}{!}{\input{figure/user_chosen/combination_target/trade-off_100}}
      \label{fig:uc:to100:combo-tgt}
    \end{subfigure}
    \\[-2.5ex]
    \begin{subfigure}[b]{\textwidth}
      \setlength\figureheight{2in}
        \centering
        \resizebox{!}{1.5em}{\input{xlabel.tex}}
    \end{subfigure}
    \vspace{1ex}
    \\
    \begin{subfigure}[b]{0.179\textwidth}
      \setlength\figureheight{1.9in}
      \centering
      \caption{\CGPT}
      \vspace{1.25ex}
      \resizebox{\textwidth}{!}{\input{figure/user_chosen/gpt/trade-off_100}}
      \label{fig:uc:to100:gpt}
    \end{subfigure}
    \hfill
    \begin{subfigure}[b]{0.141\textwidth}
      \setlength\figureheight{1.9in}
      \centering
      \caption{\CBT{4}}
      \vspace{2ex}
      \resizebox{\textwidth}{!}{\input{figure/user_chosen/chaffing4/trade-off_100}}
      \label{fig:uc:to100:cbt4}
    \end{subfigure}
    \hfill
    \begin{subfigure}[b]{0.141\textwidth}
      \setlength\figureheight{1.9in}
      \centering
      \caption{\CBTr}
      \vspace{2ex}
      \resizebox{\textwidth}{!}{\input{figure/user_chosen/CBTr/trade-off_100}}
      \label{fig:uc:to100:cbtr}
    \end{subfigure}
    \hfill
    \begin{subfigure}[b]{0.141\textwidth}
      \setlength\figureheight{1.9in}
      \centering
      \caption{\CHM}
      \vspace{2ex}
      \resizebox{\textwidth}{!}{\input{figure/user_chosen/CHM/trade-off_100}}
      \label{fig:uc:to100:chm-pwd}
    \end{subfigure}
    \hfill
    \begin{subfigure}[b]{0.141\textwidth}
      \setlength\figureheight{1.9in}
      \centering
      \caption{\Tweak}
      \vspace{2ex}
      \resizebox{\textwidth}{!}{\input{figure/user_chosen/Tweak/trade-off_100}}
      \label{fig:uc:to100:tweak}
    \end{subfigure}
    \hfill
    \begin{subfigure}[b]{0.141\textwidth}
      \setlength\figureheight{1.9in}
      \centering
      \caption{\PasstoPath}
      \vspace{2ex}
      \resizebox{\textwidth}{!}{\input{figure/user_chosen/pass2path/trade-off_100}}
      \label{fig:uc:to100:pass2path}
    \end{subfigure}
    \\[-2.5ex]
      \begin{subfigure}[b]{\textwidth}
        \setlength\figureheight{2in}
          \centering
          \resizebox{!}{1.5em}{\input{xlabel.tex}}
      \end{subfigure}
  \caption{$\fpProbWOParams(\alarmAttacker)$
    vs.\ $\fnProbWOParams(\breachAttacker)$ as \alarmThreshold is varied,
    for the case of user-chosen passwords ($\nmbrHoneywords=99$,
    $\attackBudget=1000$). The best $\fnProbWOParams(\breachAttacker)$
    are $0.47$ (\PasstoPath, \figref{fig:uc:to100:pass2path}), $0.49$
    (\CHM, \figref{fig:uc:to100:chm-pwd}), and $0.50$
    (\Tweak, \figref{fig:uc:to100:tweak}); all others have
    $\fnProbWOParams(\breachAttacker) > 0.50$. All suffer
    $\fpProbWOParams(\alarmAttacker) > 0.34$ at $\alarmThreshold = 1$.
    Those that reach $\fpProbWOParams(\alarmAttacker) \approx 0$ do so
    with $\fnProbWOParams(\breachAttacker) > 0.72$.}
  \label{fig:uc:to100}
  \vspace*{2ex}
  \end{figure*}

\begin{table}[ht!]
\small
\begin{center}
\begin{tabular}{@{}lrrrrrr@{}}
\toprule
& \multicolumn{3}{c}{$\nmbrHoneywords = 19$} & \multicolumn{3}{c}{$\nmbrHoneywords = 99$} \\
\multicolumn{1}{c}{$\chooseHoneywords{\nmbrHoneywords}$} &
\multicolumn{1}{c}{easy} &
\multicolumn{1}{c}{med} &
\multicolumn{1}{c}{hard} &
\multicolumn{1}{c}{easy} &
\multicolumn{1}{c}{med} &
\multicolumn{1}{c}{hard} \\
\midrule
\List      & 43.37 & 16.00 & 40.63 & 43.56 & 19.62 & 36.82 \\
\Markov    & 43.36 & 15.96 & 40.68 & 43.57 & 19.78 & 36.65 \\ 
\PCFG      & 43.36 & 15.59 & 41.05 & 43.47 & 18.40 & 38.13 \\ 
\RNN       & 43.36 & 16.01 & 40.63 & 43.59 & 19.57 & 36.84 \\
\Combo      & 43.33 & 15.98 & 40.69 & 43.55 & 19.27 & 37.18 \\
\Listpwd     & 43.37 & 16.07 & 40.56 & 43.41 & 19.39 & 37.20 \\ 
\Markovpwd   & 43.33 & 16.05 & 40.62 & 43.38 & 19.57 & 37.05 \\ 
\PCFGpwd     & 43.35 & 15.66 & 40.99 & 43.38 & 18.68 & 37.94 \\ 
\RNNpwd      & 43.34 & 15.92 & 40.74 & 43.40 & 19.46 & 37.14 \\
\Combopwd      & 43.37 & 15.94 & 40.69 & 43.38 & 19.70 & 36.92 \\
\TarList      & 44.11 & 19.25 & 36.64 & 43.55 & 15.80 & 40.65 \\
\TarMarkov    & 44.00 & 18.91 & 37.09 & 43.47 & 15.92 & 40.61 \\ 
\TarPCFG      & 43.47 & 15.80 & 40.73 & 44.00 & 19.13 & 36.87 \\ 
\TarRNN       & 43.54 & 15.64 & 40.82 & 43.98 & 19.11 & 36.91  \\
\TarCombo      & 43.49 & 15.74 & 40.77  & 44.09 & 18.65  & 37.26 \\
\CGPT    & 44.54 & 14.16 & 41.28 & 44.98 & 14.28 & 40.72 \\ 
\CBT{3}     & 43.35 & 14.87 & 41.78 & 43.34 & 15.28 & 41.38 \\ 
\CBT{4}     & 43.33 & 14.97 & 41.73 & 43.33 & 15.31 & 41.36 \\
\CBTr     & 43.53 & 14.92 & 41.55 & 43.84 & 15.56 & 40.60 \\ 
\CHM       & 43.46 & 15.33 & 41.21 & 43.91 & 15.83 & 40.26 \\
\Tweak     & 44.80 & 14.22 & 40.98 & 45.89 & 14.67 & 39.44 \\ 
\PasstoPath & 46.05 & 12.36 & 41.59 & 47.39 & 12.18 & 40.43 \\
\bottomrule
\end{tabular}
\end{center}
\caption{Percentages of accounts of different hardness for a
  false-negative attacker \breachAttacker, discussed in
  \secref{sec:user-selected:evaluation:results}}
  \vspace*{2ex}
\label{table:percentage}
\end{table}

\subsubsection{Experimental Results}
\label{sec:user-selected:evaluation:results}

We now report $\fpProbWOParams(\alarmAttacker)$ and
$\fnProbWOParams(\breachAttacker)$ for the attackers \alarmAttacker and
\breachAttacker described in
\secref{sec:user-selected:attack-strategy}.  To depict the tradeoffs
between these measurements, we plot them against one another as
\alarmThreshold is varied.  When evaluating
$\fnProbWOParams(\breachAttacker)$, we isolate three subcases, to
permit modeling of an attacker who prioritizes accounts based on
similarities between $\honeywordSet \cup \{\password\}$ and
\auxiliaryInfo per account.  We measured such similarity based on
definitions like those for password reuse introduced in previous work
(e.g.,~\cite{pearman2017:habitat}).  Specifically, ``easy'' accounts
are those for which $(\honeywordSet \cup \{\password\}) \cap
\auxiliaryInfo \neq \emptyset$; ``medium'' accounts are those for
which $(\honeywordSet \cup \{\password\}) \cap \auxiliaryInfo =
\emptyset$ but there is a sweetword in $\honeywordSet \cup
\{\password\}$ that shares a substring of length at least four
characters with some password in \auxiliaryInfo; and ``hard'' accounts
are those that are neither ``easy'' nor ``medium''.  The percentages
of accounts of different hardness are shown in
\tblref{table:percentage}.

\figref{fig:uc:to20} shows the tradeoffs between
$\fpProbWOParams(\alarmAttacker)$ and $\fnProbWOParams(\breachAttacker)$
for $\nmbrHoneywords = 19$ honeywords and $\attackBudget = 1000$, for
the various honeyword-generation algorithms described in
\secref{sec:background}. \RNN and its variants
achieved similar performance to \List, \PCFG, \Markov, \Combo, and their variants, and
thus we only show the results from \List, \PCFG, \Markov, \Combo, and their variants
in \figref{fig:uc:to20}; results for \RNN and its variants are in 
\appref{app:additional_results}.  In each plot, there
are four curves presenting the overall tradeoff (``all'') and those of
three subcases: ``easy'', ``medium'', and ``hard''. In each curve,
markers highlight the $\fpProbWOParams(\alarmAttacker)$
vs.\ $\fnProbWOParams(\breachAttacker)$ tradeoff at a specific values
of \alarmThreshold ranging from $\alarmThreshold=1$ to
\nmbrHoneywords.  Intuitively, a smaller \alarmThreshold yields lower
$\fnProbWOParams(\breachAttacker)$ but higher
$\fpProbWOParams(\alarmAttacker)$ and so a marker closer to the top
left corner. Increasing \alarmThreshold to $\nmbrHoneywords$ yields a
higher $\fnProbWOParams(\breachAttacker)$ but lower
$\fpProbWOParams(\alarmAttacker)$ and so a marker closer to the bottom
right corner.  We stress that $\attackBudget = 1000$ yields an
optimistic evaluation of $\fpProbWOParams(\alarmAttacker)$.  For
example, Flor\^{e}ncio, et al.~\cite{florencio2014:guide} recommend
that an account should withstand targeted online password-guessing
attacks of $10^6$ attempts in practice.  As such, arguably
$\attackBudget = 1000$ is $1000\times$ too small.

An ideal honeyword-generation algorithm would achieve
$\fpProbWOParams(\alarmAttacker) \approx 0$ and
$\fnProbWOParams(\breachAttacker) = \frac{1}{\nmbrHoneywords + 1}$
(which is $0.05$ when $\nmbrHoneywords=19$) at $\alarmThreshold = 1$.
Unfortunately, no known honeyword algorithm comes close.  As seen in
\figref{fig:uc:to20}, the best $\fnProbWOParams(\breachAttacker)$ that
the honeyword-generation techniques accomplish overall is $0.54$
(\PasstoPath, \figref{fig:uc:to20:pass2path}), $0.56$ (\Tweak,
\figref{fig:uc:to20:tweak}), $0.57$ (\CHM,
\figref{fig:uc:to20:chm-pwd}), and $0.58$ (\CGPT,
\figref{fig:uc:to20:gpt}); all others have
$\fnProbWOParams(\breachAttacker) > 0.59$.  When we consider the
attacker prioritizing ``easy'' accounts,
$\fnProbWOParams(\breachAttacker)$ of \PasstoPath, \CGPT, and \Tweak are
at least $0.93$, $0.95$, and $0.96$, respectively, while others have
$\fnProbWOParams(\breachAttacker) \approx 1$. This indicates that the
false-negative attacker can break at least $43\%$ accounts by
targeting the ``easy'' ones, with only \PasstoPath, \CGPT, and \Tweak
presenting any significant chance of catching the attacker. That said,
when such an attacker wants to guess more account passwords, i.e.,
targeting the ``medium'' accounts after the ``easy'' ones, the
probability of inducing an alarm will increase with number of attacked
accounts since $\fnProbWOParams(\breachAttacker) < 0.88$ for the
``medium'' subcase when $\alarmThreshold = 1$. The four most
successful algorithms (\PasstoPath, \Tweak, \CHM, and \CGPT) are
password-context-dependent techniques that generate honeywords similar
to the account password, and thus it is more challenging for
\breachAttacker to distinguish the account password from honeywords
produced by these algorithms than from those of the other methods. We
conclude that honeywords more similar to the account password yield a
lower $\fnProbWOParams(\breachAttacker)$, though one that is still far
from $\frac{1}{\nmbrHoneywords + 1}$ due to password reuse.

However, \PasstoPath has $\fpProbWOParams(\alarmAttacker) \approx 0.89$
at $\alarmThreshold=1$, where most others have lower
$\fpProbWOParams(\alarmAttacker)$.  The only exception is \CHM, which
includes a deterministic step that searches for nearest neighbors of
the account password and thus yields a high false-positive rate,
$\fpProbWOParams(\alarmAttacker) \approx 1$.  While \PasstoPath is the
best technique for generating honeywords similar to the account
password, it is almost the easiest for the false-positive attacker to
guess the generated honeywords with \password known.  Still, no
generation method achieves $\fpProbWOParams(\alarmAttacker) \le 0.27$
at $\alarmThreshold = 1$. Growing \alarmThreshold of course reduces
$\fpProbWOParams(\alarmAttacker)$ but increases
$\fnProbWOParams(\breachAttacker)$: all methods capable of reaching
$\fpProbWOParams(\alarmAttacker) \approx 0$ do so with
$\fnProbWOParams(\breachAttacker) > 0.81$ overall,
$\fnProbWOParams(\breachAttacker) \approx 1$ for the ``easy'' subcase,
and $\fnProbWOParams(\breachAttacker) > 0.91$ for the ``medium''
subcase.

A natural method to decrease $\fnProbWOParams(\breachAttacker)$ would
be to increase the number \nmbrHoneywords of honeywords, but the more
pronounced effect of doing so is increasing
$\fpProbWOParams(\alarmAttacker)$, instead.  Indeed,
\figref{fig:uc:to100} shows the impact of increasing \nmbrHoneywords
to $\nmbrHoneywords = 99$.  As seen there, an order-of-magnitude
increase in \nmbrHoneywords resulted in a slight improvement to
$\fnProbWOParams(\breachAttacker)$ in each case, but a more substantial
increase to $\fpProbWOParams(\alarmAttacker)$.
    
To summarize, honeyword-generation techniques like \TarCombo that have
been demonstrated to have good flatness in previous works
(e.g.,~\cite{wang2022:honeywords}) fail to achieve a low
false-negative rate in our threat model, particularly not at settings
of \alarmThreshold to ensure a small false-positive rate. Among the
honeyword-generation techniques we consider, \PasstoPath achieves the
best \fnProbWOParams but has a high \fpProbWOParams.  Most other methods
have lower \fpProbWOParams but a higher \fnProbWOParams.  Regardless, in
the case of user-chosen passwords, no existing algorithm achieves low
rates of both false positives and false negatives.  In addition, when
the attacker targets the ``easy'' accounts that are approximately
$43\%$ of users, all the honeyword-generation methods are ineffective
in detecting a breach at settings of \alarmThreshold achieving
$\fpProbWOParams(\alarmAttacker) \approx 0$.

%% file: alg_generated.tex
\section{Algorithmically Generated Passwords}
\label{sec:algorithmic}

The second case we consider is when \choosePassword is implemented
using a password-generating algorithm. To our knowledge, this
  case has not been considered in prior honeyword research, and with
  good reason: Under the best practice of storing only
  preimage-resistant hashes of passwords, it should be exceptionally
  difficult for an attacker who breaches a site's database to recover
  algorithmically generated passwords, due to their comparatively high
  strength.  For this reason, algorithmically generated passwords in a
  breached credential database are primarily at risk if the hash
  function is less preimage-resistant than initially thought or when
  the site's database was reversibly encrypted---which, while not best
  practice, is necessary in some use cases
  (e.g.,~\cite{microsoft2022:store})---and the false-negative attacker
  recovered the decryption key along with the database.

We assume there is a large but limited number \NumGenerators of
password generators denoted as
$\{\choosePassword{\GeneratorIndex}\}_{\GeneratorIndex =
  1}^{\NumGenerators}$, each of which is defined by an algorithm and
values of user-configurable parameters.  We assume that each user
determines \choosePassword by choosing a generator uniformly at random
from $\{\choosePassword{\GeneratorIndex}\}_{\GeneratorIndex =
  1}^{\NumGenerators}$, and that each user stays with its choice.  To
justify this assumption, in \secref{sec:discussion:configuration} we report a brief study
we did using the password policies of 20 commonly visited websites and
Tranco Top 1M websites~\cite{pochat2018:tranco}, where we found that
setting passwords at these websites in a random order would permit the
user to retain her chosen password-generation configuration for $>
6.3$ sites in expectation, before encountering a site for which the
user's configuration was inconsistent.  This finding is consistent
with Alroomi et al.~\cite{alroomi2023:measuring}, who reported that
only $15\%$ sites have character constraints on password creation.

We assume the length of the generated passwords is one parameter that
users can configure.  Some password managers permit user configuration
of allowable symbols, as well.  Similarly, password managers that
enable generation of easy-to-read passwords might avoid use of certain
characters that are ambiguous in some fonts (e.g., ``1'' vs.\ ``l'' in
sans-serif fonts).  Password managers that generate easy-to-say
passwords might restrict the symbols used in different positions of a
password.  We will see examples below.  The user's choice of these
parameters will generally be unknown to the defender, except as
revealed by the account password \password.

In this section, we analyze the contribution of honeywords for
detecting credential database breaches for accounts with
algorithmically generated passwords.  In \appref{app:existing}, we
show that honeyword-generation methods used in the user-chosen
password case fail to achieve both low false-negative rate and low
false-positive rate for algorithmically generated passwords.  Although
utilizing password-generation algorithms to generate honeywords can do
better, in this section we show that the choice of selected generator
is critical to achieving a low false-negative rate.

\subsection{Attack Strategies}
\label{sec:algorithmic:strategies}

In this section, we introduce the false-positive attacker
\alarmAttacker and the false-negative attacker \breachAttacker used in
the evaluation of $\fpProbWOParams(\alarmAttacker)$ and
$\fnProbWOParams(\breachAttacker)$, respectively, when account
passwords are generated algorithmically.

\myparagraph{False-positive attacker \alarmAttacker}
\alarmAttacker uses the same strategy used in the case of user-chosen
passwords in \secref{sec:user-selected:attack-strategy}.
Specifically, the attacker \alarmAttacker leverages the
honeyword-generation algorithm \chooseHoneywords{} to generate a set
of candidates and sorts the candidates by their assigned
probabilities, if applicable. Finally, it picks the top \attackBudget
candidates as the guessed honeywords \passwordGuessSet.

\myparagraph{False-negative attacker \breachAttacker}
\breachAttacker was implemented as follows. Given \auxiliaryInfo,
\breachAttacker leverages a classifier $\Classifier(\cdot):
\passwordDomain \rightarrow [0,1]^{\NumGenerators}$ that outputs a
confidence score per possible class. The construction of this
classifier is described in
\appref{app:algorithmic:classifier}. \breachAttacker classifies each
element of \auxiliaryInfo using \Classifier, using the highest-scored
generator for each $\passwordAlt \in \auxiliaryInfo$ as a ``vote'' for
the password generator that the user employs; the password generator
obtaining the most such votes is denoted
$\choosePassword{\AttackerPrediction{\breachAttacker}}$. Then
\breachAttacker assigns scores to the sweetwords from $\honeywordSet{}
\cup \{\password\}$ as follows: if the length of the sweetword is the
same as those in \auxiliaryInfo, \breachAttacker utilizes the
classifier $\Classifier(\cdot)$ to value the sweetword by the
confidence score of being from class
$\AttackerPrediction{\breachAttacker}$; otherwise, \breachAttacker
will value it by $0$. The attacker ranks the sweetwords based on the
assigned scores and uses the top \alarmThreshold sweetwords as
\passwordGuessSet.

\subsection{Generating Honeywords Using Algorithmic Password Generators}
\label{sec:algorithmic:generator}

The honeyword-generation methods introduced in
\secref{sec:background:algorithms} do not fare well (in terms of
false-negative probability) when the account password is generated
algorithmically.  Intuitively, the password-independent honeyword
generators fail to achieve a low $\fnProbWOParams(\breachAttacker)$
since the honeywords they generate are user-chosen passwords, which
makes it easy for \breachAttacker to distinguish the algorithmically
generated account password from the honeywords.  Many
password-dependent generators do little better, because even though
the account password is algorithmically generated, these models are
trained on artifacts of human behavior, which renders the honeywords
recognizable to \breachAttacker.  The primary exceptions are \CBT{3} and
\CBT{4}, which are not trained at all.  These can achieve a low
$\fnProbWOParams(\breachAttacker)$, though still with a too-high
$\fpProbWOParams(\alarmAttacker)$.  We have empirically demonstrated
these findings in \appref{app:existing}.

Therefore, here we consider the use of algorithmic password generators
to generate honeywords for algorithmically generated passwords
submitted by the user. Given an account password \password, the
honeyword system selects a generator from
$\{\choosePassword{\GeneratorIndex}\}_{\GeneratorIndex =
  1}^{\NumGenerators}$ and then leverages the selected generator to
generate \nmbrHoneywords honeywords. We categorize the methods based
on the selection strategy, as follows:
\begin{itemize}[nosep,leftmargin=1em,labelwidth=*,align=left]
\item \FixedGenerator: Given a fixed $\choosePassword{\FixedIndex} \in
  \{\choosePassword{\GeneratorIndex}\}_{\GeneratorIndex =
    1}^{\NumGenerators}$, \chooseHoneywords{\nmbrHoneywords} samples
  \nmbrHoneywords distinct honeywords using
  $\choosePassword{\FixedIndex}$ to build \honeywordSet{}.
\item \RandomGenerator: \chooseHoneywords{\nmbrHoneywords} samples a
  \choosePassword{\GeneratorIndex} uniformly from
  $\{\choosePassword{\GeneratorIndex}\}_{\GeneratorIndex =
    1}^{\NumGenerators}$ and builds \honeywordSet{} by sampling
  \nmbrHoneywords distinct honeywords using
  \choosePassword{\GeneratorIndex}.
\item \ClassifiedGenerator: \chooseHoneywords{\nmbrHoneywords}
  classifies the account password into one of \NumGenerators classes,
  indicating the generator \choosePassword{\GeneratorIndex} most
  likely to have generated it.  \chooseHoneywords{\nmbrHoneywords}
  then builds \honeywordSet{} by sampling \nmbrHoneywords distinct
  honeywords using \choosePassword{\GeneratorIndex}.
\end{itemize}

\subsection{Evaluation}
\label{sec:algorithmic:eval}

\subsubsection{Dataset}
\label{sec:algorithmic:eval:dataset}    

The datasets we used to evaluate honeyword-generation strategies in
the case of algorithmically generated passwords were synthetically
produced by querying online password generators.\footnote{We used
PyAutoGui (\url{https://pyautogui.readthedocs.io/en/latest/}) to
automate interactions with the password managers like 1Password and
LastPass.  That is, we automated generating random passwords, copying
them into the clipboard, and storing them in a local file
interactively.}  Specifically, after browser-integrated password
managers (Google Password Manager and iCloud Keychain), LastPass and
1Password are two of the most widely used password
managers/generators~\cite{vigderman2023:managers}.  LastPass permits
the user to select one of three password-generation algorithms, namely
``Easy-to-say'', ``Easy-to-read'', or ``All-characters''.  For each
type, users can further specify the generator by checking or
unchecking ``Uppercase'', ``Lowercase'', ``Numbers'', or ``Symbols'',
though the ``Easy-to-say'' generator does not permit inclusion of
Symbols or Numbers.  1Password allows users to select a type of
password from among ``Random Password'', ``Memorable Password'', and
``Pin''. The Random Password generator includes Lowercase and
Uppercase letters always, but users can check or uncheck Numbers and
Symbols.  The Memorable Password algorithm generates memorable
passwords, each of which is a sequence of word fragments connected by
separators. In this option, users can select separators among
``Hyphens'', ``Spaces'', ``Periods'', ``Commas'', ``Underscores'',
``Numbers'', and ``Numbers and Symbols''. In addition, users could
check or uncheck ``Full Words'' and ``Capitalize'' to specify the
``Memorable Password'' generator. In this work, we used all the
configurations from LastPass and 1Password's Random Password, and
selected configurations for 1Password's Memorable Password. We
consider passwords generated from each specification as one class,
yielding $38$ classes in total.  These classes are shown in
\tblref{table:generators}. We set the fixed
$\choosePassword{\FixedIndex}$ to be the ``All characters'' generator
from LastPass with ``U'', ``L'', ``S'', and ``N'' checked
($\FixedIndex = 32$).

\begin{table}[ht!]
\footnotesize
\begin{center}
\begin{tabular}{@{}ccccccc@{}}
\toprule
\multicolumn{1}{@{}c}{Class} & \multirow{2}{*}{Manager} & \multirow{2}{*}{Type} & \multicolumn{4}{c}{Alphabet} \\
\multicolumn{1}{@{}c}{index}
   &           &                & U       & L       & S       & N       \\ \midrule
 1 & LastPass  & Easy to say    &         & \chkmrk &         &         \\ 
 2 & LastPass  & Easy to say    & \chkmrk &         &         &         \\ 
 3 & LastPass  & Easy to say    & \chkmrk & \chkmrk &         &         \\ 
 4 & LastPass  & Easy to read   &         &         & \chkmrk &         \\ 
 5 & LastPass  & Easy to read   &         &         &         & \chkmrk \\ 
 6 & LastPass  & Easy to read   &         &         & \chkmrk & \chkmrk \\ 
 7 & LastPass  & Easy to read   &         & \chkmrk &         &         \\ 
 8 & LastPass  & Easy to read   &         & \chkmrk & \chkmrk &         \\ 
 9 & LastPass  & Easy to read   &         & \chkmrk &         & \chkmrk \\ 
10 & LastPass  & Easy to read   &         & \chkmrk & \chkmrk & \chkmrk \\
11 & LastPass  & Easy to read   & \chkmrk &         &         &         \\ 
12 & LastPass  & Easy to read   & \chkmrk &         & \chkmrk &         \\ 
13 & LastPass  & Easy to read   & \chkmrk &         &         & \chkmrk \\ 
14 & LastPass  & Easy to read   & \chkmrk &         & \chkmrk & \chkmrk \\ 
15 & LastPass  & Easy to read   & \chkmrk & \chkmrk &         &         \\ 
16 & LastPass  & Easy to read   & \chkmrk & \chkmrk & \chkmrk &         \\ 
17 & LastPass  & Easy to read   & \chkmrk & \chkmrk &         & \chkmrk \\ 
18 & LastPass  & Easy to read   & \chkmrk & \chkmrk & \chkmrk & \chkmrk \\ 
19 & LastPass  & All characters &         &         &         & \chkmrk \\ 
20 & LastPass  & All characters &         &         & \chkmrk & \chkmrk \\ 
21 & LastPass  & All characters &         & \chkmrk &         &         \\ 
22 & LastPass  & All characters &         & \chkmrk & \chkmrk &         \\ 
23 & LastPass  & All characters &         & \chkmrk &         & \chkmrk \\ 
24 & LastPass  & All characters &         & \chkmrk & \chkmrk & \chkmrk \\ 
25 & LastPass  & All characters & \chkmrk &         &         &         \\
26 & LastPass  & All characters & \chkmrk &         & \chkmrk &         \\
27 & LastPass  & All characters & \chkmrk &         &         & \chkmrk \\ 
28 & LastPass  & All characters & \chkmrk &         & \chkmrk & \chkmrk \\
29 & LastPass  & All characters & \chkmrk & \chkmrk &         &         \\ 
30 & LastPass  & All characters & \chkmrk & \chkmrk & \chkmrk &         \\ 
31 & LastPass  & All characters & \chkmrk & \chkmrk &         & \chkmrk \\ 
32 & LastPass  & All characters & \chkmrk & \chkmrk & \chkmrk & \chkmrk \\ 
33 & 1Password & Random Password & \chkmrk & \chkmrk & \chkmrk &         \\ 
34 & 1Password & Random Password & \chkmrk & \chkmrk &         &         \\ 
35 & 1Password & Random Password & \chkmrk & \chkmrk & \chkmrk & \chkmrk \\ 
36 & 1Password & Random Password & \chkmrk & \chkmrk &         & \chkmrk \\ 
37 & 1Password & Memorable Password &  & \chkmrk &         & \chkmrk \\ 
38 & 1Password & Memorable Password &  & \chkmrk & \chkmrk & \chkmrk \\\bottomrule 
\end{tabular}
\end{center}
\caption{Classes of algorithmically generated passwords used in our
  experiments}
\label{table:generators}
\vspace{2ex}
\end{table}

Using these online generators, we generated three datasets, denoted
\Dataset{\TrainTag}{\AlgGenTag}, \Dataset{\EvalTag}{\AlgGenTag}, and
\Dataset{\TestTag}{\AlgGenTag}, all consisting of passwords of length
$14$ only.  We used \Dataset{\TrainTag}{\AlgGenTag} to train a
classifier to classify random passwords and evaluated it on
\Dataset{\EvalTag}{\AlgGenTag}. To assemble
\Dataset{\TrainTag}{\AlgGenTag} and \Dataset{\EvalTag}{\AlgGenTag}, we
generated $8{,}000$ and $2{,}000$ passwords from each class, yielding
$304{,}000$ and $76{,}000$ passwords in total, respectively. We
applied \Dataset{\TestTag}{\AlgGenTag} in the evaluation of
\fpProbWOParams and \fnProbWOParams. In \Dataset{\TestTag}{\AlgGenTag},
there were $38$ classes of passwords, each containing $10{,}000$ sets
(corresponding to $10{,}000$ users) with $100$ passwords of that
class.  When evaluating \fpProbWOParams and \fnProbWOParams, we
implemented \choosePassword by sampling \password and \auxiliaryInfo
without replacement from a set (user) chosen uniformly at random from
\Dataset{\TestTag}{\AlgGenTag}.

\begin{figure}[t]
\vspace{2ex}
\begin{subfigure}[b]{0.45\columnwidth}
\caption{Honeyword generator \chooseHoneywords{\nmbrHoneywords}}
\vspace{2ex}
\resizebox{\textwidth}{!}{
\setcounter{MinNumber}{0}%
\setcounter{MaxNumber}{1}%
\setlength\tabcolsep{0pt}
\begin{tabular}{r*{38}{X}}
\toprule
& \multicolumn{1}{c}{1}
& \multicolumn{1}{c}{2}
& \multicolumn{1}{c}{3}
& \multicolumn{1}{c}{4}
& \multicolumn{1}{c}{5}
& \multicolumn{1}{c}{6}
& \multicolumn{1}{c}{7}
& \multicolumn{1}{c}{8}
& \multicolumn{1}{c}{9}
& \multicolumn{1}{c}{10}
& \multicolumn{1}{c}{11}
& \multicolumn{1}{c}{12}
& \multicolumn{1}{c}{13}
& \multicolumn{1}{c}{14}
& \multicolumn{1}{c}{15}
& \multicolumn{1}{c}{16}
& \multicolumn{1}{c}{17}
& \multicolumn{1}{c}{18}
& \multicolumn{1}{c}{19}
& \multicolumn{1}{c}{20}
& \multicolumn{1}{c}{21}
& \multicolumn{1}{c}{22}
& \multicolumn{1}{c}{23}
& \multicolumn{1}{c}{24}
& \multicolumn{1}{c}{25}
& \multicolumn{1}{c}{26}
& \multicolumn{1}{c}{27}
& \multicolumn{1}{c}{28}
& \multicolumn{1}{c}{29}
& \multicolumn{1}{c}{30}
& \multicolumn{1}{c}{31}
& \multicolumn{1}{c}{32}
& \multicolumn{1}{c}{33}
& \multicolumn{1}{c}{34}
& \multicolumn{1}{c}{35}
& \multicolumn{1}{c}{36}
& \multicolumn{1}{c}{37}
& \multicolumn{1}{c}{38} \\
1 & 0.992 & 0.0 & 0.0 & 0.0 & 0.0 & 0.0 & 0.0 & 0.0 & 0.0 & 0.0 & 0.0 & 0.0 & 0.0 & 0.0 & 0.0 & 0.0 & 0.0 & 0.0 & 0.0 & 0.0 & 0.008 & 0.0 & 0.0 & 0.0 & 0.0 & 0.0 & 0.0 & 0.0 & 0.0 & 0.0 & 0.0 & 0.0 & 0.0 & 0.0 & 0.0 & 0.0 & 0.0 & 0.0 \\ 
2 & 0.0 & 0.993 & 0.0 & 0.0 & 0.0 & 0.0 & 0.0 & 0.0 & 0.0 & 0.0 & 0.002 & 0.0 & 0.0 & 0.0 & 0.0 & 0.0 & 0.0 & 0.0 & 0.0 & 0.0 & 0.0 & 0.0 & 0.0 & 0.0 & 0.005 & 0.0 & 0.0 & 0.0 & 0.0 & 0.0 & 0.0 & 0.0 & 0.0 & 0.0 & 0.0 & 0.0 & 0.0 & 0.0 \\ 
3 & 0.002 & 0.0 & 0.986 & 0.0 & 0.0 & 0.0 & 0.0 & 0.0 & 0.0 & 0.0 & 0.0 & 0.0 & 0.0 & 0.0 & 0.003 & 0.0 & 0.0 & 0.0 & 0.0 & 0.0 & 0.0 & 0.0 & 0.0 & 0.0 & 0.0 & 0.0 & 0.0 & 0.0 & 0.008 & 0.0 & 0.0 & 0.0 & 0.0 & 0.001 & 0.0 & 0.0 & 0.0 & 0.0 \\ 
4 & 0.0 & 0.0 & 0.0 & 1.0 & 0.0 & 0.0 & 0.0 & 0.0 & 0.0 & 0.0 & 0.0 & 0.0 & 0.0 & 0.0 & 0.0 & 0.0 & 0.0 & 0.0 & 0.0 & 0.0 & 0.0 & 0.0 & 0.0 & 0.0 & 0.0 & 0.0 & 0.0 & 0.0 & 0.0 & 0.0 & 0.0 & 0.0 & 0.0 & 0.0 & 0.0 & 0.0 & 0.0 & 0.0 \\ 
5 & 0.0 & 0.0 & 0.0 & 0.0 & 1.0 & 0.0 & 0.0 & 0.0 & 0.0 & 0.0 & 0.0 & 0.0 & 0.0 & 0.0 & 0.0 & 0.0 & 0.0 & 0.0 & 0.0 & 0.0 & 0.0 & 0.0 & 0.0 & 0.0 & 0.0 & 0.0 & 0.0 & 0.0 & 0.0 & 0.0 & 0.0 & 0.0 & 0.0 & 0.0 & 0.0 & 0.0 & 0.0 & 0.0 \\ 
6 & 0.0 & 0.0 & 0.0 & 0.0 & 0.0 & 1.0 & 0.0 & 0.0 & 0.0 & 0.0 & 0.0 & 0.0 & 0.0 & 0.0 & 0.0 & 0.0 & 0.0 & 0.0 & 0.0 & 0.0 & 0.0 & 0.0 & 0.0 & 0.0 & 0.0 & 0.0 & 0.0 & 0.0 & 0.0 & 0.0 & 0.0 & 0.0 & 0.0 & 0.0 & 0.0 & 0.0 & 0.0 & 0.0 \\ 
7 & 0.001 & 0.0 & 0.0 & 0.0 & 0.0 & 0.0 & 0.999 & 0.0 & 0.0 & 0.0 & 0.0 & 0.0 & 0.0 & 0.0 & 0.0 & 0.0 & 0.0 & 0.0 & 0.0 & 0.0 & 0.0 & 0.0 & 0.0 & 0.0 & 0.0 & 0.0 & 0.0 & 0.0 & 0.0 & 0.0 & 0.0 & 0.0 & 0.0 & 0.0 & 0.0 & 0.0 & 0.0 & 0.0 \\ 
8 & 0.0 & 0.0 & 0.0 & 0.0 & 0.0 & 0.0 & 0.0 & 1.0 & 0.0 & 0.0 & 0.0 & 0.0 & 0.0 & 0.0 & 0.0 & 0.0 & 0.0 & 0.0 & 0.0 & 0.0 & 0.0 & 0.0 & 0.0 & 0.0 & 0.0 & 0.0 & 0.0 & 0.0 & 0.0 & 0.0 & 0.0 & 0.0 & 0.0 & 0.0 & 0.0 & 0.0 & 0.0 & 0.0 \\ 
9 & 0.0 & 0.0 & 0.0 & 0.0 & 0.0 & 0.0 & 0.0 & 0.0 & 1.0 & 0.0 & 0.0 & 0.0 & 0.0 & 0.0 & 0.0 & 0.0 & 0.0 & 0.0 & 0.0 & 0.0 & 0.0 & 0.0 & 0.0 & 0.0 & 0.0 & 0.0 & 0.0 & 0.0 & 0.0 & 0.0 & 0.0 & 0.0 & 0.0 & 0.0 & 0.0 & 0.0 & 0.0 & 0.0 \\ 
10 & 0.0 & 0.0 & 0.0 & 0.0 & 0.0 & 0.0 & 0.0 & 0.0 & 0.0 & 1.0 & 0.0 & 0.0 & 0.0 & 0.0 & 0.0 & 0.0 & 0.0 & 0.0 & 0.0 & 0.0 & 0.0 & 0.0 & 0.0 & 0.0 & 0.0 & 0.0 & 0.0 & 0.0 & 0.0 & 0.0 & 0.0 & 0.0 & 0.0 & 0.0 & 0.0 & 0.0 & 0.0 & 0.0 \\ 
11 & 0.0 & 0.001 & 0.0 & 0.0 & 0.0 & 0.0 & 0.0 & 0.0 & 0.0 & 0.0 & 0.999 & 0.0 & 0.0 & 0.0 & 0.0 & 0.0 & 0.0 & 0.0 & 0.0 & 0.0 & 0.0 & 0.0 & 0.0 & 0.0 & 0.0 & 0.0 & 0.0 & 0.0 & 0.0 & 0.0 & 0.0 & 0.0 & 0.0 & 0.0 & 0.0 & 0.0 & 0.0 & 0.0 \\ 
12 & 0.0 & 0.0 & 0.0 & 0.0 & 0.0 & 0.0 & 0.0 & 0.0 & 0.0 & 0.0 & 0.0 & 1.0 & 0.0 & 0.0 & 0.0 & 0.0 & 0.0 & 0.0 & 0.0 & 0.0 & 0.0 & 0.0 & 0.0 & 0.0 & 0.0 & 0.0 & 0.0 & 0.0 & 0.0 & 0.0 & 0.0 & 0.0 & 0.0 & 0.0 & 0.0 & 0.0 & 0.0 & 0.0 \\ 
13 & 0.0 & 0.0 & 0.0 & 0.0 & 0.0 & 0.0 & 0.0 & 0.0 & 0.0 & 0.0 & 0.0 & 0.0 & 1.0 & 0.0 & 0.0 & 0.0 & 0.0 & 0.0 & 0.0 & 0.0 & 0.0 & 0.0 & 0.0 & 0.0 & 0.0 & 0.0 & 0.0 & 0.0 & 0.0 & 0.0 & 0.0 & 0.0 & 0.0 & 0.0 & 0.0 & 0.0 & 0.0 & 0.0 \\ 
14 & 0.0 & 0.0 & 0.0 & 0.0 & 0.0 & 0.0 & 0.0 & 0.0 & 0.0 & 0.0 & 0.0 & 0.0 & 0.0 & 1.0 & 0.0 & 0.0 & 0.0 & 0.0 & 0.0 & 0.0 & 0.0 & 0.0 & 0.0 & 0.0 & 0.0 & 0.0 & 0.0 & 0.0 & 0.0 & 0.0 & 0.0 & 0.0 & 0.0 & 0.0 & 0.0 & 0.0 & 0.0 & 0.0 \\ 
15 & 0.0 & 0.0 & 0.0 & 0.0 & 0.0 & 0.0 & 0.0 & 0.0 & 0.0 & 0.0 & 0.0 & 0.0 & 0.0 & 0.0 & 1.0 & 0.0 & 0.0 & 0.0 & 0.0 & 0.0 & 0.0 & 0.0 & 0.0 & 0.0 & 0.0 & 0.0 & 0.0 & 0.0 & 0.0 & 0.0 & 0.0 & 0.0 & 0.0 & 0.0 & 0.0 & 0.0 & 0.0 & 0.0 \\ 
16 & 0.0 & 0.0 & 0.0 & 0.0 & 0.0 & 0.0 & 0.0 & 0.0 & 0.0 & 0.0 & 0.0 & 0.0 & 0.0 & 0.0 & 0.0 & 1.0 & 0.0 & 0.0 & 0.0 & 0.0 & 0.0 & 0.0 & 0.0 & 0.0 & 0.0 & 0.0 & 0.0 & 0.0 & 0.0 & 0.0 & 0.0 & 0.0 & 0.0 & 0.0 & 0.0 & 0.0 & 0.0 & 0.0 \\ 
17 & 0.0 & 0.0 & 0.0 & 0.0 & 0.0 & 0.0 & 0.0 & 0.0 & 0.0 & 0.0 & 0.0 & 0.0 & 0.0 & 0.0 & 0.0 & 0.0 & 1.0 & 0.0 & 0.0 & 0.0 & 0.0 & 0.0 & 0.0 & 0.0 & 0.0 & 0.0 & 0.0 & 0.0 & 0.0 & 0.0 & 0.0 & 0.0 & 0.0 & 0.0 & 0.0 & 0.0 & 0.0 & 0.0 \\ 
18 & 0.0 & 0.0 & 0.0 & 0.0 & 0.0 & 0.0 & 0.0 & 0.0 & 0.0 & 0.0 & 0.0 & 0.0 & 0.0 & 0.0 & 0.0 & 0.0 & 0.0 & 1.0 & 0.0 & 0.0 & 0.0 & 0.0 & 0.0 & 0.0 & 0.0 & 0.0 & 0.0 & 0.0 & 0.0 & 0.0 & 0.0 & 0.0 & 0.0 & 0.0 & 0.0 & 0.0 & 0.0 & 0.0 \\ 
19 & 0.0 & 0.0 & 0.0 & 0.0 & 0.035 & 0.0 & 0.0 & 0.0 & 0.0 & 0.0 & 0.0 & 0.0 & 0.0 & 0.0 & 0.0 & 0.0 & 0.0 & 0.0 & 0.965 & 0.0 & 0.0 & 0.0 & 0.0 & 0.0 & 0.0 & 0.0 & 0.0 & 0.0 & 0.0 & 0.0 & 0.0 & 0.0 & 0.0 & 0.0 & 0.0 & 0.0 & 0.0 & 0.0 \\ 
20 & 0.0 & 0.0 & 0.0 & 0.0 & 0.0 & 0.175 & 0.0 & 0.0 & 0.0 & 0.0 & 0.0 & 0.0 & 0.0 & 0.0 & 0.0 & 0.0 & 0.0 & 0.0 & 0.0 & 0.825 & 0.0 & 0.0 & 0.0 & 0.0 & 0.0 & 0.0 & 0.0 & 0.0 & 0.0 & 0.0 & 0.0 & 0.0 & 0.0 & 0.0 & 0.0 & 0.0 & 0.0 & 0.0 \\ 
21 & 0.008 & 0.0 & 0.0 & 0.0 & 0.0 & 0.0 & 0.18 & 0.0 & 0.0 & 0.0 & 0.0 & 0.0 & 0.0 & 0.0 & 0.0 & 0.0 & 0.0 & 0.0 & 0.0 & 0.0 & 0.812 & 0.0 & 0.0 & 0.0 & 0.0 & 0.0 & 0.0 & 0.0 & 0.0 & 0.0 & 0.0 & 0.0 & 0.0 & 0.0 & 0.0 & 0.0 & 0.0 & 0.0 \\ 
22 & 0.0 & 0.0 & 0.0 & 0.0 & 0.0 & 0.0 & 0.0 & 0.309 & 0.0 & 0.0 & 0.0 & 0.0 & 0.0 & 0.0 & 0.0 & 0.0 & 0.0 & 0.0 & 0.0 & 0.0 & 0.0 & 0.691 & 0.0 & 0.0 & 0.0 & 0.0 & 0.0 & 0.0 & 0.0 & 0.0 & 0.0 & 0.0 & 0.0 & 0.0 & 0.0 & 0.0 & 0.0 & 0.0 \\ 
23 & 0.0 & 0.0 & 0.0 & 0.0 & 0.0 & 0.0 & 0.0 & 0.0 & 0.103 & 0.0 & 0.0 & 0.0 & 0.0 & 0.0 & 0.0 & 0.0 & 0.0 & 0.0 & 0.0 & 0.0 & 0.0 & 0.0 & 0.897 & 0.0 & 0.0 & 0.0 & 0.0 & 0.0 & 0.0 & 0.0 & 0.0 & 0.0 & 0.0 & 0.0 & 0.0 & 0.0 & 0.0 & 0.0 \\ 
24 & 0.0 & 0.0 & 0.0 & 0.0 & 0.0 & 0.0 & 0.0 & 0.0 & 0.0 & 0.128 & 0.0 & 0.0 & 0.0 & 0.0 & 0.0 & 0.0 & 0.0 & 0.0 & 0.0 & 0.0 & 0.0 & 0.0 & 0.0 & 0.872 & 0.0 & 0.0 & 0.0 & 0.0 & 0.0 & 0.0 & 0.0 & 0.0 & 0.0 & 0.0 & 0.0 & 0.0 & 0.0 & 0.0 \\ 
25 & 0.0 & 0.009 & 0.0 & 0.0 & 0.0 & 0.0 & 0.0 & 0.0 & 0.0 & 0.0 & 0.21 & 0.0 & 0.0 & 0.0 & 0.0 & 0.0 & 0.0 & 0.0 & 0.0 & 0.0 & 0.0 & 0.0 & 0.0 & 0.0 & 0.781 & 0.0 & 0.0 & 0.0 & 0.0 & 0.0 & 0.0 & 0.0 & 0.0 & 0.0 & 0.0 & 0.0 & 0.0 & 0.0 \\ 
26 & 0.0 & 0.0 & 0.0 & 0.0 & 0.0 & 0.0 & 0.0 & 0.0 & 0.0 & 0.0 & 0.0 & 0.269 & 0.0 & 0.0 & 0.0 & 0.0 & 0.0 & 0.0 & 0.0 & 0.0 & 0.0 & 0.0 & 0.0 & 0.0 & 0.0 & 0.731 & 0.0 & 0.0 & 0.0 & 0.0 & 0.0 & 0.0 & 0.0 & 0.0 & 0.0 & 0.0 & 0.0 & 0.0 \\ 
27 & 0.0 & 0.0 & 0.0 & 0.0 & 0.0 & 0.0 & 0.0 & 0.0 & 0.0 & 0.0 & 0.0 & 0.0 & 0.085 & 0.0 & 0.0 & 0.0 & 0.0 & 0.0 & 0.0 & 0.0 & 0.0 & 0.0 & 0.0 & 0.0 & 0.0 & 0.0 & 0.915 & 0.0 & 0.0 & 0.0 & 0.0 & 0.0 & 0.0 & 0.0 & 0.0 & 0.0 & 0.0 & 0.0 \\ 
28 & 0.0 & 0.0 & 0.0 & 0.0 & 0.0 & 0.0 & 0.0 & 0.0 & 0.0 & 0.0 & 0.0 & 0.0 & 0.0 & 0.152 & 0.0 & 0.0 & 0.0 & 0.0 & 0.0 & 0.0 & 0.0 & 0.0 & 0.0 & 0.0 & 0.0 & 0.0 & 0.0 & 0.848 & 0.0 & 0.0 & 0.0 & 0.0 & 0.0 & 0.0 & 0.0 & 0.0 & 0.0 & 0.0 \\ 
29 & 0.0 & 0.0 & 0.004 & 0.0 & 0.0 & 0.0 & 0.0 & 0.0 & 0.0 & 0.0 & 0.0 & 0.0 & 0.0 & 0.0 & 0.184 & 0.0 & 0.0 & 0.0 & 0.0 & 0.0 & 0.0 & 0.0 & 0.0 & 0.0 & 0.0 & 0.0 & 0.0 & 0.0 & 0.62 & 0.0 & 0.0 & 0.0 & 0.0 & 0.192 & 0.0 & 0.0 & 0.0 & 0.0 \\
30 & 0.0 & 0.0 & 0.0 & 0.0 & 0.0 & 0.0 & 0.0 & 0.0 & 0.0 & 0.0 & 0.0 & 0.0 & 0.0 & 0.0 & 0.0 & 0.241 & 0.0 & 0.0 & 0.0 & 0.0 & 0.0 & 0.0 & 0.0 & 0.0 & 0.0 & 0.0 & 0.0 & 0.0 & 0.0 & 0.758 & 0.0 & 0.0 & 0.001 & 0.0 & 0.0 & 0.0 & 0.0 & 0.0 \\ 
31 & 0.0 & 0.0 & 0.0 & 0.0 & 0.0 & 0.0 & 0.0 & 0.0 & 0.0 & 0.0 & 0.0 & 0.0 & 0.0 & 0.0 & 0.0 & 0.0 & 0.112 & 0.0 & 0.0 & 0.0 & 0.0 & 0.0 & 0.0 & 0.0 & 0.0 & 0.0 & 0.0 & 0.0 & 0.0 & 0.0 & 0.774 & 0.0 & 0.0 & 0.0 & 0.0 & 0.114 & 0.0 & 0.0 \\ 
32 & 0.0 & 0.0 & 0.0 & 0.0 & 0.0 & 0.0 & 0.0 & 0.0 & 0.0 & 0.0 & 0.0 & 0.0 & 0.0 & 0.0 & 0.0 & 0.0 & 0.0 & 0.145 & 0.0 & 0.0 & 0.0 & 0.0 & 0.0 & 0.0 & 0.0 & 0.0 & 0.0 & 0.0 & 0.0 & 0.0 & 0.0 & 0.847 & 0.0 & 0.0 & 0.008 & 0.0 & 0.0 & 0.0 \\ 
33 & 0.0 & 0.0 & 0.0 & 0.0 & 0.0 & 0.0 & 0.0 & 0.0 & 0.0 & 0.0 & 0.0 & 0.0 & 0.0 & 0.0 & 0.0 & 0.015 & 0.0 & 0.0 & 0.0 & 0.0 & 0.0 & 0.0 & 0.0 & 0.0 & 0.0 & 0.0 & 0.0 & 0.0 & 0.0 & 0.018 & 0.0 & 0.0 & 0.967 & 0.0 & 0.0 & 0.0 & 0.0 & 0.0 \\ 
34 & 0.0 & 0.0 & 0.004 & 0.0 & 0.0 & 0.0 & 0.0 & 0.0 & 0.0 & 0.0 & 0.0 & 0.0 & 0.0 & 0.0 & 0.416 & 0.0 & 0.0 & 0.0 & 0.0 & 0.0 & 0.0 & 0.0 & 0.0 & 0.0 & 0.0 & 0.0 & 0.0 & 0.0 & 0.0 & 0.0 & 0.0 & 0.0 & 0.0 & 0.58 & 0.0 & 0.0 & 0.0 & 0.0 \\ 
35 & 0.0 & 0.0 & 0.0 & 0.0 & 0.0 & 0.0 & 0.0 & 0.0 & 0.0 & 0.0 & 0.0 & 0.0 & 0.0 & 0.0 & 0.0 & 0.0 & 0.0 & 0.013 & 0.0 & 0.0 & 0.0 & 0.0 & 0.0 & 0.0 & 0.0 & 0.0 & 0.0 & 0.0 & 0.0 & 0.0 & 0.0 & 0.024 & 0.0 & 0.0 & 0.963 & 0.0 & 0.0 & 0.0 \\ 
36 & 0.0 & 0.0 & 0.0 & 0.0 & 0.0 & 0.0 & 0.0 & 0.0 & 0.0 & 0.0 & 0.0 & 0.0 & 0.0 & 0.0 & 0.0 & 0.0 & 0.097 & 0.0 & 0.0 & 0.0 & 0.0 & 0.0 & 0.001 & 0.0 & 0.0 & 0.0 & 0.0 & 0.0 & 0.0 & 0.0 & 0.13 & 0.0 & 0.0 & 0.0 & 0.0 & 0.772 & 0.0 & 0.0 \\ 
37 & 0.0 & 0.0 & 0.0 & 0.0 & 0.0 & 0.0 & 0.0 & 0.0 & 0.0 & 0.0 & 0.0 & 0.0 & 0.0 & 0.0 & 0.0 & 0.0 & 0.0 & 0.0 & 0.0 & 0.0 & 0.0 & 0.0 & 0.0 & 0.0 & 0.0 & 0.0 & 0.0 & 0.0 & 0.0 & 0.0 & 0.0 & 0.0 & 0.0 & 0.0 & 0.0 & 0.0 & 1.0 & 0.0 \\ 
38 & 0.0 & 0.0 & 0.0 & 0.0 & 0.0 & 0.0 & 0.0 & 0.0 & 0.0 & 0.0 & 0.0 & 0.0 & 0.0 & 0.0 & 0.0 & 0.0 & 0.0 & 0.0 & 0.0 & 0.0 & 0.0 & 0.0 & 0.0 & 0.0 & 0.0 & 0.0 & 0.0 & 0.0 & 0.0 & 0.0 & 0.0 & 0.0 & 0.0 & 0.0 & 0.0 & 0.0 & 0.0 & 1.0 \\
\bottomrule 
\end{tabular}
}
\label{fig:cm:hw}
\end{subfigure}
\hfill
\begin{subfigure}[b]{0.45\columnwidth}
\caption{Attacker \breachAttacker}
\vspace{2ex}
\resizebox{\textwidth}{!}{
\setcounter{MinNumber}{0}%
\setcounter{MaxNumber}{1}%
\setlength\tabcolsep{0pt}
\begin{tabular}{r*{38}{X}}
\toprule
& \multicolumn{1}{c}{1}
& \multicolumn{1}{c}{2}
& \multicolumn{1}{c}{3}
& \multicolumn{1}{c}{4}
& \multicolumn{1}{c}{5}
& \multicolumn{1}{c}{6}
& \multicolumn{1}{c}{7}
& \multicolumn{1}{c}{8}
& \multicolumn{1}{c}{9}
& \multicolumn{1}{c}{10}
& \multicolumn{1}{c}{11}
& \multicolumn{1}{c}{12}
& \multicolumn{1}{c}{13}
& \multicolumn{1}{c}{14}
& \multicolumn{1}{c}{15}
& \multicolumn{1}{c}{16}
& \multicolumn{1}{c}{17}
& \multicolumn{1}{c}{18}
& \multicolumn{1}{c}{19}
& \multicolumn{1}{c}{20}
& \multicolumn{1}{c}{21}
& \multicolumn{1}{c}{22}
& \multicolumn{1}{c}{23}
& \multicolumn{1}{c}{24}
& \multicolumn{1}{c}{25}
& \multicolumn{1}{c}{26}
& \multicolumn{1}{c}{27}
& \multicolumn{1}{c}{28}
& \multicolumn{1}{c}{29}
& \multicolumn{1}{c}{30}
& \multicolumn{1}{c}{31}
& \multicolumn{1}{c}{32}
& \multicolumn{1}{c}{33}
& \multicolumn{1}{c}{34}
& \multicolumn{1}{c}{35}
& \multicolumn{1}{c}{36}
& \multicolumn{1}{c}{37}
& \multicolumn{1}{c}{38} \\
1 & 0.997 & 0.0 & 0.0 & 0.0 & 0.0 & 0.0 & 0.0 & 0.0 & 0.0 & 0.0 & 0.0 & 0.0 & 0.0 & 0.0 & 0.0 & 0.0 & 0.0 & 0.0 & 0.0 & 0.0 & 0.002 & 0.0 & 0.0 & 0.0 & 0.0 & 0.0 & 0.0 & 0.0 & 0.0 & 0.0 & 0.0 & 0.0 & 0.0 & 0.0 & 0.0 & 0.0 & 0.0 & 0.0 \\ 
2 & 0.0 & 1.0 & 0.0 & 0.0 & 0.0 & 0.0 & 0.0 & 0.0 & 0.0 & 0.0 & 0.0 & 0.0 & 0.0 & 0.0 & 0.0 & 0.0 & 0.0 & 0.0 & 0.0 & 0.0 & 0.0 & 0.0 & 0.0 & 0.0 & 0.0 & 0.0 & 0.0 & 0.0 & 0.0 & 0.0 & 0.0 & 0.0 & 0.0 & 0.0 & 0.0 & 0.0 & 0.0 & 0.0 \\ 
3 & 0.0 & 0.0 & 0.999 & 0.0 & 0.0 & 0.0 & 0.0 & 0.0 & 0.0 & 0.0 & 0.0 & 0.0 & 0.0 & 0.0 & 0.0 & 0.0 & 0.0 & 0.0 & 0.0 & 0.0 & 0.0 & 0.0 & 0.0 & 0.0 & 0.0 & 0.0 & 0.0 & 0.0 & 0.0 & 0.0 & 0.0 & 0.0 & 0.0 & 0.0 & 0.0 & 0.0 & 0.0 & 0.0 \\ 
4 & 0.0 & 0.0 & 0.0 & 1.0 & 0.0 & 0.0 & 0.0 & 0.0 & 0.0 & 0.0 & 0.0 & 0.0 & 0.0 & 0.0 & 0.0 & 0.0 & 0.0 & 0.0 & 0.0 & 0.0 & 0.0 & 0.0 & 0.0 & 0.0 & 0.0 & 0.0 & 0.0 & 0.0 & 0.0 & 0.0 & 0.0 & 0.0 & 0.0 & 0.0 & 0.0 & 0.0 & 0.0 & 0.0 \\ 
5 & 0.0 & 0.0 & 0.0 & 0.0 & 1.0 & 0.0 & 0.0 & 0.0 & 0.0 & 0.0 & 0.0 & 0.0 & 0.0 & 0.0 & 0.0 & 0.0 & 0.0 & 0.0 & 0.0 & 0.0 & 0.0 & 0.0 & 0.0 & 0.0 & 0.0 & 0.0 & 0.0 & 0.0 & 0.0 & 0.0 & 0.0 & 0.0 & 0.0 & 0.0 & 0.0 & 0.0 & 0.0 & 0.0 \\ 
6 & 0.0 & 0.0 & 0.0 & 0.0 & 0.0 & 1.0 & 0.0 & 0.0 & 0.0 & 0.0 & 0.0 & 0.0 & 0.0 & 0.0 & 0.0 & 0.0 & 0.0 & 0.0 & 0.0 & 0.0 & 0.0 & 0.0 & 0.0 & 0.0 & 0.0 & 0.0 & 0.0 & 0.0 & 0.0 & 0.0 & 0.0 & 0.0 & 0.0 & 0.0 & 0.0 & 0.0 & 0.0 & 0.0 \\ 
7 & 0.0 & 0.0 & 0.0 & 0.0 & 0.0 & 0.0 & 1.0 & 0.0 & 0.0 & 0.0 & 0.0 & 0.0 & 0.0 & 0.0 & 0.0 & 0.0 & 0.0 & 0.0 & 0.0 & 0.0 & 0.0 & 0.0 & 0.0 & 0.0 & 0.0 & 0.0 & 0.0 & 0.0 & 0.0 & 0.0 & 0.0 & 0.0 & 0.0 & 0.0 & 0.0 & 0.0 & 0.0 & 0.0 \\ 
8 & 0.0 & 0.0 & 0.0 & 0.0 & 0.0 & 0.0 & 0.0 & 1.0 & 0.0 & 0.0 & 0.0 & 0.0 & 0.0 & 0.0 & 0.0 & 0.0 & 0.0 & 0.0 & 0.0 & 0.0 & 0.0 & 0.0 & 0.0 & 0.0 & 0.0 & 0.0 & 0.0 & 0.0 & 0.0 & 0.0 & 0.0 & 0.0 & 0.0 & 0.0 & 0.0 & 0.0 & 0.0 & 0.0 \\ 
9 & 0.0 & 0.0 & 0.0 & 0.0 & 0.0 & 0.0 & 0.0 & 0.0 & 1.0 & 0.0 & 0.0 & 0.0 & 0.0 & 0.0 & 0.0 & 0.0 & 0.0 & 0.0 & 0.0 & 0.0 & 0.0 & 0.0 & 0.0 & 0.0 & 0.0 & 0.0 & 0.0 & 0.0 & 0.0 & 0.0 & 0.0 & 0.0 & 0.0 & 0.0 & 0.0 & 0.0 & 0.0 & 0.0 \\ 
10 & 0.0 & 0.0 & 0.0 & 0.0 & 0.0 & 0.0 & 0.0 & 0.0 & 0.0 & 1.0 & 0.0 & 0.0 & 0.0 & 0.0 & 0.0 & 0.0 & 0.0 & 0.0 & 0.0 & 0.0 & 0.0 & 0.0 & 0.0 & 0.0 & 0.0 & 0.0 & 0.0 & 0.0 & 0.0 & 0.0 & 0.0 & 0.0 & 0.0 & 0.0 & 0.0 & 0.0 & 0.0 & 0.0 \\ 
11 & 0.0 & 0.0 & 0.0 & 0.0 & 0.0 & 0.0 & 0.0 & 0.0 & 0.0 & 0.0 & 0.999 & 0.0 & 0.0 & 0.0 & 0.0 & 0.0 & 0.0 & 0.0 & 0.0 & 0.0 & 0.0 & 0.0 & 0.0 & 0.0 & 0.0 & 0.0 & 0.0 & 0.0 & 0.0 & 0.0 & 0.0 & 0.0 & 0.0 & 0.0 & 0.0 & 0.0 & 0.0 & 0.0 \\ 
12 & 0.0 & 0.0 & 0.0 & 0.0 & 0.0 & 0.0 & 0.0 & 0.0 & 0.0 & 0.0 & 0.0 & 1.0 & 0.0 & 0.0 & 0.0 & 0.0 & 0.0 & 0.0 & 0.0 & 0.0 & 0.0 & 0.0 & 0.0 & 0.0 & 0.0 & 0.0 & 0.0 & 0.0 & 0.0 & 0.0 & 0.0 & 0.0 & 0.0 & 0.0 & 0.0 & 0.0 & 0.0 & 0.0 \\ 
13 & 0.0 & 0.0 & 0.0 & 0.0 & 0.0 & 0.0 & 0.0 & 0.0 & 0.0 & 0.0 & 0.0 & 0.0 & 1.0 & 0.0 & 0.0 & 0.0 & 0.0 & 0.0 & 0.0 & 0.0 & 0.0 & 0.0 & 0.0 & 0.0 & 0.0 & 0.0 & 0.0 & 0.0 & 0.0 & 0.0 & 0.0 & 0.0 & 0.0 & 0.0 & 0.0 & 0.0 & 0.0 & 0.0 \\ 
14 & 0.0 & 0.0 & 0.0 & 0.0 & 0.0 & 0.0 & 0.0 & 0.0 & 0.0 & 0.0 & 0.0 & 0.0 & 0.0 & 1.0 & 0.0 & 0.0 & 0.0 & 0.0 & 0.0 & 0.0 & 0.0 & 0.0 & 0.0 & 0.0 & 0.0 & 0.0 & 0.0 & 0.0 & 0.0 & 0.0 & 0.0 & 0.0 & 0.0 & 0.0 & 0.0 & 0.0 & 0.0 & 0.0 \\ 
15 & 0.0 & 0.0 & 0.0 & 0.0 & 0.0 & 0.0 & 0.0 & 0.0 & 0.0 & 0.0 & 0.0 & 0.0 & 0.0 & 0.0 & 1.0 & 0.0 & 0.0 & 0.0 & 0.0 & 0.0 & 0.0 & 0.0 & 0.0 & 0.0 & 0.0 & 0.0 & 0.0 & 0.0 & 0.0 & 0.0 & 0.0 & 0.0 & 0.0 & 0.0 & 0.0 & 0.0 & 0.0 & 0.0 \\ 
16 & 0.0 & 0.0 & 0.0 & 0.0 & 0.0 & 0.0 & 0.0 & 0.0 & 0.0 & 0.0 & 0.0 & 0.0 & 0.0 & 0.0 & 0.0 & 1.0 & 0.0 & 0.0 & 0.0 & 0.0 & 0.0 & 0.0 & 0.0 & 0.0 & 0.0 & 0.0 & 0.0 & 0.0 & 0.0 & 0.0 & 0.0 & 0.0 & 0.0 & 0.0 & 0.0 & 0.0 & 0.0 & 0.0 \\ 
17 & 0.0 & 0.0 & 0.0 & 0.0 & 0.0 & 0.0 & 0.0 & 0.0 & 0.0 & 0.0 & 0.0 & 0.0 & 0.0 & 0.0 & 0.0 & 0.0 & 0.999 & 0.0 & 0.0 & 0.0 & 0.0 & 0.0 & 0.0 & 0.0 & 0.0 & 0.0 & 0.0 & 0.0 & 0.0 & 0.0 & 0.0 & 0.0 & 0.0 & 0.0 & 0.0 & 0.0 & 0.0 & 0.0 \\ 
18 & 0.0 & 0.0 & 0.0 & 0.0 & 0.0 & 0.0 & 0.0 & 0.0 & 0.0 & 0.0 & 0.0 & 0.0 & 0.0 & 0.0 & 0.0 & 0.0 & 0.0 & 1.0 & 0.0 & 0.0 & 0.0 & 0.0 & 0.0 & 0.0 & 0.0 & 0.0 & 0.0 & 0.0 & 0.0 & 0.0 & 0.0 & 0.0 & 0.0 & 0.0 & 0.0 & 0.0 & 0.0 & 0.0 \\ 
19 & 0.0 & 0.0 & 0.0 & 0.0 & 0.001 & 0.0 & 0.0 & 0.0 & 0.0 & 0.0 & 0.0 & 0.0 & 0.0 & 0.0 & 0.0 & 0.0 & 0.0 & 0.0 & 0.998 & 0.0 & 0.0 & 0.0 & 0.0 & 0.0 & 0.0 & 0.0 & 0.0 & 0.0 & 0.0 & 0.0 & 0.0 & 0.0 & 0.0 & 0.0 & 0.0 & 0.0 & 0.0 & 0.0 \\ 
20 & 0.0 & 0.0 & 0.0 & 0.0 & 0.0 & 0.013 & 0.0 & 0.0 & 0.0 & 0.0 & 0.0 & 0.0 & 0.0 & 0.0 & 0.0 & 0.0 & 0.0 & 0.0 & 0.0 & 0.986 & 0.0 & 0.0 & 0.0 & 0.0 & 0.0 & 0.0 & 0.0 & 0.0 & 0.0 & 0.0 & 0.0 & 0.0 & 0.0 & 0.0 & 0.0 & 0.0 & 0.0 & 0.0 \\ 
21 & 0.002 & 0.0 & 0.0 & 0.0 & 0.0 & 0.0 & 0.029 & 0.0 & 0.0 & 0.0 & 0.0 & 0.0 & 0.0 & 0.0 & 0.0 & 0.0 & 0.0 & 0.0 & 0.0 & 0.0 & 0.967 & 0.0 & 0.0 & 0.0 & 0.0 & 0.0 & 0.0 & 0.0 & 0.0 & 0.0 & 0.0 & 0.0 & 0.0 & 0.0 & 0.0 & 0.0 & 0.0 & 0.0 \\ 
22 & 0.0 & 0.0 & 0.0 & 0.0 & 0.0 & 0.0 & 0.0 & 0.092 & 0.0 & 0.0 & 0.0 & 0.0 & 0.0 & 0.0 & 0.0 & 0.0 & 0.0 & 0.0 & 0.0 & 0.0 & 0.0 & 0.907 & 0.0 & 0.0 & 0.0 & 0.0 & 0.0 & 0.0 & 0.0 & 0.0 & 0.0 & 0.0 & 0.0 & 0.0 & 0.0 & 0.0 & 0.0 & 0.0 \\ 
23 & 0.0 & 0.0 & 0.0 & 0.0 & 0.0 & 0.0 & 0.0 & 0.0 & 0.003 & 0.0 & 0.0 & 0.0 & 0.0 & 0.0 & 0.0 & 0.0 & 0.0 & 0.0 & 0.0 & 0.0 & 0.0 & 0.0 & 0.996 & 0.0 & 0.0 & 0.0 & 0.0 & 0.0 & 0.0 & 0.0 & 0.0 & 0.0 & 0.0 & 0.0 & 0.0 & 0.0 & 0.0 & 0.0 \\ 
24 & 0.0 & 0.0 & 0.0 & 0.0 & 0.0 & 0.0 & 0.0 & 0.0 & 0.0 & 0.02 & 0.0 & 0.0 & 0.0 & 0.0 & 0.0 & 0.0 & 0.0 & 0.0 & 0.0 & 0.0 & 0.0 & 0.0 & 0.0 & 0.98 & 0.0 & 0.0 & 0.0 & 0.0 & 0.0 & 0.0 & 0.0 & 0.0 & 0.0 & 0.0 & 0.0 & 0.0 & 0.0 & 0.0 \\ 
25 & 0.0 & 0.0 & 0.0 & 0.0 & 0.0 & 0.0 & 0.0 & 0.0 & 0.0 & 0.0 & 0.099 & 0.0 & 0.0 & 0.0 & 0.0 & 0.0 & 0.0 & 0.0 & 0.0 & 0.0 & 0.0 & 0.0 & 0.0 & 0.0 & 0.901 & 0.0 & 0.0 & 0.0 & 0.0 & 0.0 & 0.0 & 0.0 & 0.0 & 0.0 & 0.0 & 0.0 & 0.0 & 0.0 \\ 
26 & 0.0 & 0.0 & 0.0 & 0.0 & 0.0 & 0.0 & 0.0 & 0.0 & 0.0 & 0.0 & 0.0 & 0.034 & 0.0 & 0.0 & 0.0 & 0.0 & 0.0 & 0.0 & 0.0 & 0.0 & 0.0 & 0.0 & 0.0 & 0.0 & 0.0 & 0.965 & 0.0 & 0.0 & 0.0 & 0.0 & 0.0 & 0.0 & 0.0 & 0.0 & 0.0 & 0.0 & 0.0 & 0.0 \\ 
27 & 0.0 & 0.0 & 0.0 & 0.0 & 0.0 & 0.0 & 0.0 & 0.0 & 0.0 & 0.0 & 0.0 & 0.0 & 0.0 & 0.0 & 0.0 & 0.0 & 0.0 & 0.0 & 0.0 & 0.0 & 0.0 & 0.0 & 0.0 & 0.0 & 0.0 & 0.0 & 0.999 & 0.0 & 0.0 & 0.0 & 0.0 & 0.0 & 0.0 & 0.0 & 0.0 & 0.0 & 0.0 & 0.0 \\ 
28 & 0.0 & 0.0 & 0.0 & 0.0 & 0.0 & 0.0 & 0.0 & 0.0 & 0.0 & 0.0 & 0.0 & 0.0 & 0.0 & 0.062 & 0.0 & 0.0 & 0.0 & 0.0 & 0.0 & 0.0 & 0.0 & 0.0 & 0.0 & 0.0 & 0.0 & 0.0 & 0.0 & 0.938 & 0.0 & 0.0 & 0.0 & 0.0 & 0.0 & 0.0 & 0.0 & 0.0 & 0.0 & 0.0 \\ 
29 & 0.0 & 0.0 & 0.001 & 0.0 & 0.0 & 0.0 & 0.0 & 0.0 & 0.0 & 0.0 & 0.0 & 0.0 & 0.0 & 0.0 & 0.084 & 0.0 & 0.0 & 0.0 & 0.0 & 0.0 & 0.0 & 0.0 & 0.0 & 0.0 & 0.0 & 0.0 & 0.0 & 0.0 & 0.818 & 0.0 & 0.0 & 0.0 & 0.0 & 0.095 & 0.0 & 0.0 & 0.0 & 0.0 \\ 
30 & 0.0 & 0.0 & 0.0 & 0.0 & 0.0 & 0.0 & 0.0 & 0.0 & 0.0 & 0.0 & 0.0 & 0.0 & 0.0 & 0.0 & 0.0 & 0.063 & 0.0 & 0.0 & 0.0 & 0.0 & 0.0 & 0.0 & 0.0 & 0.0 & 0.0 & 0.0 & 0.0 & 0.0 & 0.0 & 0.936 & 0.0 & 0.0 & 0.0 & 0.0 & 0.0 & 0.0 & 0.0 & 0.0 \\ 
31 & 0.0 & 0.0 & 0.0 & 0.0 & 0.0 & 0.0 & 0.0 & 0.0 & 0.0 & 0.0 & 0.0 & 0.0 & 0.0 & 0.0 & 0.0 & 0.0 & 0.054 & 0.0 & 0.0 & 0.0 & 0.0 & 0.0 & 0.0 & 0.0 & 0.0 & 0.0 & 0.0 & 0.0 & 0.0 & 0.0 & 0.908 & 0.0 & 0.0 & 0.0 & 0.0 & 0.036 & 0.0 & 0.0 \\ 
32 & 0.0 & 0.0 & 0.0 & 0.0 & 0.0 & 0.0 & 0.0 & 0.0 & 0.0 & 0.0 & 0.0 & 0.0 & 0.0 & 0.0 & 0.0 & 0.0 & 0.0 & 0.061 & 0.0 & 0.0 & 0.0 & 0.0 & 0.0 & 0.0 & 0.0 & 0.0 & 0.0 & 0.0 & 0.0 & 0.0 & 0.0 & 0.935 & 0.0 & 0.0 & 0.003 & 0.0 & 0.0 & 0.0 \\ 
33 & 0.0 & 0.0 & 0.0 & 0.0 & 0.0 & 0.0 & 0.0 & 0.0 & 0.0 & 0.0 & 0.0 & 0.0 & 0.0 & 0.0 & 0.0 & 0.0 & 0.0 & 0.0 & 0.0 & 0.0 & 0.0 & 0.0 & 0.0 & 0.0 & 0.0 & 0.0 & 0.0 & 0.0 & 0.0 & 0.001 & 0.0 & 0.0 & 0.998 & 0.0 & 0.0 & 0.0 & 0.0 & 0.0 \\ 
34 & 0.0 & 0.0 & 0.0 & 0.0 & 0.0 & 0.0 & 0.0 & 0.0 & 0.0 & 0.0 & 0.0 & 0.0 & 0.0 & 0.0 & 0.189 & 0.0 & 0.0 & 0.0 & 0.0 & 0.0 & 0.0 & 0.0 & 0.0 & 0.0 & 0.0 & 0.0 & 0.0 & 0.0 & 0.0 & 0.0 & 0.0 & 0.0 & 0.0 & 0.81 & 0.0 & 0.0 & 0.0 & 0.0 \\ 
35 & 0.0 & 0.0 & 0.0 & 0.0 & 0.0 & 0.0 & 0.0 & 0.0 & 0.0 & 0.0 & 0.0 & 0.0 & 0.0 & 0.0 & 0.0 & 0.0 & 0.0 & 0.0 & 0.0 & 0.0 & 0.0 & 0.0 & 0.0 & 0.0 & 0.0 & 0.0 & 0.0 & 0.0 & 0.0 & 0.0 & 0.0 & 0.001 & 0.0 & 0.0 & 0.998 & 0.0 & 0.0 & 0.0 \\ 
36 & 0.0 & 0.0 & 0.0 & 0.0 & 0.0 & 0.0 & 0.0 & 0.0 & 0.0 & 0.0 & 0.0 & 0.0 & 0.0 & 0.0 & 0.0 & 0.0 & 0.04 & 0.0 & 0.0 & 0.0 & 0.0 & 0.0 & 0.0 & 0.0 & 0.0 & 0.0 & 0.0 & 0.0 & 0.0 & 0.0 & 0.049 & 0.0 & 0.0 & 0.0 & 0.0 & 0.909 & 0.0 & 0.0 \\ 
37 & 0.0 & 0.0 & 0.0 & 0.0 & 0.0 & 0.0 & 0.0 & 0.0 & 0.0 & 0.0 & 0.0 & 0.0 & 0.0 & 0.0 & 0.0 & 0.0 & 0.0 & 0.0 & 0.0 & 0.0 & 0.0 & 0.0 & 0.0 & 0.0 & 0.0 & 0.0 & 0.0 & 0.0 & 0.0 & 0.0 & 0.0 & 0.0 & 0.0 & 0.0 & 0.0 & 0.0 & 1.0 & 0.0 \\ 
38 & 0.0 & 0.0 & 0.0 & 0.0 & 0.0 & 0.0 & 0.0 & 0.0 & 0.0 & 0.0 & 0.0 & 0.0 & 0.0 & 0.0 & 0.0 & 0.0 & 0.0 & 0.0 & 0.0 & 0.0 & 0.0 & 0.0 & 0.0 & 0.0 & 0.0 & 0.0 & 0.0 & 0.0 & 0.0 & 0.0 & 0.0 & 0.0 & 0.0 & 0.0 & 0.0 & 0.0 & 0.0 & 1.0 \\ \bottomrule 
\end{tabular}
}
\label{fig:cm:attacker}
\end{subfigure}
\vspace{2ex}
\caption{Confusion matrices: Probability with which a password of one
  class (row) is classified as another class (column) by
  \chooseHoneywords{\nmbrHoneywords} (\figref{fig:cm:hw}) or
  \breachAttacker with $\setSize{\auxiliaryInfo} > 1$
  (\figref{fig:cm:attacker}).  Box shading is scaled linearly between
  0.0 (white) and 1.0 (black).}
\label{fig:cm}
\vspace{2ex}
\end{figure}

\subsubsection{Experimental Results}
\label{sec:algorithmic:eval:results}

    \begin{figure}[t]
      \vspace{2ex}
      \begin{subfigure}[b]{0.233\textwidth}
        \setlength\figureheight{1.9in}
        \centering
        \caption{$\nmbrHoneywords=19$}
        \vspace{2ex}
        \resizebox{\textwidth}{!}{\input{figure/lastpass+1pass/fn_20.tex}}
        \label{fig:lastpass+1pass:fn19}
      \end{subfigure}
      \hfill
      \begin{subfigure}[b]{0.21\textwidth}
        \setlength\figureheight{1.9in}
        \centering
        \caption{$\nmbrHoneywords=99$}
        \vspace{2.75ex}
        \resizebox{\textwidth}{!}{\input{figure/lastpass+1pass/fn_100.tex}}
        \label{fig:lastpass+1pass:fn99}
      \end{subfigure}
      \vspace{-3ex}
      \caption{$\fnProbWOParams(\breachAttacker)$ for honeywords by
        algorithmic password generators}
      \label{fig:lastpass+1pass} 
      \vspace*{2ex}
    \end{figure}

    \begin{figure*}[t!]
      \vspace{0.1in}
      \begin{subfigure}[b]{0.170\textwidth}
      \setlength\figureheight{1.9in}
        \centering
        \caption{Class 22}
        \vspace{1.25ex}
        \resizebox{\textwidth}{!}{\input{figure/lastpass+1pass/21/fn_20.tex}}
        \label{fig:lastpass+1pass:someclasses:fn19:21}
      \end{subfigure}
      \hfill
       \begin{subfigure}[b]{0.145\textwidth}
      \setlength\figureheight{1.9in}
        \centering
        \caption{Class 26}
        \vspace{2ex}
        \resizebox{\textwidth}{!}{\input{figure/lastpass+1pass/25/fn_20.tex}}
        \label{fig:lastpass+1pass:someclasses:fn19:25}
      \end{subfigure}
      \hfill
      \begin{subfigure}[b]{0.145\textwidth}
        \setlength\figureheight{1.9in}
        \centering
        \caption{Class 29}
        \vspace{2ex}
        \resizebox{\textwidth}{!}{\input{figure/lastpass+1pass/28/fn_20.tex}}
        \label{fig:lastpass+1pass:someclasses:fn19:28}
      \end{subfigure}
      \hfill
      \begin{subfigure}[b]{0.145\textwidth}
        \setlength\figureheight{1.9in}
        \centering
        \caption{Class 30}
        \vspace{2ex}
        \resizebox{\textwidth}{!}{\input{figure/lastpass+1pass/29/fn_20.tex}}
        \label{fig:lastpass+1pass:someclasses:fn19:29}
      \end{subfigure}
      \hfill
      \begin{subfigure}[b]{0.145\textwidth}
        \setlength\figureheight{1.9in}
        \centering
        \caption{Class 34}
        \vspace{2ex}
        \resizebox{\textwidth}{!}{\input{figure/lastpass+1pass/33/fn_20.tex}}
        \label{fig:lastpass+1pass:someclasses:fn19:33}
      \end{subfigure}
      \hfill
      \begin{subfigure}[b]{0.145\textwidth}
        \setlength\figureheight{1.9in}
        \centering
        \caption{Class 36}
        \vspace{2ex}
        \resizebox{\textwidth}{!}{\input{figure/lastpass+1pass/35/fn_20.tex}}
        \label{fig:lastpass+1pass:someclasses:fn19:35}
      \end{subfigure}
      \vspace{-2ex}
       \caption{$\fnProbWOParams(\breachAttacker)$ for
         \chooseHoneywords{\nmbrHoneywords} of \ClassifiedGenerator
         ($\nmbrHoneywords=19$), for \setSize{\auxiliaryInfo}
         distributed as in \figref{fig:distribution_auxiliaryInfo}
         ($\mbox{--}$) and for $\setSize{\auxiliaryInfo} = 99$
         ($\mbox{-}\mbox{-}$)}
      \label{fig:lastpass+1pass:someclasses:fn19} 
      \vspace{4ex}
    \end{figure*}

    \begin{figure*}[t!]
      \vspace*{0.25em}
      \begin{subfigure}[b]{0.170\textwidth}
        \setlength\figureheight{1.9in}
        \centering
        \caption{Class 22}
        \vspace{1.25ex}
        \resizebox{\textwidth}{!}{\input{figure/lastpass+1pass/21/fn_100.tex}}
        \label{fig:lastpass+1pass:someclasses:fn99:21}
      \end{subfigure}
      \hfill
      \begin{subfigure}[b]{0.145\textwidth}
        \setlength\figureheight{1.9in}
        \centering
        \caption{Class 26}
        \vspace{2ex}
        \resizebox{\textwidth}{!}{\input{figure/lastpass+1pass/25/fn_100.tex}}
        \label{fig:lastpass+1pass:someclasses:fn99:25}
      \end{subfigure}
      \hfill
      \begin{subfigure}[b]{0.145\textwidth}
        \setlength\figureheight{1.9in}
        \centering
        \caption{Class 29}
        \vspace{2ex}
        \resizebox{\textwidth}{!}{\input{figure/lastpass+1pass/28/fn_100.tex}}
        \label{fig:lastpass+1pass:someclasses:fn99:28}
      \end{subfigure}
      \hfill
      \begin{subfigure}[b]{0.145\textwidth}
        \setlength\figureheight{1.9in}
        \centering
        \caption{Class 30}
        \vspace{2ex}
        \resizebox{\textwidth}{!}{\input{figure/lastpass+1pass/29/fn_100.tex}}
        \label{fig:lastpass+1pass:someclasses:fn99:29}
      \end{subfigure}
      \hfill
      \begin{subfigure}[b]{0.145\textwidth}
        \setlength\figureheight{1.9in}
        \centering
        \caption{Class 34}
        \vspace{2ex}
        \resizebox{\textwidth}{!}{\input{figure/lastpass+1pass/33/fn_100.tex}}
        \label{fig:lastpass+1pass:someclasses:fn99:33}
      \end{subfigure}
      \hfill
      \begin{subfigure}[b]{0.145\textwidth}
        \setlength\figureheight{1.9in}
        \centering
        \caption{Class 36}
        \vspace{2ex}
        \resizebox{\textwidth}{!}{\input{figure/lastpass+1pass/35/fn_100.tex}}
        \label{fig:lastpass+1pass:someclasses:fn99:35}
      \end{subfigure}
      \vspace{-2ex} 
      \caption{$\fnProbWOParams(\breachAttacker)$ for
        \chooseHoneywords{\nmbrHoneywords} of \ClassifiedGenerator
        ($\nmbrHoneywords=99$), for \setSize{\auxiliaryInfo}
        distributed as in \figref{fig:distribution_auxiliaryInfo}
        ($\mbox{--}$) and for $\setSize{\auxiliaryInfo} = 99$
        ($\mbox{-}\mbox{-}$)}
      \label{fig:lastpass+1pass:someclasses:fn99} 
      \vspace*{2ex}
    \end{figure*}

We evaluated the honeyword-generation methods described in
\secref{sec:algorithmic:generator}, though we plot only
$\fnProbWOParams(\breachAttacker)$ since
$\fpProbWOParams(\alarmAttacker)$ was essentially perfect.  We plot
$\fnProbWOParams(\breachAttacker)$ against \alarmThreshold in
\figref{fig:lastpass+1pass}.  As seen there, both the \FixedGenerator and
\RandomGenerator methods had a high $\fnProbWOParams(\breachAttacker)$. Even
when $\alarmThreshold=1$, they had $\fnProbWOParams(\breachAttacker) >
0.94$ for $\nmbrHoneywords = 19$ and $\fnProbWOParams(\breachAttacker)
> 0.93$ for $\nmbrHoneywords = 99$.

In contrast, the \ClassifiedGenerator method achieves nearly perfect
$\fnProbWOParams(\breachAttacker)$. This method selects the most
plausible algorithmic password generator based on the account password
to generate honeywords.  The confusion matrix experienced by
\chooseHoneywords{\nmbrHoneywords} (i.e., using \password) are shown
in \figref{fig:cm:hw}.  When $\setSize{\auxiliaryInfo} = 1$, the
confusion experienced by \breachAttacker is virtually identical, of
course, but the confusion experienced by \breachAttacker when
$\setSize{\auxiliaryInfo} > 1$ is notably less, as shown in
\figref{fig:cm:attacker}.  As this figure shows, when
$\setSize{\auxiliaryInfo} > 1$, \breachAttacker has greater ability to
classify the user's password generator based on \auxiliaryInfo than
\chooseHoneywords{\nmbrHoneywords} does based on \password, at least
for certain classes. Since our dataset is dominated by accounts
for which the number of passwords known by \breachAttacker numbers
$\setSize{\auxiliaryInfo} = 1$
(\figref{fig:distribution_auxiliaryInfo}), the confusion shown in
\figref{fig:cm:hw} (where $\setSize{\auxiliaryInfo} > 1$) cannot
effectively be exploited by \breachAttacker.

However, if the fraction of accounts for which \breachAttacker holds
$\setSize{\auxiliaryInfo} > 1$ passwords were larger, the better
classification accuracy this would enable (\figref{fig:cm:attacker})
would permit an average increase in $\fnProbWOParams(\breachAttacker)$.
To illustrate this, in \figref{fig:lastpass+1pass:someclasses:fn19}
and \figref{fig:lastpass+1pass:someclasses:fn99} we show the effect on
$\fnProbWOParams(\breachAttacker)$ of increasing
\setSize{\auxiliaryInfo} from its original distribution to
$\setSize{\auxiliaryInfo} = 99$ always, for $\nmbrHoneywords = 19$ or
$\nmbrHoneywords = 99$, respectively.  Each subfigure shows
$\fnProbWOParams(\breachAttacker)$ for certain classes of the actual
password \password; e.g.,
\figref{fig:lastpass+1pass:someclasses:fn19:21} shows this effect when
$\password \gets \choosePassword{22}()$.  As can be seen in these
figures, increasing \setSize{\auxiliaryInfo} to
$\setSize{\auxiliaryInfo} = 99$ enables \breachAttacker to improve
$\fnProbWOParams(\breachAttacker)$, for these classes noticeably.

In conclusion, in the case of algorithmically generated passwords, it
is critical for \chooseHoneywords{\nmbrHoneywords} to identify the
algorithmic password generator used by each user in order to achieve
low $\fnProbWOParams(\breachAttacker)$.  Even then, as the number of
passwords \setSize{\auxiliaryInfo} grows, this measure will decay.

%% file: discussion.tex
\section{Discussion} \label{sec:discussion}

\subsection{False-Negative Attacks with Less Auxiliary Information}
\label{sec:discussion:auxiliary}

In \secref{sec:user-selected}, we assumed that the false-negative
attacker has knowledge of the passwords used by the same user at
other sites.  We additionally explored a setting where the
false-negative attacker had minimal auxiliary information about the
users, specifically only one password used by the same user at
another site ($\setSize{\auxiliaryInfo}=1$).  To do so, we
implemented the algorithm \choosePassword by choosing two passwords
without replacement from a single multiset \passwordSet{\emailAddr}
chosen uniformly at random from
$\Dataset{\TestTag}{\UserChosenTag}$, returning one as \password and
the other as the only element of \auxiliaryInfo. The experimental
results in this setting are shown in \figref{fig:uc:to20:varyingX}. 
When the false-negative attacker
\breachAttacker had little information about users, his success rate
of guessing the account passwords from sweetwords slightly dropped,
yielding a smaller $\fnProbWOParams(\breachAttacker)$. However, all
the honeyword-generation methods remained insufficiently resilient
to these attacks. The best $\fnProbWOParams(\breachAttacker)$ that
the honeyword-generation techniques accomplished for all accounts
was $0.48$ (\PasstoPath,
\figref{fig:uc:to20:varyingX:pass2path}). No existing algorithm
achieved low rates of both false positives and false negatives.

  \begin{figure*}[t!]
    \begin{subfigure}[b]{\textwidth}
      \centering
      \resizebox{!}{2em}{\input{varyingX_trade-off-legend.tex}}
    \end{subfigure}
    
    \begin{subfigure}[b]{0.179\textwidth}
      \setlength\figureheight{1.9in}
      \centering
      \caption{\List}
      \vspace{1.25ex}
      \resizebox{\textwidth}{!}{\input{figure/user_chosen/List/trade-off_varyingX_20}}
      \label{fig:uc:to20:varyingX:list}
    \end{subfigure}
    \hfill
    \begin{subfigure}[b]{0.141\textwidth}
      \setlength\figureheight{1.9in}
      \centering
      \caption{\PCFG}
      \vspace{2ex}
      \resizebox{\textwidth}{!}{\input{figure/user_chosen/PCFG/trade-off_varyingX_20}}
      \label{fig:uc:to20:varyingX:pcfg}
    \end{subfigure}
    \hfill
    \begin{subfigure}[b]{0.141\textwidth}
      \setlength\figureheight{1.9in}
      \centering
      \caption{\Markov}
      \vspace{2ex}
      \resizebox{\textwidth}{!}{\input{figure/user_chosen/Markov/trade-off_varyingX_20}}
      \label{fig:uc:to20:varyingX:markov}
    \end{subfigure}
    \hfill
    \begin{subfigure}[b]{0.141\textwidth}
      \setlength\figureheight{1.9in}
      \centering
      \caption{\Combo}
      \vspace{2ex}
      \resizebox{\textwidth}{!}{\input{figure/user_chosen/combination/trade-off_varyingX_20}}
      \label{fig:uc:to20:varyingX:combo}
    \end{subfigure}
    \hfill
    \begin{subfigure}[b]{0.141\textwidth}
      \setlength\figureheight{1.9in}
      \centering
      \caption{\Listpwd}
      \vspace{2ex}
      \resizebox{\textwidth}{!}{\input{figure/user_chosen/List_/trade-off_varyingX_20}}
      \label{fig:uc:to20:varyingX:list-pwd}
    \end{subfigure}
    \hfill
    \begin{subfigure}[b]{0.141\textwidth}
      \setlength\figureheight{1.9in}
      \centering
      \caption{\PCFGpwd}
      \vspace{2ex}
      \resizebox{\textwidth}{!}{\input{figure/user_chosen/PCFG_/trade-off_varyingX_20}}
      \label{fig:uc:to20:varyingX:pcfg-pwd}
    \end{subfigure}
    \\[-2.5ex]
    \begin{subfigure}[b]{\textwidth}
      \setlength\figureheight{2in}
        \centering
        \resizebox{!}{1.5em}{\input{xlabel.tex}}
    \end{subfigure}
    \vspace{1ex}
    \\
    \begin{subfigure}[b]{0.179\textwidth}
      \setlength\figureheight{1.9in}
      \centering
      \caption{\Markovpwd}
      \vspace{1.25ex}
      \resizebox{\textwidth}{!}{\input{figure/user_chosen/Markov_/trade-off_varyingX_20}}
      \label{fig:uc:to20:varyingX:markov-pwd}
    \end{subfigure}
    \hfill
    \begin{subfigure}[b]{0.141\textwidth}
      \setlength\figureheight{1.9in}
      \centering
      \caption{\Combopwd}
      \vspace{2ex}
      \resizebox{\textwidth}{!}{\input{figure/user_chosen/combination_/trade-off_varyingX_20}}
      \label{fig:uc:to20:varyingX:combo-pwd}
    \end{subfigure}
    \hfill
    \begin{subfigure}[b]{0.141\textwidth}
      \setlength\figureheight{1.9in}
      \centering
      \caption{\TarList}
      \vspace{2ex}
      \resizebox{\textwidth}{!}{\input{figure/user_chosen/List_target/trade-off_varyingX_20}}
      \label{fig:uc:to20:varyingX:list-tgt}
    \end{subfigure}
    \hfill
    \begin{subfigure}[b]{0.141\textwidth}
      \setlength\figureheight{1.9in}
      \centering
      \caption{\TarPCFG}
      \vspace{2ex}
      \resizebox{\textwidth}{!}{\input{figure/user_chosen/PCFG_target/trade-off_varyingX_20}}
      \label{fig:uc:to20:varyingX:pcfg-tgt}
    \end{subfigure}
    \hfill
    \begin{subfigure}[b]{0.141\textwidth}
      \setlength\figureheight{1.9in}
      \centering
      \caption{\TarMarkov}
      \vspace{2ex}
      \resizebox{\textwidth}{!}{\input{figure/user_chosen/Markov_target/trade-off_varyingX_20}}
      \label{fig:uc:to20:varyingX:markov-tgt}
    \end{subfigure}
    \hfill
    \begin{subfigure}[b]{0.141\textwidth}
      \setlength\figureheight{1.9in}
      \centering
      \caption{\TarCombo}
      \vspace{2ex}
      \resizebox{\textwidth}{!}{\input{figure/user_chosen/combination_target/trade-off_varyingX_20}}
      \label{fig:uc:to20:varyingX:combo-tgt}
    \end{subfigure}
    \\[-2.5ex]
    \begin{subfigure}[b]{\textwidth}
      \setlength\figureheight{2in}
        \centering
        \resizebox{!}{1.5em}{\input{xlabel.tex}}
    \end{subfigure}
    \vspace{1ex}
    \\
    \begin{subfigure}[b]{0.179\textwidth}
      \setlength\figureheight{1.9in}
      \centering
      \caption{\CGPT}
      \vspace{1.25ex}
      \resizebox{\textwidth}{!}{\input{figure/user_chosen/gpt/trade-off_varyingX_20}}
      \label{fig:uc:to20:varyingX:gpt}
    \end{subfigure}
    \hfill
    \begin{subfigure}[b]{0.141\textwidth}
      \setlength\figureheight{1.9in}
      \centering
      \caption{\CBT{4}}
      \vspace{2ex}
      \resizebox{\textwidth}{!}{\input{figure/user_chosen/chaffing4/trade-off_varyingX_20}}
      \label{fig:uc:to20:varyingX:cbt4}
    \end{subfigure}
    \hfill
    \begin{subfigure}[b]{0.141\textwidth}
      \setlength\figureheight{1.9in}
      \centering
      \caption{\CBTr}
      \vspace{2ex}
      \resizebox{\textwidth}{!}{\input{figure/user_chosen/CBTr/trade-off_varyingX_20}}
      \label{fig:uc:to20:varyingX:cbtr}
    \end{subfigure}
    \hfill
    \begin{subfigure}[b]{0.141\textwidth}
      \setlength\figureheight{1.9in}
      \centering
      \caption{\CHM}
      \vspace{2ex}
      \resizebox{\textwidth}{!}{\input{figure/user_chosen/CHM/trade-off_varyingX_20}}
      \label{fig:uc:to20:varyingX:chm-pwd}
    \end{subfigure}
    \hfill
    \begin{subfigure}[b]{0.141\textwidth}
      \setlength\figureheight{1.9in}
      \centering
      \caption{\Tweak}
      \vspace{2ex}
      \resizebox{\textwidth}{!}{\input{figure/user_chosen/Tweak/trade-off_varyingX_20}}
      \label{fig:uc:to20:varyingX:tweak}
    \end{subfigure}
    \hfill
    \begin{subfigure}[b]{0.141\textwidth}
      \setlength\figureheight{1.9in}
      \centering
      \caption{\PasstoPath}
      \vspace{2ex}
      \resizebox{\textwidth}{!}{\input{figure/user_chosen/pass2path/trade-off_varyingX_20}}
      \label{fig:uc:to20:varyingX:pass2path}
    \end{subfigure}
    \\[-2.5ex]
      \begin{subfigure}[b]{\textwidth}
        \setlength\figureheight{2in}
          \centering
          \resizebox{!}{1.5em}{\input{xlabel.tex}}
      \end{subfigure}
      
   \caption{$\fpProbWOParams(\alarmAttacker)$ vs.\
    $\fnProbWOParams(\breachAttacker)$ as \setSize{\auxiliaryInfo} is
    varied, for the case of user-chosen passwords ($\nmbrHoneywords=19$,
    $\attackBudget=1000$) and all accounts.}  \label{fig:uc:to20:varyingX}
  \vspace*{2ex}
  \end{figure*}

\subsection{Countermeasures to False-Positive Attacks} \label{sec:discussion:countermeasures}

False-positive attacks can be very costly to sites, if they
induce an investigation into the possibility of a breach and/or a
password reset for every site account. Moreover, repeated false
positives might eventually result in the defense being ignored
or disabled outright.  Despite the consequences of false positives,
only a few previous works on honeywords have briefly discussed how to
prevent them~\cite{juels2013:honeywords, wang2022:honeywords}.  Wang,
et al.~\cite{wang2022:honeywords} suggested that applying a blocklist
of popular passwords to honeyword selection can reduce false
positives, since the honeyword-generation methods considered in their
work generate honeywords by sampling from a public password
distribution (e.g., leveraging a password model like \List).  As such,
a blocklist would avoid using popular passwords as honeywords, which
can mitigate the guessability of honeywords by their proposed methods.
However, a blocklist of popular passwords is much less effective when
considering password-dependent honeyword-generation algorithms (e.g.,
\CBT, \CHM, \Tweak, and \PasstoPath), since these methods assign more
likelihood to those candidates similar to the account password.  A way
to mitigate false positives of these methods is to avoid using
passwords similar to the account password as honeywords, which makes
them suffer a high false-negative rate.

Another countermeasure to reduce false positives, as mentioned by
Juels and Rivest~\cite{juels2013:honeywords}, is to select
\nmbrHoneywords honeywords uniformly at random from a large pool of
candidate honeywords that are similar to the account password. In
order to achieve a small false-positive rate, the size of the pool
should be much larger than \nmbrHoneywords.  However, it is
challenging to generate such a large pool of candidates that are
sufficiently similar to the account password to ensure a small
false-negative rate via this process.  As such, an interesting
direction is to explore how to generate such a large candidate pool to
achieve a target false-negative rate.

Ultimately, a site might find it most palatable to address the
risk of false positives by adopting a lenient policy toward
honeyword-induced breach alarms.  Previous works
(e.g.,~\cite[\secrefstatic{2.4}]{juels2013:honeywords}) outlined a
range of possible reactions to honeyword-induced alarms, ranging
from severe (e.g., shutting down the system and forcing all users to
reset their passwords) to lenient (e.g., allow the login to proceed
as usual).  Whatever the policy, however, it will apply to both
false and true positives alike, and so a policy can be relaxed only
so far as is acceptable when a breach actually occurs.

\subsection{Balancing Attention to False Positives and False Negatives} \label{sec:discussion:balancing_FP_FN}

Since honeywords' proposal, a challenge has been to design good
honeyword-generation methods that achieve both low false-positives and
low false-negatives, i.e., $\fpProbWOParams \approx 0$ and
$\fnProbWOParams = \frac{\alarmThreshold}{\nmbrHoneywords + 1}$.
However, our experimental results in
\secref{sec:user-selected:evaluation:results} show that no existing
method achieves this goal in a threat model in which passwords from
the same user at other sites are exposed to the attacker.  While this
leaves us skeptical that a perfect honeyword-generation method exists
for this threat model (at least not when passwords are user-chosen,
versus algorithmically generated), we do not mean to suggest that
research in this direction should end.  However, we do advocate that
new honeyword-generation methods should be investigated with balanced
attention to false positives and false negatives in this threat model,
rather than more narrowly focusing on false negatives, as has been
typical in most prior research.

In the absence of improved honeyword-generation algorithms that
more effectively balance false positives and false negatives, we
believe that a change in perspective on the use of honeywords might
be warranted.  One such perspective, reflected in recent
work~\cite{wang2024:bernoulli}, is to tightly constrain false
positives to enable a honeyword-induced alarm to confidently be
taken seriously and dealt with severely, and then to rely on a
false-negative attacker's interests in harvesting \textit{many}
accounts to likely trigger a breach alarm even if the true-positive
probability \textit{per account} is modest.  Such an approach will
likely permit breaching attackers to harvest selected accounts
without triggering a breach alarm, but greedy breaching attackers
will still trigger a breach alarm with high probability.

\begin{table}[ht!]
  \small
  \vspace{0.1in}
\begin{center}
\begin{tabular}{@{}lp{0.7\columnwidth}@{}}
\toprule
Site & Password composition policy \\ \midrule
google.com      &     $\geq 8$ characters including a letter, a symbol, and a number        \\ 
youtube.com      &    $\geq 8$ characters including a letter, a symbol, and a number     \\ 
facebook.com     &   $\geq 6$ characters including a letter, a symbol, and a number  \\ 
twitter.com   &    $\geq 8$ characters including a letter, a symbol, and a number     \\
instagram.com   &    $\geq 6$ characters including a letter, a symbol, and a number  \\ 
baidu.com    &   $\geq 8$ and $\leq 14$ characters of at least two types from uppercase letter, lowercase letter, symbol, and number  \\ 
wikipedia.org    &     $\geq 8$ characters    \\ 
yandex.ru   &   $\geq 6$ characters including an uppercase letter, a lowercase letter, and a number    \\ 
yahoo.com    &    $\geq 9$ characters \\ 
xvideos.com    &   No requirement \\
pornhub.com  &   $\geq 6$ characters \\
amazon.com   &  $\geq 6$ characters  \\ 
tiktok.com   &  $\geq 8$ and $\leq 20$ characters including a letter, a symbol, and a number  \\ 
live.com  &  $\geq 8$ characters of at least two types from uppercase letter, lowercase letter, symbol, and number \\ 
openai.com  &  $\geq 8$ characters  \\ 
reddit.com   &   $\geq 8$ characters  \\ 
linkedin.com   &  $\geq 6$ characters \\ 
netflix.com  & $\geq 6$ characters  \\ 
office.com   & $\geq 8$ characters of at least two types from uppercase letter, lowercase letter, symbol, and number \\ 
twitch.tv  &   $\geq 8$ characters \\   \bottomrule 
\end{tabular}
\end{center}
\caption{Most visited websites~\cite{similarweb:most} and their
  password composition policies retrieved in May 2023.}
\label{table:policy}
\vspace*{2ex}
\end{table}

\begin{table}[ht!]
  \small
  \vspace{0.1in}
  \begin{center}
  \begin{tabular}{@{}ccccc@{}}
  \toprule
  Tranco Top & U & L & S & N \\ \midrule
  10K & $81.7\%$ & $78.4\%$ & $80.3\%$ & $71.2\%$ \\
  100K & $79.4\%$ & $79.2\%$ & $76.3\%$ & $82.5\%$ \\
  1M & $83.7\%$ & $84.1\%$ & $86.3\%$ & $82.0\%$ \\
  \bottomrule 
  \end{tabular}
  \end{center}
  \caption{Percentages of sites from
  Tranco Top 10K, 100K, and 1M that do not require some types of characters 
  (U: uppercase letter, L: lowercase letter, S: special symbol, N: number). 
  These statistics were obtained on Dec. 2021~\cite{alroomi2023:measuring}.}
  \label{table:policy_statistics}
  \end{table}

\begin{table*}[ht!]
  \small
  \vspace{0.1in}
  \begin{center}
  \begin{tabular}{@{}cccc@{}}
  \toprule
  \multicolumn{1}{c}{\multirow{2}{*}{Sites}} & \multicolumn{1}{c}{\multirow{2}{*}{Number of conflicts}} & Probability of conflict & Average number of non-conflicting \\
  &  &  with the next site & sites before a conflict \\ \midrule
  20 commonly visited websites & $2.143$ & $0.1127$ & $8.864$ \\
  101 simulated sites from Tranco Top 10K & $15.829$ & $0.1529$ & $6.317$  \\
  101 simulated sites from Tranco Top 100K & $14.450$ & $0.1445$ & $6.920$  \\
  101 simulated sites from Tranco Top 1M & $11.825$ & $0.1182$ & $8.456$  \\
  \bottomrule 
  \end{tabular}
  \end{center}
  \caption{Evaluation results on random walking at websites.}
  \label{table:policy_results}
  \end{table*}

\subsection{Assumptions on Algorithmic Password Generator Configuration} \label{sec:discussion:configuration}

A limitation of our analysis in \secref{sec:algorithmic} is that it
was conducted assuming that the user uniformly randomly selects a
configuration for her algorithmic password generator and, once
adopting a configuration, does not change it.

\myparagraph{Changes of password manager configuration} To justify the
assumption that users rarely change password manager configurations,
we performed a study of the password-creation policies of twenty
commonly accessed websites and those from Tranco Top 1M
websites~\cite{pochat2018:tranco} to show that users are rarely
\textit{required} to change configurations.  Specifically, we sought
to determine the frequency with which a user who sets passwords at
these sites in a random order will be required to change his
password-creation algorithm configuration to comply with the next site
in the sequence.

To perform this evaluation, we retrieved the password requirements
from twenty commonly visited websites~\cite{similarweb:most}, shown in
\tblref{table:policy}, or simulated password composition policies
based on the statistics from a recent large-scale
study~\cite{alroomi2023:measuring}, shown in
\tblref{table:policy_statistics}.  For a sequence of password policies
that is a permutation of password requirements from the twenty
commonly visited websites or a sequence of $101$ simulated
password-composition policies randomly constructed from the
statistics, we evaluated the number of times that the current
password-generator configuration conflicted with the password-creation
policy of the next website in the sequence, starting from a
configuration initialized by the minimum password requirement of the
first site. To ensure a conservative evaluation, when a conflict
occurred, the current configuration was replaced with the minimum
password requirements of the conflicting site. For each type of
password policy sequence, we performed this analysis for $10^{6}$
times. The evaluation results are shown in
\tblref{table:policy_results}.  We found that the numbers of conflicts
ranged from an average of $2.143$ in sequences of 20 websites to
$15.829$ in sequences of 101 simulated sites drawn from Tranco Top
10,000 sites, averaged over the $10^{6}$ sequences. These implied a
probability of conflict with the next site in the sequence
between $0.1127$ and $0.1529$. In expectation, then, the probability of
exactly \resets consecutive resets with no conflicts, followed by a
site that conflicts, is $<(1-0.1529)^{\resets}(0.1529)$, and the
average number of non-conflicting sites before a conflict was $>6.3$.
Given the conservative nature of our evaluation, we believe this
result justifies our assumption that a user would rarely change its
password-generator configuration.

However, an interesting direction for future work
would be to confirm or refute this assumption more broadly, since as
shown in \secref{sec:algorithmic:eval:results}, the assumption
somewhat diminishes the effectiveness of honeywords generated for
accounts with algorithmically generated passwords.  Alternatively, an
algorithmic password generator could be designed to encourage changing
these configuration settings regularly, in which case an interesting
research direction would be to explore the acceptability of this
practice for users.

\myparagraph{Tendency to use default password configuration}
In reality, we expect people to generally defer to the default
password-manager configuration until forced to change it by a site's
password policy. However, imposing a non-uniform distribution on
$\{\choosePassword{\GeneratorIndex}\}_{\GeneratorIndex =
  1}^{\NumGenerators}$ to reflect this tendency should not
qualitatively change the results of our study: First, analogous to our
results in \appref{app:existing}, the existing honeyword-generation
methods would still fail to provide a low false-negative probability
since the false-negative attacker would even more easily distinguish
the algorithmically generated password from the honeywords if he knows
how the selection of configuration is biased.  Second, as shown in
\secref{sec:algorithmic:eval:results}, when the honeyword system can
correctly predict the configuration used by the user---which should
only become easier when the distribution is biased toward the
default---and use that configuration to generate the honeyword, the
generated honeywords can provide sufficient security.

\subsection{A Mixed Case Study}  \label{sec:discussion:mixed_study}

In this work, we studied two representative cases where users create
user-chosen passwords (\secref{sec:user-selected}) or where users
generate their passwords algorithmically using a password manager
(\secref{sec:algorithmic}).  To assess the efficacy of honeywords when
users employ mixed strategies (i.e., chose some passwords themselves
and algorithmically generate others), we further constructed two test
datasets by mixing \Dataset{\TestTag}{\UserChosenTag} and the
algorithmically generated dataset. Then we generated honeywords based
on the type of the account password, i.e., applying
honeyword-generation methods described in
\secref{sec:background:algorithms} to generate honeywords for
user-chosen passwords and password managers to generate honeywords for
algorithmically generated passwords. Our study showed that increased
use of password managers in password creation can ease the tensions
brought on by password reuse and thus make better trade-offs between
false-positive and false-negative rates of honeywords. More details on
the experiments and results are shown in \appref{app:mixed}.

\subsection{Password Reuse}  \label{sec:discussion:password_reuse}

Our findings that password reuse across sites is so detrimental to
honeyword false-negative rates
(\secref{sec:user-selected:evaluation:results}) provides yet more
evidence that moving more users toward password managers would be good
policy (notwithstanding the risk of password-manager breaches,
e.g.,~\cite{toubba2023:incident}).  That said, a recent university
survey~\cite{mayer2022:managers} found that though a large majority
(77\%) of respondents reported using a password manager, another large
majority (again, 77\%) also reported still reusing passwords across
accounts.  So, while a step in the right direction, password managers
are evidently not a panacea.  A potentially more effective approach
might be explicitly hindering attempts to reuse passwords, either
through adoption of intentionally conflicting password requirements at
websites (which is not commonplace, see \secref{sec:discussion:configuration}) or through
explicit interventions during the password (re)setting process to
interfere with reusing the same or similar passwords
(e.g.,~\cite{wang2019:reuse}).

%% file: conclusion.tex
\section{Conclusion}
\label{sec:conclusion}

In this paper, we have conducted the first critical analysis of
honeyword-generation techniques for users who have suffered exposed
passwords for their accounts at other sites. We formalized the
false-positive rate and false-negative rate of honeywords in a model
where the attacker has access to passwords for the same users at other
sites or, in the case of false-positive attackers, even passwords for
users at the defending site (as the real users would). Using these
formalized definitions and a large dataset of leaked passwords, we
experimentally demonstrated that existing honeyword-generation
algorithms exhibit poor tradeoffs between false positives and false
negatives when the account password is chosen by an average human
user. Then we studied the case where the account password is
algorithmically generated and used passwords from popular password
managers to show that the existing honeyword-generation methods offer
modest protection against false-negative attackers. We further
explored the use of algorithmic password generators in honeyword
generation and determined that seemingly the only effective strategy
is to generate honeywords using the same password generator that the
user does, if it can determine what that password generator is.  In
total, we believe our results paint a cautionary picture for the state
of honeyword-generation algorithms to date, though they also set forth
new research challenges for the field.

%% file: appendix.tex
\appendix

\section{Honeyword-Generation Algorithms} \label{app:honeyword-generation}

\subsection{Password-Independent Honeyword\\ Generation}
\label{app:honeyword-generation:independent}

Password models used in password-independent honeyword
generation are pretrained on a multiset of
passwords \Dataset{\TrainTag}{\UserChosenTag}.  Let
\multiplicity{\regexp} be the multiplicity of passwords in
\Dataset{\TrainTag}{\UserChosenTag} matching regular expression
\regexp.  In particular, for passwords with characters drawn from the
alphabet \pwdAlphabet, $\multiplicity{\pwdAlphabet^{\ast}} =
\setSize{\Dataset{\TrainTag}{\UserChosenTag}}$.

\myparagraph{List model (\List)~\cite{wang2018:honeywords}}
The list model estimates the probability of password \password by
$\frac{\multiplicity{\password}}{\multiplicity{\pwdAlphabet^{\ast}}}$.
It generates a honeyword
by sampling a candidate from the list based on its estimated probability.

\myparagraph{Probabilistic context-free grammar model
  (\PCFG)~\cite{weir2009:pcfg-cracking}}
This algorithm treats a password as a sequence of segments, each a
digit segment, letter segment, or symbol segment.  Denoting the
alphabet $\pwdAlphabet = \letters \cup \digits \cup \symbols$ where
\begin{align*}
    \letters & = \{a, \ldots, z, A, \ldots, Z\} & \overline{\letters} & = \pwdAlphabet\setminus\letters \\
    \digits & = \{0, \ldots, 9\} & \overline{\digits} & = \pwdAlphabet\setminus\digits \\
    \symbols & = \{!, @, \#, \ldots\} & \overline{\symbols} & = \pwdAlphabet\setminus\symbols 
\end{align*}
PCFG models the probability of a password ``bike123'', for
example, as the product $\prob{\LetterSeg{4}\DigitSeg{3}}
\cprob{\big}{\text{bike}}{\LetterSeg{4}}
\cprob{\big}{\text{123}}{\DigitSeg{3}}$ where
{\small
\begin{align*}
  \prob{\LetterSeg{4}\DigitSeg{3}} & = \frac{\multiplicity{\letters^4 \digits^3}}{\multiplicity{\pwdAlphabet^{\ast}}} \\
  \cprob{\big}{\text{bike}}{\LetterSeg{4}} & = \frac{\multiplicity{\text{bike} ~\cup~ \text{bike}\nonletters\pwdAlphabet^{\ast} ~\cup~ \pwdAlphabet^{\ast}\nonletters\text{bike}\nonletters\pwdAlphabet^{\ast} ~\cup~ \pwdAlphabet^{\ast}\nonletters\text{bike}}}{\multiplicity{\letters^4 ~\cup~ \letters^4\nonletters\pwdAlphabet^{\ast} ~\cup~ \pwdAlphabet^{\ast}\nonletters\letters^4\nonletters\pwdAlphabet^{\ast} ~\cup~ \pwdAlphabet^{\ast}\nonletters\letters^4}} \\
  \cprob{\big}{\text{123}}{\DigitSeg{3}} & = \frac{\multiplicity{\text{123} ~\cup~ \text{123}\nondigits\pwdAlphabet^{\ast} ~\cup~ \pwdAlphabet^{\ast}\nondigits\text{123}\nondigits\pwdAlphabet^{\ast} ~\cup~ \pwdAlphabet^{\ast}\nondigits\text{123}}}{\multiplicity{\digits^3 ~\cup~ \digits^3\nondigits\pwdAlphabet^{\ast} ~\cup~ \pwdAlphabet^{\ast}\nondigits\digits^3\nondigits\pwdAlphabet^{\ast} ~\cup~ \pwdAlphabet^{\ast}\nondigits\digits^3}}
\end{align*}
}

We followed the implementations used in previous works
(e.g.,~\cite{weir2009:pcfg-cracking}) in the
training of \PCFG. We trained the \PCFG model by assigning the 
conditional probability of each
production rule such that the likelihood 
of passwords in \Dataset{\TrainTag}{\UserChosenTag} 
is maximized.

\myparagraph{Markov model (\Markov)~\cite{ma2014:markov-cracking}}
This model is a Markov chain of order \MarkovOrder such
that the prediction of character at the current position only depends
on the characters at the previous \MarkovOrder positions.  That is,
for a password $\password = \character{1}\ldots\character{\PasswordLength}$,
$\prob{\password} = \prod_{\PositionIndex=1}^{\PasswordLength}
\prob{\character{\PositionIndex} \mid \character{\PositionIndex-\min\{\PositionIndex-1, \MarkovOrder\} \dots \character{\PositionIndex-1}}}$
where
\begin{align*}
 \cprob{\big}{\character{\PositionIndex}}{\character{\PositionIndex-\min\{\PositionIndex-1,\MarkovOrder\}} \dots \character{\PositionIndex-1}}
=\frac{\multiplicity{\pwdAlphabet^{\ast}\character{\PositionIndex-\min\{\PositionIndex-1,\MarkovOrder\}} \dots \character{\PositionIndex-1} \character{\PositionIndex} \pwdAlphabet^{\ast}}}
     {\multiplicity{\pwdAlphabet^{\ast}\character{\PositionIndex-\min\{\PositionIndex-1,\MarkovOrder\}} \dots \character{\PositionIndex-1} \pwdAlphabet^{\ast}}}
\end{align*}

We trained the \Markov model to maximize the likelihood of passwords in 
\Dataset{\TrainTag}{\UserChosenTag}.

\myparagraph{Recurrent neural network (\RNN)~\cite{melicher2016:fla-pwdmeter}}
An RNN is a type of deep neural network, where nodes are connected to
form cycles. This property enables a recurrent neural network to
process ordered information, which makes it one of commonly used deep
neural networks used in text
generation~\cite{sutskever2011:textgen}. Previous works
(e.g.,~\cite{melicher2016:fla-pwdmeter}) have demonstrated
state-of-the-art performance by RNNs on password cracking and password
strength measurement, essentially by estimating the probability of a
password much like a Markov model does, but without the constraint of
the Markov assumption.  Specifically, the RNN $\Model(\cdot):
\pwdAlphabet^{\ast} \rightarrow [0,1]^{\setSize{\pwdAlphabet }+2}$
models a password distribution. That is, for a shifted password
$\password =
\character{0}\character{1}\ldots\character{\PasswordLength}$ with the
start-of-sequence $\character{0} = \SymbolOfStart$ added at the start,
the RNN defines the probability of a password as:
\begin{align*}
\prob{\password}
& = \prod_{\PositionIndex=1}^{\PasswordLength} \cprob{\big}{\character{\PositionIndex}}{\character{0} \dots \character{\PositionIndex-1}} 
 = \prod_{\PositionIndex=1}^{\PasswordLength} \arrComponent{\Model(\character{0} \dots \character{\PositionIndex-1})}{\character{\PositionIndex}}
\end{align*}
where \arrComponent{\arr}{\arrIdx} denotes the \arrIdx-th component of
vector \arr and we assume each $\character{} \in \pwdAlphabet \cup
\{\SymbolOfStart,\SymbolOfEnd\}$ is encoded as a unique $\arrIdx \in
\{1, \ldots, \setSize{\pwdAlphabet}+2\}$, where $\SymbolOfEnd$ denotes
the end-of-sequence. 

In this work, we referred to previous
work~\cite{melicher2016:fla-pwdmeter} to design a RNN model, which
consists of one embedding layer, one Long Short-Term Memory (LSTM)
network~\cite{hochreiter1997:lstm}, and a fully-connected layer. The
embedding layer works like a lookup table converting each character in
$\pwdAlphabet \cup \{\SymbolOfStart,\SymbolOfEnd\}$ into a fixed
length vector of 128 dimensions. The LSTM network is a stack of three
LSTM layers, each of which is a set of recurrently connected memory
blocks used to process information; it accepts 128 input features and
has 128 hidden features.  The fully connected layer is a linear layer
with 128-dimension inputs and 96-dimension outputs. The RNN model is trained by minimizing the cross-entropy loss in
\textit{batches}, each a multiset $\modelBatch \subseteq
\passwordDomain$. The cross-entropy loss is averaged over batches as:
\begin{align}
 \avg_{\password \in \modelBatch}~\sum_{\PositionIndex=1}^{\PasswordLength} -\log \arrComponent{\Model(\character{0} \dots \character{\PositionIndex-1})}{\character{\PositionIndex}},
 \label{eqn:rnn_loss}
\end{align}
where $\password =
\character{0}\character{1}\ldots\character{\PasswordLength}$.

We trained the RNN model by minimizing \eqref{eqn:rnn_loss} using
the Adam algorithm~\cite{kingma2014:adam} with a learning rate of
$10^{-3}$. We set the batch size to be $8192$ and the number of
training epochs to be $30$. To constitute a data batch used in this
model training, we merged all the password multisets in
\Dataset{\TrainTag}{\UserChosenTag} and then created each batch as a
consecutive window of $8192$ passwords (with start-of-sequence
added at the beginning) from the merged multiset.

The RNN model generates the honeywords as follows: at the
\PositionIndex-th step, given the start-of-sequence \character{0} and a
sequence of previously generated characters $\character{1} \dots
\character{\PositionIndex-1}$, the RNN model predicts the next
character \character{\PositionIndex} by randomly sampling a character
from $\pwdAlphabet \cup \{\SymbolOfEnd\}$ according to the
distribution $\Model(\character{0} \dots
\character{\PositionIndex-1})$. The process of generating a honeyword
terminates when it generates \PasswordLength characters or samples the
end-of-sequence.\footnote{All the hyperparameters of the models
considered in this work were fine-tuned based on the training dataset,
independent of the test dataset.}

\myparagraph{A combined method (\Combo)~\cite{wang2022:honeywords}}
This method is a combination of several password models, namely \List, \PCFG, and \Markov. 
It generates a honeyword as follows: with a probability of
$\frac{1}{3}$ each, it employs
\List, \PCFG, or \Markov to generate a honeyword.

\subsection{Password-Dependent Honeyword\\ Generation}

\label{app:honeyword-generation:dependent}

\myparagraph{Chaffing-by-Tweaking
  (\CBT{\TweakParameter}~\cite{juels2013:honeywords} and
 \CBTr~\cite{dionysiou2021:honeygen})} \CBT{\TweakParameter} tweaks the
input password by randomly changing the last \TweakParameter
characters into other characters of the same types (i.e., symbols,
letters, or digits). For example, ``bike123z'' can be tweaked by \CBT{3}
method into ``bike164T''. \CBTr is an improved method by Dionysiou,
et al.~\cite{dionysiou2021:honeygen}, which randomly replaces all the
symbols of the input password based on the following strategy:
lower-case each upper-case letter with probability 0.3, capitalize
each lower-case letter with probability 0.03, and change each digit to
a different, uniformly chosen digit with probability 0.05.

\myparagraph{Chaffing-with-a-hybrid-model (\CHM)~\cite{dionysiou2021:honeygen}}
This method leverages a FastText model~\cite{bojanowski2017:fasttext}
to search for $9$ nearest neighbors of the input password. Then, for
each output from FastText model (including the input password), it
applies\CBTr to \Tweak the password to generate
$(\nmbrHoneywords+1)/10$ passwords. As such, totally \nmbrHoneywords
passwords are returned as honeywords.

We followed the implementations used in previous works
(e.g.,~\cite{dionysiou2021:honeygen}) in the training of FastText. We
used the \texttt{train\_unsupervised} function of the FastText
library\footnote{https://fasttext.cc/docs/en/python-module.html} to
perform unsupervised training of the FastText ``skipgram'' model, with
the minimal number of word occurrences set to be $1$, the minimal
length of character-grams to be $2$, and the number of epochs to be
$500$.

\myparagraph{Chunk-level GPT-3 (\CGPT)~\cite{yu2023:honeygpt}}
This technique includes two steps.  First, taking the account password as
input, a password-specific segmentation technique called
PwdSegment~\cite{xu2021:chunk} is used to return the chunks contained
in the account password. For example, given an input password
``bike2000'', PwdSegment returns two chunks: ``bike'' and
``2000''. Second, a prompt containing information of the account
password and its chunks is provided to the GPT-3
model~\cite{brown2020:gpt3}, e.g.,
\begin{quote}
  Derive \nmbrHoneywords passwords that are similar to ``bike2000''
  and contain ``bike'' and ``2000''. The length of the passwords
  should be at most 8. Do not add digits at the end of the passwords.
\end{quote}
Then, the GPT-3 model returns a list of passwords, used as
honeywords. The temperature of GPT-3 is set to one in order to
guarantee diversity of the honeywords generated, as suggested by Yu
and Martin~\cite{yu2023:honeygpt}. However, the number of honeywords
output by GPT-3 is usually unequal to \nmbrHoneywords, especially when
\nmbrHoneywords is large (e.g., $\nmbrHoneywords = 99$).  If the
number of returned passwords is more than \nmbrHoneywords, we use the
top \nmbrHoneywords as honeywords. If the number is less than
\nmbrHoneywords, we use \CBTr to tweak the returned passwords
one-by-one until the number of honeywords reaches \nmbrHoneywords.

\myparagraph{Deep tweak model (\Tweak)~\cite{he2022:passtrans}}
This method applies a deep neural network that takes a password as
input and returns a tweaked password. Specifically, it leverages a
deep neural network $\Model(\cdot):\pwdAlphabet^{\ast} \times
\passwordDomain \rightarrow [0,1]^{\setSize{\pwdAlphabet}+2}$ to model a
conditional distribution.  That is, given a password input \password,
for a shifted tweaked password $\passwordAlt =
\character{0}\character{1}\ldots\character{\PasswordLength}$ with the
start-of-sequence $\character{0}$ added at the beginning, the deep
neural network models:
\begin{align*}
\cprob{\big}{\passwordAlt}{\password}
 = & \prod_{\PositionIndex=1}^{\PasswordLength} \cprob{\big}{\character{\PositionIndex}}{\character{0} \dots \character{\PositionIndex-1}, \password}  \\
 = & \prod_{\PositionIndex=1}^{\PasswordLength} \arrComponent{\Model(\character{0} \dots \character{\PositionIndex-1}, \password)}{\character{\PositionIndex}}
\end{align*}

This DNN model is a transformer, which is one of the widely used DNNs
for text representation learning in natural language
processing~\cite{vaswani2017:attention}. 
This model consists of an
embedding layer, a transformer network, and a fully connected
layer. The embedding layer works like a lookup table converting each
character of a password into a 64-dimension vector.  The transformer
network is a stack of three transformer encoder layers and three
decoder layers, each of which is the composition of a four-head
self-attention mechanism calculating the weighted average of
64-dimension representations and a fully-connected feed-forward
network outputting 256-dimension outputs.  The fully connected layer
is a linear layer taking 64-dimension inputs to 96-dimension outputs.
Such a model is trained by minimizing the cross-entropy loss performed
in \textit{batches}, each a multiset $\modelBatch \subseteq
\passwordDomain \times \auxiliaryInfoDomain$, which takes the form of:
\begin{align}
 \avg_{(\password, \passwordAlt) \in \modelBatch}~\sum_{\PositionIndex=1}^{\PasswordLength} -\log \arrComponent{\Model(\character{0} \dots \character{\PositionIndex-1}, \password)}{\character{\PositionIndex}}
 \label{eqn:tweak_loss}
\end{align}
where $\passwordAlt =
\character{0}\character{1}\ldots\character{\PasswordLength}$.

We trained the \Tweak model by minimizing \eqref{eqn:tweak_loss}
using the Adam algorithm with a learning rate of $10^{-3}$. The batch
size and the number of epochs are set to be $8192$ and $30$,
respectively. When training this model, each batch was created as a
consecutive window of $8192$ multisets of passwords
from \Dataset{\TrainTag}{\UserChosenTag} and then
sampling a pair of passwords $(\password, \passwordAlt)$ from each
selected multiset, where \passwordAlt is padded by a start-of-sequence.

Leveraging a trained \Tweak model, this method generates tweaked
passwords as follows: at the \PositionIndex-th step, given a password
input \password to be tweaked, the start-of-sequence \character{0}, and
a sequence of previously generated characters $\character{1} \dots
\character{\PositionIndex-1}$, the \Tweak model predicts the next
character \character{\PositionIndex} by randomly sampling a character
from $\pwdAlphabet \cup \{\SymbolOfEnd\}$ according to the conditional
distribution $\Model(\character{0} \dots
\character{\PositionIndex-1}, \password)$. The process of generating a
tweaked password terminates when it generates \PasswordLength
characters or samples the end-of-sequence.

\myparagraph{Tweaking path model (\PasstoPath)~\cite{pal2019:credential-stuffing}}
This method utilizes a DNN that takes password as input and returns an
\textit{edit path} $\TransformationSeq{\password}{\passwordAlt}
\subseteq \TransformationSet^{\ast}$, which is a sequence of
transformations from the password $\password$ to $\passwordAlt$. Here
\TransformationSet is a global set of transformation units, each of
which takes the form of $\TransformationUnit =
(\TransformationOperation, \TransformationCharacter,
\PositionIndex)$. \TransformationOperation is an edit operation
selected among insertion \Insertion, substitute \Substitute, and
deletion \Deletion. $\TransformationCharacter \in \pwdAlphabet \cup
\{\EmptySymbol\}$ is a character from \pwdAlphabet to insert or to
substitute, or a empty symbol \EmptySymbol used in
deletion. $\PositionIndex$ is the position index of the input password
to be edited. For example, (\Substitute, ``a'', 1) denotes the
transformation that substitutes the first character $\character{1}$ of
the password with ``a''.

Specifically, this method utilizes a DNN
$\Model(\cdot): \TransformationSet^{\ast} \times
\auxiliaryInfoDomain \rightarrow [0,1]^{\setSize{\TransformationSet}+2}$
to model a conditional
distribution.  That is, given a password input \password, for a
shifted edit path $\TransformationSeq{\password}{\passwordAlt} =
\TransformationUnit{0} \TransformationUnit{1} \dots
\TransformationUnit{\TransformationLength}$ with the start-of-sequence
$\TransformationUnit{0} = \SymbolOfStart$ added at the beginning, the
DNN models the conditional distribution as:
\begin{align*}
\cprob{\big}{\TransformationSeq{\password}{\passwordAlt}}{\password}
 = & \prod_{\TransformationIndex=1}^{\TransformationLength} \cprob{\big}{\TransformationUnit{\TransformationIndex}}{\TransformationUnit{0} \dots \TransformationUnit{\TransformationIndex-1}, \password} \\
 = & \prod_{\TransformationIndex=1}^{\TransformationLength} \arrComponent{\Model(\TransformationUnit{0} \dots \TransformationUnit{\TransformationIndex-1}, \password)}{\TransformationUnit{\TransformationIndex}},
\end{align*}
where we assume each $\TransformationUnit \in \TransformationSet \cup
\{\SymbolOfStart, \SymbolOfEnd\}$ is encoded as a unique $\arrIdx \in
\{1, \ldots, \setSize{\TransformationSet}+2\}$.

We designed the \PasstoPath model based on a similar design for cracking
passwords~\cite{pal2019:credential-stuffing}.
This model consists of
two embedding layers, two recurrent neural networks, and a fully
connected layer.  The first embedding layer converts each character of
a password (96 dimensions) into a 200-dimension vector, while the
second one converts each token of a path (5861 dimensions) into a
200-dimension vector. The two recurrent neural networks are used as an
encoder and an decoder, both of which are stacks of three LSTM layers
each with $128$ hidden units. The fully connected layer is a linear
layer with 128-dimension inputs and 5861-dimension outputs.
This model is trained by minimizing the
cross entropy loss performed in \textit{batches}, each a multiset
$\modelBatch \subseteq \passwordDomain \times
\TransformationSet^{\ast}$, which takes the form of:
\begin{align}
 \avg_{(\password, \TransformationSeq{\password}{\passwordAlt}) \in \modelBatch}~\sum_{\TransformationIndex=1}^{\TransformationLength} -\log \arrComponent{\Model(\TransformationUnit{0} \dots \TransformationUnit{\TransformationIndex-1}, \password)}{\TransformationUnit{\TransformationIndex}},
 \label{eqn:p2p_loss}
\end{align}
where $\TransformationSeq{\password}{\passwordAlt} =
\TransformationUnit{0} \TransformationUnit{1} \dots
\TransformationUnit{\TransformationLength}$.

We trained the \PasstoPath model using the Adam algorithm with a
learning rate of $3 \times 10^{-4}$, a drop rate of $0.4$, a batch
size of $2048$, and $3$ epochs, as suggested by previous
work~\cite{pal2019:credential-stuffing}. In this model training, each
data batch consisted of pairs of password and edit path used as input
and label, respectively. To constitute such a data batch, we first select a consecutive window of $2048$ password multisets from the training sets and then sample a pair of passwords from each selected multiset. Then for each pair of passwords $(\password,
\passwordAlt)$, we calculated the minimal edit path
$\TransformationSeq{\password}{\passwordAlt}$ by dynamic programming
and backtracking~\cite{pal2019:credential-stuffing}.

After training, the \PasstoPath method generates honeywords as follows:
at the $\TransformationIndex$-th step, given a password input
$\password$, the start-of-sequence $\TransformationUnit{0}$, and a
sequence of previously generated transformation units
$\TransformationUnit{1}\dots\TransformationUnit{\TransformationIndex-1}$,
\PasstoPath generates the next transformation unit
$\TransformationUnit{\TransformationIndex}$ by randomly sampling a
candidate from $\TransformationSet \cup \{\SymbolOfEnd\}$ based on the
conditional distribution defined by
$\Model(\TransformationUnit{0} \dots
\TransformationUnit{\TransformationIndex-1}, \password)$. The edit
path generation process stops when it completes
$\TransformationLength$ transformation units or it samples the
end-of-sequence. Then the method produces a honeyword by applying the
generated edit path on the password $\password$.

\section{Similarity Model Design and Training} \label{app:sim_model}

We designed the similarity model (described in \secref{sec:user-selected:sim-model}) by a transformer. The
transformer-based similarity model \Model consists of an embedding
layer, a transformer encoder, and a fully connected layer.
The embedding layer works like a lookup table converting each character
into a 64-dimension vector.  The transformer encoder network is a
stack of three transformer encoder layers, each of which is composed
by a four-head self-attention mechanism calculating the weighted
average of 64-dimension representations and a fully-connected
feed-forward network outputting 256-dimension outputs.  The fully
connected layer is a linear layer with 1920-dimension inputs and
256-dimension outputs.

We trained the similarity model to minimize \eqnref{eqn:loss} using the Adam
algorithm with learning rate $10^{-3}$. When training the model, each
batch was created as a consecutive window of samples from the
training sets. We found that with a batch size of
$\setSize{\modelBatch} = 256$, the loss \eqnref{eqn:loss} converged
and the performance of the model became stable after only a few
training epochs. Therefore, in all the results we report below, we
trained the model for only $10$ epochs.

\section{Additional Experimental Results in the Case of
User-Chosen Password}  \label{app:additional_results}

The experimental results from \RNN,
\RNNpwd, \TarRNN, and \CBT{3} in 
the case of user-chosen password are shown in
\figref{fig:uc:to20:additional} and \figref{fig:uc:to100:additional}.

\begin{figure}[t!]
  \vspace{0.1in}
  \begin{subfigure}[b]{0.47\textwidth}
    \centering
    \resizebox{!}{2em}{\input{user_chosen_trade-off_legend.tex}}
  \end{subfigure}
  
  \begin{subfigure}[b]{0.232\textwidth}
    \setlength\figureheight{1.9in}
    \centering
    \caption{\RNN}
    \vspace{1.25ex}
    \resizebox{\textwidth}{!}{\input{figure/user_chosen/RNN/trade-off_20}}
    \label{fig:uc:to20:rnn}
  \end{subfigure}
  \hfill
  \begin{subfigure}[b]{0.181\textwidth}
    \setlength\figureheight{1.9in}
    \centering
    \caption{\RNNpwd}
    \vspace{2ex}
    \resizebox{\textwidth}{!}{\input{figure/user_chosen/RNN_/trade-off_20}}
    \label{fig:uc:to20:rnn-pwd}
  \end{subfigure}
  \\[-2.5ex]
  \begin{subfigure}[b]{0.47\textwidth}
    \setlength\figureheight{2in}
      \centering
      \resizebox{!}{1.5em}{\input{xlabel.tex}}
  \end{subfigure}
  \vspace{1ex}
  \\
  \begin{subfigure}[b]{0.232\textwidth}
    \setlength\figureheight{1.9in}
    \centering
    \caption{\TarRNN}
    \vspace{1.25ex}
    \resizebox{\textwidth}{!}{\input{figure/user_chosen/RNN_target/trade-off_20}}
    \label{fig:uc:to20:rnn-tgt}
  \end{subfigure}
  \hfill
  \begin{subfigure}[b]{0.181\textwidth}
    \setlength\figureheight{1.9in}
    \centering
    \caption{\CBT{3}}
    \vspace{2ex}
    \resizebox{\textwidth}{!}{\input{figure/user_chosen/chaffing3/trade-off_20}}
    \label{fig:uc:to20:cbt3}
  \end{subfigure}
  \\[-2.5ex]
    \begin{subfigure}[b]{0.47\textwidth}
      \setlength\figureheight{2in}
        \centering
        \resizebox{!}{1.5em}{\input{xlabel.tex}}
    \end{subfigure}
    
  \caption{$\fpProbWOParams(\alarmAttacker)$
    vs.\ $\fnProbWOParams(\breachAttacker)$ as \alarmThreshold is varied,
    for the case of user-chosen passwords ($\nmbrHoneywords=19$,
    $\attackBudget=1000$).}
  \label{fig:uc:to20:additional} 
  \end{figure}

  \begin{figure}[t!]
    \vspace{0.1in}
    \begin{subfigure}[b]{0.47\textwidth}
      \centering
      \resizebox{!}{2em}{\input{user_chosen_trade-off_legend.tex}}
    \end{subfigure}
    
    \begin{subfigure}[b]{0.232\textwidth}
      \setlength\figureheight{1.9in}
      \centering
      \caption{\RNN}
      \vspace{1.25ex}
      \resizebox{\textwidth}{!}{\input{figure/user_chosen/RNN/trade-off_100}}
      \label{fig:uc:to100:rnn}
    \end{subfigure}
    \hfill
    \begin{subfigure}[b]{0.181\textwidth}
      \setlength\figureheight{1.9in}
      \centering
      \caption{\RNNpwd}
      \vspace{2ex}
      \resizebox{\textwidth}{!}{\input{figure/user_chosen/RNN_/trade-off_100}}
      \label{fig:uc:to100:rnn-pwd}
    \end{subfigure}
    \\[-2.5ex]
    \begin{subfigure}[b]{0.47\textwidth}
      \setlength\figureheight{2in}
        \centering
        \resizebox{!}{1.5em}{\input{xlabel.tex}}
    \end{subfigure}
    \vspace{1ex}
    \\
    \begin{subfigure}[b]{0.232\textwidth}
      \setlength\figureheight{1.9in}
      \centering
      \caption{\TarRNN}
      \vspace{1.25ex}
      \resizebox{\textwidth}{!}{\input{figure/user_chosen/RNN_target/trade-off_100}}
      \label{fig:uc:to100:rnn-tgt}
    \end{subfigure}
    \hfill
    \begin{subfigure}[b]{0.181\textwidth}
      \setlength\figureheight{1.9in}
      \centering
      \caption{\CBT{3}}
      \vspace{2ex}
      \resizebox{\textwidth}{!}{\input{figure/user_chosen/chaffing3/trade-off_100}}
      \label{fig:uc:to100:cbt3}
    \end{subfigure}
    \\[-2.5ex]
      \begin{subfigure}[b]{0.47\textwidth}
        \setlength\figureheight{2in}
          \centering
          \resizebox{!}{1.5em}{\input{xlabel.tex}}
      \end{subfigure}
  \caption{$\fpProbWOParams(\alarmAttacker)$
    vs.\ $\fnProbWOParams(\breachAttacker)$ as \alarmThreshold is varied,
    for the case of user-chosen passwords ($\nmbrHoneywords=99$,
    $\attackBudget=1000$).}
  \label{fig:uc:to100:additional} 
  \end{figure}

\section{Training Classifier of Password Generators}
\label{app:algorithmic:classifier}

We trained a classifier to classify algorithmically generated passwords
into $\{1, 2, \dots, \NumGenerators\}$. Such a classifier can be used
by the false-negative attacker to distinguish the account password from the honeywords or
by the honeyword system to select a proper password generator to generate 
honeywords. The classifier consisted of
one embedding layer, a LSTM network, and two fully connected
layers. The embedding layer converted each character into a vector of
$128$ dimensions. The LSTM network was a stack of three LSTM layers,
accepting $128$ input features and $128$ hidden features. The first
fully connected layer was a linear layer with $128$-dimension inputs
and $64$-dimension outputs.  The second had $64$-dimension inputs
and $96$-dimension outputs.

We trained the model by minimizing the cross-entropy loss using the
Adam algorithm~\cite{kingma2014:adam} with a learning rate of
$10^{-4}$. We set the batch size to $1024$ and the training epochs
to $300$. To constitute each training batch, we select a consecutive window of $1024$
passwords from \Dataset{\TrainTag}{\AlgGenTag}. We
used classification accuracy as the evaluation metric. After each
epoch, we evaluated the trained classifier on
\Dataset{\EvalTag}{\AlgGenTag} and saved the model that had the best
performance on evaluation. The best model achieved evaluation accuracy
of $90.91\%$.  Increasing the amount of training data did not
materially improve the accuracy.

\begin{figure*}[ht!]
  \vspace{0.1in}
\begin{subfigure}[b]{0.179\textwidth}
  \setlength\figureheight{1.9in}
  \centering
  \caption{\List}
  \vspace{1.25ex}
  \resizebox{\textwidth}{!}{\input{figure/random_generated/List/trade-off_20}}
  \label{fig:ag:to20:list}
\end{subfigure}
\hfill
\begin{subfigure}[b]{0.141\textwidth}
  \setlength\figureheight{1.9in}
  \centering
  \caption{\Markov}
  \vspace{2ex}
  \resizebox{\textwidth}{!}{\input{figure/random_generated/Markov/trade-off_20}}
  \label{fig:ag:to20:markov}
\end{subfigure}
\hfill
\begin{subfigure}[b]{0.141\textwidth}
  \setlength\figureheight{1.9in}
  \centering
  \caption{\PCFG}
  \vspace{2ex}
  \resizebox{\textwidth}{!}{\input{figure/random_generated/PCFG/trade-off_20}}
  \label{fig:ag:to20:pcfg}
\end{subfigure}
\hfill
\begin{subfigure}[b]{0.141\textwidth}
  \setlength\figureheight{1.9in}
  \centering
  \caption{\Combo}
  \vspace{2ex}
  \resizebox{\textwidth}{!}{\input{figure/random_generated/combination/trade-off_20}}
  \label{fig:ag:to20:combo}
\end{subfigure}
\hfill
\begin{subfigure}[b]{0.141\textwidth}
  \setlength\figureheight{1.9in}
  \centering
  \caption{\CBT{3}}
  \vspace{2ex}
  \resizebox{\textwidth}{!}{\input{figure/random_generated/chaffing3/trade-off_20}}
  \label{fig:ag:to20:cbt3}
\end{subfigure}
\\[-2.5ex]
  \begin{subfigure}[b]{\textwidth}
    \setlength\figureheight{2in}
      \centering
      \resizebox{!}{1.5em}{\input{xlabel.tex}}
  \end{subfigure}
  \vspace{1ex}
\\ 
\begin{subfigure}[b]{0.179\textwidth}
  \setlength\figureheight{1.9in}
  \centering
  \caption{\Listpwd}
  \vspace{1.25ex}
  \resizebox{\textwidth}{!}{\input{figure/random_generated/List_/trade-off_20}}
  \label{fig:ag:to20:list-pwd}
\end{subfigure}
\hfill
\begin{subfigure}[b]{0.141\textwidth}
  \setlength\figureheight{1.9in}
  \centering
  \caption{\Markovpwd}
  \vspace{2ex}
  \resizebox{\textwidth}{!}{\input{figure/random_generated/Markov_/trade-off_20}}
  \label{fig:ag:to20:markov-pwd}
\end{subfigure}
\hfill
\begin{subfigure}[b]{0.141\textwidth}
  \setlength\figureheight{1.9in}
  \centering
  \caption{\PCFGpwd}
  \vspace{2ex}
  \resizebox{\textwidth}{!}{\input{figure/random_generated/PCFG_/trade-off_20}}
  \label{fig:ag:to20:pcfg-pwd}
\end{subfigure}
\hfill
\begin{subfigure}[b]{0.141\textwidth}
  \setlength\figureheight{1.9in}
  \centering
  \caption{\Combopwd}
  \vspace{2ex}
  \resizebox{\textwidth}{!}{\input{figure/random_generated/combination_/trade-off_20}}
  \label{fig:ag:to20:combo-pwd}
\end{subfigure}
\hfill
\begin{subfigure}[b]{0.141\textwidth}
  \setlength\figureheight{1.9in}
  \centering
  \caption{\CBT{4}}
  \vspace{2ex}
  \resizebox{\textwidth}{!}{\input{figure/random_generated/chaffing4/trade-off_20}}
  \label{fig:ag:to20:cbt4}
\end{subfigure}
\\[-2.5ex]
  \begin{subfigure}[b]{\textwidth}
    \setlength\figureheight{2in}
      \centering
      \resizebox{!}{1.5em}{\input{xlabel.tex}}
  \end{subfigure}
  \vspace{1ex}
\\
  \begin{subfigure}[b]{0.179\textwidth}
  \setlength\figureheight{1.9in}
  \centering
  \caption{\CBTr}
  \vspace{1.25ex}
  \resizebox{\textwidth}{!}{\input{figure/random_generated/CBTr/trade-off_20}}
  \label{fig:ag:to20:cbtr}
\end{subfigure}
\hfill
\begin{subfigure}[b]{0.141\textwidth}
  \setlength\figureheight{1.9in}
  \centering
  \caption{\CHM}
  \vspace{2ex}
  \resizebox{\textwidth}{!}{\input{figure/random_generated/CHM/trade-off_20}}
  \label{fig:ag:to20:chm}
\end{subfigure}
\hfill
\begin{subfigure}[b]{0.141\textwidth}
  \setlength\figureheight{1.9in}
  \centering
  \caption{\CGPT}
  \vspace{2ex}
  \resizebox{\textwidth}{!}{\input{figure/random_generated/gpt/trade-off_20}}
  \label{fig:ag:to20:gpt}
\end{subfigure}
\hfill
\begin{subfigure}[b]{0.141\textwidth}
  \setlength\figureheight{1.9in}
  \centering
  \caption{\Tweak}
  \vspace{2ex}
  \resizebox{\textwidth}{!}{\input{figure/random_generated/Tweak/trade-off_20}}
  \label{fig:ag:to20:tweak}
\end{subfigure}
\hfill
\begin{subfigure}[b]{0.141\textwidth}
  \setlength\figureheight{1.9in}
  \centering
  \caption{\PasstoPath}
  \vspace{2ex}
  \resizebox{\textwidth}{!}{\input{figure/random_generated/pass2path/trade-off_20}}
  \label{fig:ag:to20:pass2path}
\end{subfigure}
\\[-2.5ex]
  \begin{subfigure}[b]{\textwidth}
    \setlength\figureheight{2in}
      \centering
      \resizebox{!}{1.5em}{\input{xlabel.tex}}
  \end{subfigure}

\caption{$\fpProbWOParams(\alarmAttacker)$
  vs.\ $\fnProbWOParams(\breachAttacker)$ as \alarmThreshold is varied,
  for algorithmically generated passwords ($\nmbrHoneywords=19$,
  $\attackBudget=1000$). The best $\fnProbWOParams(\breachAttacker)$
  are $0.27$ (\CBT{3}, \figref{fig:ag:to20:cbt3}), $0.37$ (\CBT{4},
  \figref{fig:ag:to20:cbt4}), $0.55$ (\CBTr,
  \figref{fig:ag:to20:cbtr}), and $0.59$ (\PasstoPath,
  \figref{fig:ag:to20:pass2path}); all others have
  $\fnProbWOParams(\breachAttacker) > 0.69$. All suffer
  $\fpProbWOParams(\alarmAttacker) > 0.09$ at $\alarmThreshold = 1$.
  Those that reach $\fpProbWOParams(\alarmAttacker) \approx 0$ do so
  with $\fnProbWOParams(\breachAttacker) > 0.80$.}
\label{fig:ag:to20} 
\vspace{0.1in}
\end{figure*}

\begin{figure*}[ht!]
  \vspace{0.1in}
\begin{subfigure}[b]{0.179\textwidth}
  \setlength\figureheight{1.9in}
  \centering
  \caption{\List}
  \vspace{1.25ex}
  \resizebox{\textwidth}{!}{\input{figure/random_generated/List/trade-off_100}}
  \label{fig:ag:to100:list}
\end{subfigure}
\hfill
\begin{subfigure}[b]{0.141\textwidth}
  \setlength\figureheight{1.9in}
  \centering
  \caption{\Markov}
  \vspace{2ex}
  \resizebox{\textwidth}{!}{\input{figure/random_generated/Markov/trade-off_100}}
  \label{fig:ag:to100:markov}
\end{subfigure}
\hfill
\begin{subfigure}[b]{0.141\textwidth}
  \setlength\figureheight{1.9in}
  \centering
  \caption{\PCFG}
  \vspace{2ex}
  \resizebox{\textwidth}{!}{\input{figure/random_generated/PCFG/trade-off_100}}
  \label{fig:ag:to100:pcfg}
\end{subfigure}
\hfill
\begin{subfigure}[b]{0.141\textwidth}
  \setlength\figureheight{1.9in}
  \centering
  \caption{\Combo}
  \vspace{2ex}
  \resizebox{\textwidth}{!}{\input{figure/random_generated/combination/trade-off_100}}
  \label{fig:ag:to100:combo}
\end{subfigure}
\hfill
\begin{subfigure}[b]{0.141\textwidth}
  \setlength\figureheight{1.9in}
  \centering
  \caption{\CBT{3}}
  \vspace{2ex}
  \resizebox{\textwidth}{!}{\input{figure/random_generated/chaffing3/trade-off_100}}
  \label{fig:ag:to100:cbt3}
\end{subfigure}
\\[-2.5ex]
  \begin{subfigure}[b]{\textwidth}
    \setlength\figureheight{2in}
      \centering
      \resizebox{!}{1.5em}{\input{xlabel.tex}}
  \end{subfigure}
  \vspace{1ex}
\\  
\begin{subfigure}[b]{0.179\textwidth}
  \setlength\figureheight{1.9in}
  \centering
  \caption{\Listpwd}
  \vspace{1.25ex}
  \resizebox{\textwidth}{!}{\input{figure/random_generated/List_/trade-off_100}}
  \label{fig:ag:to100:list-pwd}
\end{subfigure}
\hfill
\begin{subfigure}[b]{0.141\textwidth}
  \setlength\figureheight{1.9in}
  \centering
  \caption{\Markovpwd}
  \vspace{2ex}
  \resizebox{\textwidth}{!}{\input{figure/random_generated/Markov_/trade-off_100}}
  \label{fig:ag:to100:markov-pwd}
\end{subfigure}
\hfill
\begin{subfigure}[b]{0.141\textwidth}
  \setlength\figureheight{1.9in}
  \centering
  \caption{\PCFGpwd}
  \vspace{2ex}
  \resizebox{\textwidth}{!}{\input{figure/random_generated/PCFG_/trade-off_100}}
  \label{fig:ag:to100:pcfg-pwd}
\end{subfigure}
\hfill
\begin{subfigure}[b]{0.141\textwidth}
  \setlength\figureheight{1.9in}
  \centering
  \caption{\Combopwd}
  \vspace{2ex}
  \resizebox{\textwidth}{!}{\input{figure/random_generated/combination_/trade-off_100}}
  \label{fig:ag:to100:combo-pwd}
\end{subfigure}
\hfill
\begin{subfigure}[b]{0.141\textwidth}
  \setlength\figureheight{1.9in}
  \centering
  \caption{\CBT{4}}
  \vspace{2ex}
  \resizebox{\textwidth}{!}{\input{figure/random_generated/chaffing4/trade-off_100}}
  \label{fig:ag:to100:cbt4}
\end{subfigure}
\\[-2.5ex]
  \begin{subfigure}[b]{\textwidth}
    \setlength\figureheight{2in}
      \centering
      \resizebox{!}{1.5em}{\input{xlabel.tex}}
  \end{subfigure}
  \vspace{1ex}
\\
\begin{subfigure}[b]{0.179\textwidth}
  \setlength\figureheight{1.9in}
  \centering
  \caption{\CBTr}
  \vspace{1.25ex}
  \resizebox{\textwidth}{!}{\input{figure/random_generated/CBTr/trade-off_100}}
  \label{fig:ag:to100:cbtr}
\end{subfigure}
\hfill
\begin{subfigure}[b]{0.141\textwidth}
  \setlength\figureheight{1.9in}
  \centering
  \caption{\CHM}
  \vspace{2ex}
  \resizebox{\textwidth}{!}{\input{figure/random_generated/CHM/trade-off_100}}
  \label{fig:ag:to100:chm}
\end{subfigure}
\hfill
\begin{subfigure}[b]{0.141\textwidth}
  \setlength\figureheight{1.9in}
  \centering
  \caption{\CGPT}
  \vspace{2ex}
  \resizebox{\textwidth}{!}{\input{figure/random_generated/gpt/trade-off_100}}
  \label{fig:ag:to100:gpt}
\end{subfigure}
\hfill
\begin{subfigure}[b]{0.141\textwidth}
  \setlength\figureheight{1.9in}
  \centering
  \caption{\Tweak}
  \vspace{2ex}
  \resizebox{\textwidth}{!}{\input{figure/random_generated/Tweak/trade-off_100}}
  \label{fig:ag:to100:tweak}
\end{subfigure}
\hfill
\begin{subfigure}[b]{0.141\textwidth}
  \setlength\figureheight{1.9in}
  \centering
  \caption{\PasstoPath}
  \vspace{2ex}
  \resizebox{\textwidth}{!}{\input{figure/random_generated/pass2path/trade-off_100}}
  \label{fig:ag:to100:pass2path}
\end{subfigure}
\\[-2.5ex]
  \begin{subfigure}[b]{\textwidth}
    \setlength\figureheight{2in}
      \centering
      \resizebox{!}{1.5em}{\input{xlabel.tex}}
  \end{subfigure}

\caption{$\fpProbWOParams(\alarmAttacker)$
  vs.\ $\fnProbWOParams(\breachAttacker)$ as \alarmThreshold is varied,
  for algorithmically generated passwords ($\nmbrHoneywords=99$,
  $\attackBudget=1000$). The best $\fnProbWOParams(\breachAttacker)$
  are $0.07$ (\CBT{3}, \figref{fig:ag:to100:cbt3}), $0.14$ (\CBT{4},
  \figref{fig:ag:to100:cbt4}), and $0.25$ (\PasstoPath,
  \figref{fig:ag:to100:pass2path}); all others have
  $\fnProbWOParams(\breachAttacker) > 0.40$. All suffer
  $\fpProbWOParams(\alarmAttacker) > 0.17$ at $\alarmThreshold = 1$.
  Those that reach $\fpProbWOParams(\alarmAttacker) \approx 0$ do so
  with $\fnProbWOParams(\breachAttacker) > 0.70$.}
\label{fig:ag:to100} 
\vspace{0.1in}
\end{figure*}

\section{Existing Honeyword Generators with \\ Algorithmically Generated Passwords}
\label{app:existing}

In this section, we show experimental results for existing
honeyword-generation methods described in
\secref{sec:background} in the case when passwords are
algorithmically-generated.  We report
$\fpProbWOParams(\alarmAttacker)$ and $\fnProbWOParams(\breachAttacker)$
on those honeyword-generation algorithms for the attackers \alarmAttacker and
\breachAttacker described in \secref{sec:algorithmic:strategies}. To
depict the tradeoffs between $\fpProbWOParams(\alarmAttacker)$ and
$\fnProbWOParams(\breachAttacker)$, we plot them against one another as
\alarmThreshold is varied.  \figref{fig:ag:to20} shows these tradeoffs
when $\nmbrHoneywords = 19$ honeywords and $\attackBudget = 1000$,
where circles ($\bullet$) mark the $\fpProbWOParams(\alarmAttacker)$
vs.\ $\fnProbWOParams(\breachAttacker)$ tradeoff at specific values of
\alarmThreshold ranging from $\alarmThreshold = 1$ to \nmbrHoneywords
when $\auxiliaryInfo=\passwordSet{\emailAddr} \setminus \{\password\}$
in each plot. Again, we stress that $\attackBudget = 1000$ yields an
optimistic evaluation of $\fpProbWOParams(\alarmAttacker)$ since
$\attackBudget = 1000$ is $1000\times$ too small compared with the
number $10^6$ recommended by Flor\^{e}ncio, et
al.~\cite{florencio2014:guide}.

As seen in \figref{fig:ag:to20}, the best
$\fnProbWOParams(\breachAttacker)$ that the honeyword-generation
techniques accomplish is $0.27$ (\CBT{3}, \figref{fig:ag:to20:cbt3}), $0.37$ (\CBT{4},
\figref{fig:ag:to20:cbt4}), $0.55$ (\CBTr,
\figref{fig:ag:to20:cbtr}), and $0.59$ (\PasstoPath, \figref{fig:ag:to20:pass2path}). All others suffer from
$\fnProbWOParams(\breachAttacker) > 0.79$, and password-independent
methods including \List, \Markov, \PCFG, \RNN, and \Combo have
$\fnProbWOParams(\breachAttacker) \approx 1$. Intuitively, those
methods, except \CBT, generate honeywords based on a
pretrained password models. Most of their generated honeywords do not
fit the pattern of algorithmically generated passwords, and thus it is
easy for \breachAttacker to distinguish the account password from the
honeywords. In contrast, \CBT generates honeywords by
random replacement. Although random replacement still breaks the
pattern from an algorithmic password generator with some probability,
it produces honeywords looking more like algorithmically generated
passwords, making it challenging for \breachAttacker to distinguish
them. In the evaluation of $\fpProbWOParams(\alarmAttacker)$, none of
the honeyword-generation methods achieved
$\fpProbWOParams(\alarmAttacker) \approx 0$ at $\alarmThreshold =
1$. \CBT{3}, which accomplished the best
$\fnProbWOParams(\breachAttacker)$, has a high
$\fpProbWOParams(\alarmAttacker) = 0.37$. The best method in
$\fpProbWOParams(\alarmAttacker)$ is \CBT{4}, which achieves $0.09$,
still much larger than $0$. Growing \alarmThreshold reduces
$\fpProbWOParams(\alarmAttacker)$ but increases
$\fnProbWOParams(\breachAttacker)$.  Those methods that achieve
$\fpProbWOParams(\alarmAttacker) \approx 0$ do so with
$\fnProbWOParams(\breachAttacker) > 0.80$.

A natural method to attempt to decrease
$\fnProbWOParams(\breachAttacker)$ would be to increase the number
\nmbrHoneywords of honeywords, but the more pronounced effect of doing
so is increasing $\fpProbWOParams(\alarmAttacker)$, instead.  Indeed,
\figref{fig:ag:to100} shows the impact of increasing \nmbrHoneywords
to $\nmbrHoneywords = 99$.  As seen there, an order-of-magnitude
increase in \nmbrHoneywords resulted in a slight improvement to
$\fnProbWOParams(\breachAttacker)$ in each case, but a more substantial
increase to $\fpProbWOParams(\alarmAttacker)$.  All
honeyword-generation techniques described in \secref{sec:background}
failed to achieve low $\fpProbWOParams(\alarmAttacker)$ and
$\fnProbWOParams(\breachAttacker)$ simultaneously when the user employs
an algorithmic password generator.

\section{A Mixed Case Study} \label{app:mixed}

\begin{figure*}[t!]
  \vspace{0.1in}
  \begin{subfigure}[b]{\textwidth}
    \centering
    \resizebox{!}{2em}{\input{user_chosen_trade-off_legend.tex}}
  \end{subfigure}
  
  \begin{subfigure}[b]{0.188\textwidth}
    \setlength\figureheight{1.9in}
    \centering
    \caption{\PCFG+\ClassifiedGenerator}
    \vspace{1.25ex}
    \resizebox{\textwidth}{!}{\input{figure/mixed_0.33/PCFG/trade-off_20}}
    \label{fig:uc:mix0.33:pcfg}
  \end{subfigure}
  \hfill
  \begin{subfigure}[b]{0.148\textwidth}
    \setlength\figureheight{1.9in}
    \centering
    \caption{\Combo+\ClassifiedGenerator}
    \vspace{2ex}
    \resizebox{\textwidth}{!}{\input{figure/mixed_0.33/combination/trade-off_20}}
    \label{fig:uc:mix0.33:combo}
  \end{subfigure}
  \hfill
  \begin{subfigure}[b]{0.148\textwidth}
    \setlength\figureheight{1.9in}
    \centering
    \caption{\PCFGpwd+\ClassifiedGenerator}
    \vspace{2ex}
    \resizebox{\textwidth}{!}{\input{figure/mixed_0.33/PCFG_/trade-off_20}}
    \label{fig:uc:mix0.33:pcfg-pwd}
  \end{subfigure}
  \hfill
  \begin{subfigure}[b]{0.148\textwidth}
    \setlength\figureheight{1.9in}
    \centering
    \caption{\Combopwd+\ClassifiedGenerator}
    \vspace{2ex}
    \resizebox{\textwidth}{!}{\input{figure/mixed_0.33/combination_/trade-off_20}}
    \label{fig:uc:mix0.33:combo-pwd}
  \end{subfigure}
  \hfill
  \begin{subfigure}[b]{0.148\textwidth}
    \setlength\figureheight{1.9in}
    \centering
    \caption{\TarPCFG+\ClassifiedGenerator}
    \vspace{2ex}
    \resizebox{\textwidth}{!}{\input{figure/mixed_0.33/PCFG_target/trade-off_20}}
    \label{fig:uc:mix0.33:pcfg-tgt}
  \end{subfigure}
  \hfill
  \begin{subfigure}[b]{0.148\textwidth}
    \setlength\figureheight{1.9in}
    \centering
    \caption{\TarCombo+\ClassifiedGenerator}
    \vspace{2ex}
    \resizebox{\textwidth}{!}{\input{figure/mixed_0.33/combination_target/trade-off_20}}
    \label{fig:uc:mix0.33:combo-tgt}
  \end{subfigure}
  \\[-2.5ex]
  \begin{subfigure}[b]{\textwidth}
    \setlength\figureheight{2in}
      \centering
      \resizebox{!}{1.5em}{\input{xlabel.tex}}
  \end{subfigure}
  \vspace{1ex}
  \\
  \begin{subfigure}[b]{0.188\textwidth}
    \setlength\figureheight{1.9in}
    \centering
    \caption{\CGPT+\ClassifiedGenerator}
    \vspace{1.25ex}
    \resizebox{\textwidth}{!}{\input{figure/mixed_0.33/gpt/trade-off_20}}
    \label{fig:uc:mix0.33:gpt}
  \end{subfigure}
  \hfill
  \begin{subfigure}[b]{0.148\textwidth}
    \setlength\figureheight{1.9in}
    \centering
    \caption{\CBT{4}+\ClassifiedGenerator}
    \vspace{2ex}
    \resizebox{\textwidth}{!}{\input{figure/mixed_0.33/chaffing4/trade-off_20}}
    \label{fig:uc:mix0.33:cbt4}
  \end{subfigure}
  \hfill
  \begin{subfigure}[b]{0.148\textwidth}
    \setlength\figureheight{1.9in}
    \centering
    \caption{\CBTr+\ClassifiedGenerator}
    \vspace{2ex}
    \resizebox{\textwidth}{!}{\input{figure/mixed_0.33/CBTr/trade-off_20}}
    \label{fig:uc:mix0.33:cbtr}
  \end{subfigure}
  \hfill
  \begin{subfigure}[b]{0.148\textwidth}
    \setlength\figureheight{1.9in}
    \centering
    \caption{\CHM+\ClassifiedGenerator}
    \vspace{2ex}
    \resizebox{\textwidth}{!}{\input{figure/mixed_0.33/CHM/trade-off_20}}
    \label{fig:uc:mix0.33:chm-pwd}
  \end{subfigure}
  \hfill
  \begin{subfigure}[b]{0.148\textwidth}
    \setlength\figureheight{1.9in}
    \centering
    \caption{\Tweak+\ClassifiedGenerator}
    \vspace{2ex}
    \resizebox{\textwidth}{!}{\input{figure/mixed_0.33/Tweak/trade-off_20}}
    \label{fig:uc:mix0.33:tweak}
  \end{subfigure}
  \hfill
  \begin{subfigure}[b]{0.148\textwidth}
    \setlength\figureheight{1.9in}
    \centering
    \caption{\PasstoPath+\ClassifiedGenerator}
    \vspace{2ex}
    \resizebox{\textwidth}{!}{\input{figure/mixed_0.33/pass2path/trade-off_20}}
    \label{fig:uc:mix0.33:pass2path}
  \end{subfigure}
  \\[-2.5ex]
    \begin{subfigure}[b]{\textwidth}
      \setlength\figureheight{2in}
        \centering
        \resizebox{!}{1.5em}{\input{xlabel.tex}}
    \end{subfigure}
    
  \caption{$\fpProbWOParams(\alarmAttacker)$
    vs.\ $\fnProbWOParams(\breachAttacker)$ as \alarmThreshold is varied,
    for the case of mixed passwords (\PercentRandom = 0.33) ($\nmbrHoneywords=19$,
    $\attackBudget=1000$).}
  \label{fig:uc:mix0.33} 
  \end{figure*}

  \begin{figure*}[t!]
    \vspace{0.1in}
    \begin{subfigure}[b]{\textwidth}
      \centering
      \resizebox{!}{2em}{\input{user_chosen_trade-off_legend.tex}}
    \end{subfigure}
    
    \begin{subfigure}[b]{0.188\textwidth}
      \setlength\figureheight{1.9in}
      \centering
      \caption{\PCFG+\ClassifiedGenerator}
      \vspace{1.25ex}
      \resizebox{\textwidth}{!}{\input{figure/mixed_0.67/PCFG/trade-off_20}}
      \label{fig:uc:mix0.67:pcfg}
    \end{subfigure}
    \hfill
    \begin{subfigure}[b]{0.148\textwidth}
      \setlength\figureheight{1.9in}
      \centering
      \caption{\Combo+\ClassifiedGenerator}
      \vspace{2ex}
      \resizebox{\textwidth}{!}{\input{figure/mixed_0.67/combination/trade-off_20}}
      \label{fig:uc:mix0.67:combo}
    \end{subfigure}
    \hfill
    \begin{subfigure}[b]{0.148\textwidth}
      \setlength\figureheight{1.9in}
      \centering
      \caption{\PCFGpwd+\ClassifiedGenerator}
      \vspace{2ex}
      \resizebox{\textwidth}{!}{\input{figure/mixed_0.67/PCFG_/trade-off_20}}
      \label{fig:uc:mix0.67:pcfg-pwd}
    \end{subfigure}
    \hfill
    \begin{subfigure}[b]{0.148\textwidth}
      \setlength\figureheight{1.9in}
      \centering
      \caption{\Combopwd+\ClassifiedGenerator}
      \vspace{2ex}
      \resizebox{\textwidth}{!}{\input{figure/mixed_0.67/combination_/trade-off_20}}
      \label{fig:uc:mix0.67:combo-pwd}
    \end{subfigure}
    \hfill
    \begin{subfigure}[b]{0.148\textwidth}
      \setlength\figureheight{1.9in}
      \centering
      \caption{\TarPCFG+\ClassifiedGenerator}
      \vspace{2ex}
      \resizebox{\textwidth}{!}{\input{figure/mixed_0.67/PCFG_target/trade-off_20}}
      \label{fig:uc:mix0.67:pcfg-tgt}
    \end{subfigure}
    \hfill
    \begin{subfigure}[b]{0.148\textwidth}
      \setlength\figureheight{1.9in}
      \centering
      \caption{\TarCombo+\ClassifiedGenerator}
      \vspace{2ex}
      \resizebox{\textwidth}{!}{\input{figure/mixed_0.67/combination_target/trade-off_20}}
      \label{fig:uc:mix0.67:combo-tgt}
    \end{subfigure}
    \\[-2.5ex]
    \begin{subfigure}[b]{\textwidth}
      \setlength\figureheight{2in}
        \centering
        \resizebox{!}{1.5em}{\input{xlabel.tex}}
    \end{subfigure}
    \vspace{1ex}
    \\
    \begin{subfigure}[b]{0.188\textwidth}
      \setlength\figureheight{1.9in}
      \centering
      \caption{\CGPT+\ClassifiedGenerator}
      \vspace{1.25ex}
      \resizebox{\textwidth}{!}{\input{figure/mixed_0.67/gpt/trade-off_20}}
      \label{fig:uc:mix0.67:gpt}
    \end{subfigure}
    \hfill
    \begin{subfigure}[b]{0.148\textwidth}
      \setlength\figureheight{1.9in}
      \centering
      \caption{\CBT{4}+\ClassifiedGenerator}
      \vspace{2ex}
      \resizebox{\textwidth}{!}{\input{figure/mixed_0.67/chaffing4/trade-off_20}}
      \label{fig:uc:mix0.67:cbt4}
    \end{subfigure}
    \hfill
    \begin{subfigure}[b]{0.148\textwidth}
      \setlength\figureheight{1.9in}
      \centering
      \caption{\CBTr+\ClassifiedGenerator}
      \vspace{2ex}
      \resizebox{\textwidth}{!}{\input{figure/mixed_0.67/CBTr/trade-off_20}}
      \label{fig:uc:mix0.67:cbtr}
    \end{subfigure}
    \hfill
    \begin{subfigure}[b]{0.148\textwidth}
      \setlength\figureheight{1.9in}
      \centering
      \caption{\CHM+\ClassifiedGenerator}
      \vspace{2ex}
      \resizebox{\textwidth}{!}{\input{figure/mixed_0.67/CHM/trade-off_20}}
      \label{fig:uc:mix0.67:chm-pwd}
    \end{subfigure}
    \hfill
    \begin{subfigure}[b]{0.148\textwidth}
      \setlength\figureheight{1.9in}
      \centering
      \caption{\Tweak+\ClassifiedGenerator}
      \vspace{2ex}
      \resizebox{\textwidth}{!}{\input{figure/mixed_0.67/Tweak/trade-off_20}}
      \label{fig:uc:mix0.67:tweak}
    \end{subfigure}
    \hfill
    \begin{subfigure}[b]{0.148\textwidth}
      \setlength\figureheight{1.9in}
      \centering
      \caption{\PasstoPath+\ClassifiedGenerator}
      \vspace{2ex}
      \resizebox{\textwidth}{!}{\input{figure/mixed_0.67/pass2path/trade-off_20}}
      \label{fig:uc:mix0.67:pass2path}
    \end{subfigure}
    \\[-2.5ex]
      \begin{subfigure}[b]{\textwidth}
        \setlength\figureheight{2in}
          \centering
          \resizebox{!}{1.5em}{\input{xlabel.tex}}
      \end{subfigure}
  \caption{$\fpProbWOParams(\alarmAttacker)$
    vs.\ $\fnProbWOParams(\breachAttacker)$ as \alarmThreshold is varied,
    for the case of mixed passwords (\PercentRandom = 0.67) ($\nmbrHoneywords=19$,
    $\attackBudget=1000$).}
  \label{fig:uc:mix0.67} 
  \vspace{0.1in}
  \end{figure*}

In this section, we aim to investigate how the percentage of use of password managers
in password creation affects the efficacy of honeywords under our threat model.

\myparagraph{Mixed datasets construction}
To do so,  we constructed two test datasets by mixing up  
\Dataset{\TestTag}{\UserChosenTag} and the algorithmically generated 
dataset as follows: 
for a multiset \passwordSet{\emailAddr} from the processed
test dataset \Dataset{\TestTag}{\UserChosenTag} (described in \secref{sec:user-selected:evaluation:dataset}),
we uniformly at random sampled a password generator from $\{\choosePassword{\GeneratorIndex}\}_{\GeneratorIndex =
1}^{\NumGenerators}$ and randomly replace each password 
in the multiset by a password generated from sampled password generator independently with a probability of \PercentRandom. We chose 
\PercentRandom to be $0.33$ and $0.67$. As such, we constructed two mixed datasets denoted as
\Dataset{0.33}{\mixedTag} and \Dataset{0.67}{\mixedTag}. To evaluate false-positive and false-negative of the honeyword methods, 
we implemented the algorithm \choosePassword by choosing \password and the members of \auxiliaryInfo without 
replacement from a single multiset, chosen uniformly at random from the mixed 
test datasets \Dataset{0.33}{\mixedTag} and \Dataset{0.67}{\mixedTag}.

\myparagraph{Honeyword generation}
Given an account password \password from 
\Dataset{0.33}{\mixedTag} and \Dataset{0.67}{\mixedTag}, we generated honeywords
based on the following rule: if \password is an algorithmically generated password,
we generated honeywords by \ClassifiedGenerator (described in \secref{sec:algorithmic:strategies}); if 
\password is a user-chosen one, we generated honeywords by a method described in \secref{sec:background:algorithms}.
Here we assumed that user-chosen passwords and algorithmically generated passwords
can be distinguished easily.

\myparagraph{Attack strategies}
We implemented the false-positive attacker following the one
described in \secref{sec:user-selected:attack-strategy}. We implemented the false-negative
attacker as follows: given the sweetwords $\honeywordSet \cup \{\password\}$, 
if most of sweetwords are algorithmically generated, the false-negative attacker
followed the strategy described in \secref{sec:algorithmic:strategies} to guess the account
password; if most of sweetwords are user-chosen, he followed the strategies described
in \secref{sec:user-selected:attack-strategy}.

\myparagraph{Experimental results}
We report
$\fpProbWOParams(\alarmAttacker)$ and $\fnProbWOParams(\breachAttacker)$
on those honeyword-generation algorithms for the attackers \alarmAttacker and
\breachAttacker described above. To
depict the tradeoffs between $\fpProbWOParams(\alarmAttacker)$ and
$\fnProbWOParams(\breachAttacker)$, we plot them against one another as
\alarmThreshold is varied.  \figref{fig:uc:mix0.33} (\figref{fig:uc:mix0.67}) shows these tradeoffs
when $\nmbrHoneywords = 19$ honeywords, $\attackBudget = 1000$, and $\PercentRandom=0.33$ ($\PercentRandom=0.67$)
where circles ($\bullet$) mark the $\fpProbWOParams(\alarmAttacker)$
vs.\ $\fnProbWOParams(\breachAttacker)$ tradeoff at specific values of
\alarmThreshold ranging from $\alarmThreshold = 1$ to \nmbrHoneywords
when $\auxiliaryInfo=\passwordSet{\emailAddr} \setminus \{\password\}$
in each plot. Again, we stress that $\attackBudget = 1000$ yields an
optimistic evaluation of $\fpProbWOParams(\alarmAttacker)$ since
$\attackBudget = 1000$ is $1000\times$ too small compared with the
number $10^6$ recommended by Flor\^{e}ncio, et
al.~\cite{florencio2014:guide}. Compared with those results from user-chosen
case (depicted in \figref{fig:uc:to20}), honeyword-generation methods achieved
better trade-offs between $\fpProbWOParams(\alarmAttacker)$ and
$\fnProbWOParams(\breachAttacker)$ in \Dataset{0.33}{\mixedTag} and \Dataset{0.67}{\mixedTag}.
When we set \PercentRandom to be higher, we have better trade-offs because a higher 
\PercentRandom makes less password reuse across sites. From these results,
we concluded that honeyword efficacy benefits from the tendency of using password manager in
password creation.